\documentclass{aa}  

\usepackage{graphicx}
\usepackage{color, soul}
\usepackage{txfonts}
\usepackage{natbib}
\usepackage[FIGTOPCAP, center, nooneline]{subfigure}
\begin{document} 

   \title{Yebes 40 m radio telescope and the broad band NANOCOSMOS receivers at 7 mm and 3 mm for line surveys}

   \author{F. Tercero\inst{1}
           \and J.~A. L\'opez-P\'erez\inst{\ref{inst1}}
           \and J.~D. Gallego\inst{\ref{inst1}}
           \and F. Beltr\'an\inst{\ref{inst1}}
           \and O. Garc\'ia\inst{\ref{inst1}}        
           \and M. Patino-Esteban\inst{\ref{inst1}}
           \and I. L\'opez-Fern\'andez\inst{\ref{inst1}}
           \and G. G\'omez-Molina\inst{\ref{inst1}}
           \and M. Diez\inst{\ref{inst1}}
           \and P. Garc\'\i a-Carre\~no\inst{\ref{inst1}}
           \and I. Malo\inst{\ref{inst1}}
           \and R. Amils\inst{\ref{inst1}}
           \and J.~M. Serna\inst{\ref{inst1}}
           \and C. Albo\inst{\ref{inst1}}
           \and J.~M. Hern\'andez\inst{\ref{inst1}}
           \and B. Vaquero\inst{\ref{inst1}}
           \and J. Gonz\'alez-Garc\'\i a\inst{\ref{inst1}}
           \and L. Barbas\inst{\ref{inst1}}
           \and J.~A. L\'opez-Fern\'andez\inst{\ref{inst2}}
           \and V. Bujarrabal\inst{\ref{inst3}}
           \and M. G\'omez-Garrido\inst{\ref{inst1},\ref{inst3}}
           \and J.~R. Pardo\inst{\ref{inst4}}
           \and M. Santander-Garc\'\i a\inst{\ref{inst1},\ref{inst3}}
           \and B. Tercero\inst{\ref{inst1},\ref{inst3}}
           \and J. Cernicharo\inst{\ref{inst4}}
	   \and P. de Vicente\inst{\ref{inst1}}\thanks{corresponding authors. email:p.devicente@oan.es, jose.cernicharo@csic.es}
        }
   \institute{Centro de Desarrollos Tecnol\'ogicos, Observatorio de Yebes (IGN), 19141 Yebes, Guadalajara, Spain. \label{inst1}
     \and Instituto Geogr\'afico Nacional (IGN), C/General Iba\~nez de Ibero, 3, 28003, Madrid, Spain.\label{inst2}
     \and Observatorio Astron\'omico Nacional (IGN), C/ Alfonso XII, 3, 28014, Madrid, Spain.\label{inst3}
     \and Instituto de Física Fundamental (IFF-CSIC). Group of Molecular Astrophysics, C/ Serrano 123, 28006 Madrid, Spain. \label{inst4}
             }
   \date{Received ...; accepted ...}
   
   \titlerunning{Broad band Nanocosmos receivers at Yebes 40\,m RT}
   \authorrunning{F. Tercero et al.}
   
 
  \abstract
   {Yebes 40\,m radio telescope is the main and largest observing instrument at Yebes Observatory and it is devoted to Very Long Baseline Interferometry (VLBI) and single dish observations since 2010. It has been covering frequency bands between 2\,GHz and 90\,GHz in discontinuous and narrow windows in most of the cases, to match the current needs of the European VLBI Network (EVN) and the Global Millimeter VLBI Array (GMVA).}
   {Nanocosmos project, a European Union funded synergy grant, opened the possibility to increase the instantaneous frequency coverage  to observe many molecular transitions with single tunnings in single dish mode. This reduces the observing time and maximises the output from the telescope.}
   {We present the technical specifications of the recently installed $31.5-50$\,GHz (Q band) and $72-90.5$\,GHz (W band) receivers along with the main characteristics of the telescope at these frequency ranges. We have observed IRC+10216, CRL\,2688 and CRL\,618, which harbour a rich molecular chemistry, to demonstrate the capabilities of the new instrumentation for spectral observations in single dish mode.}
   {The results 
   show the high sensitivity of the telescope in the Q band. The spectrum of IRC+10126 offers a signal to noise ratio never seen before for this source in this band. On the other hand, the spectrum normalised by the continuum flux towards CRL\,618 in the W band demonstrates that the 40~m radio telescope produces comparable results to those from the IRAM 30~m radio telescope, although with a smaller sensitivity. The new receivers fulfil one of the main goals of Nanocosmos and open the possibility to study the spectrum of different astrophysical media with unprecedented sensitivity.}
   {}


   \maketitle
%
\section{Introduction}

The 40\,m radio telescope has been observing in Q band, between 41\,GHz and 49\,GHz, with an instantaneous 500 MHz bandwidth since 2013 \citep{agundez15,devicente16}
and with a 2.5\,GHz IF bandwidth since 2016. It achieved an instantaneous band of 9\,GHz in 2017 \citep{fuente2019}
with the installation of 8 Fast Fourier Transform Spectrometers (FFTSs) within the frame of the Nanocosmos project\footnote{\texttt{https://nanocosmos.iff.csic.es/technology}}, a European Synergy Grant.
The W band receiver using a Superconductor-Isolator-Superconductor mixer, with an instantaneous band of 600\,MHz, has been in use since 2010 mainly for VLBI observations (see for example \citealt{issaoun19}).
In 2019, new Q and W band receivers were designed and manufactured at Yebes Observatory with funds from the Nanocosmos project. 
This paper summarises the changes implemented in 2019 and some astronomical results which demonstrate the capabilities of the new receiving system around 7\,mm and 3\,mm wavelengths \citep{Cernicharo2019b,Tercero2020}.

The technical section of this paper (Sect.\,\ref{sect_the_telescope}) describes the new additions to the telescope: two cryogenic receivers for the $31.5-50$\,GHz and the $72-90.5$\,GHz frequency bands, a new optical circuit for the W band receiver with its mirrors, new mirrors for the Q band receiver, and a new hot-cold load calibration system. In addition, this section offers the description of the whole system: primary reflector, tertiary optics, cryogenics, low noise amplifiers (LNAs), frequency downconversion, and backends.
Section\,\ref{sect_performance_receivers} is devoted to presenting the performance of the new receivers installed at the 40\,m.
Sections\,\ref{sect_software} and \ref{sect_calibration} describe the implemented software control and calibration procedures, respectively. In Sect.\,\ref{sect_eficiencias} we focus on the derivation of the telescope efficiencies at Q and W bands.
Finally, Sect.\,\ref{sect_survey} provides some general results of the first astronomical observations with these new receiver systems, since more specific and detailed results will be published elsewhere.

\section{The telescope}
\label{sect_the_telescope}

The 40 metre diameter radio telescope at Yebes Observatory \citep{lopezfernandez2006} is a classical Nasmyth-Cassegrain system with a main parabolic dish of 15~m focal distance and a hyperbolical 3.28~m diameter secondary mirror with 26.6\,m inter-foci distance. The edge illumination at the sub-reflector is 12\,dB at 3.621$^\circ$ and it is responsible for the highly directive feed system in the equivalent Cassegrain focus of the antenna. In order to keep a constant aperture efficiency (see Sect.\,\ref{sect_eficiencias}) the required beam waist needs to be 5.9$\lambda$, being $\lambda$ the wavelength. The usage of an additional tertiary optics allows feasible designs and practical sizes for the feed horns, and the installation of multiple receivers.

\subsection{The primary reflector}
\label{sect_holography}

The primary 40 metre reflector is composed of 420 aluminium panels arranged in 10 concentric rings. Most of the panels are supported by 4 screws, except the larger ones which are supported by 6 screws. These screws can be adjusted using a special tool that allows lifting, lowering, or tilting the panels. The root-mean-square (RMS) error for each panel is below 80\,$\mu$m, according to photogrammetric measurements performed by the manufacturer. The best surface error accuracy, according to a model from the antenna designer and when the effects of gravity, wind and thermal gradients are considered, is 150\,$\mu$m RMS at 45$^{\circ}$ elevation under good weather conditions. The back structure of the primary reflector is covered with a cladding to avoid direct Sun exposure. The cladding is not hermetic and, to create a uniform distribution of temperature all over the structure, three fans circulate the air in a tangential direction. 

The surface accuracy of the 40~m radio telescope can be measured with the help of a prime focus microwave coherent holography receiver developed at the laboratories of Yebes Observatory. The details of this procedure are described in \cite{lopezperez12} and \cite{lopezperez14}, along with observational results after surface adjustments.

\begin{figure}
  \includegraphics[width=9cm]{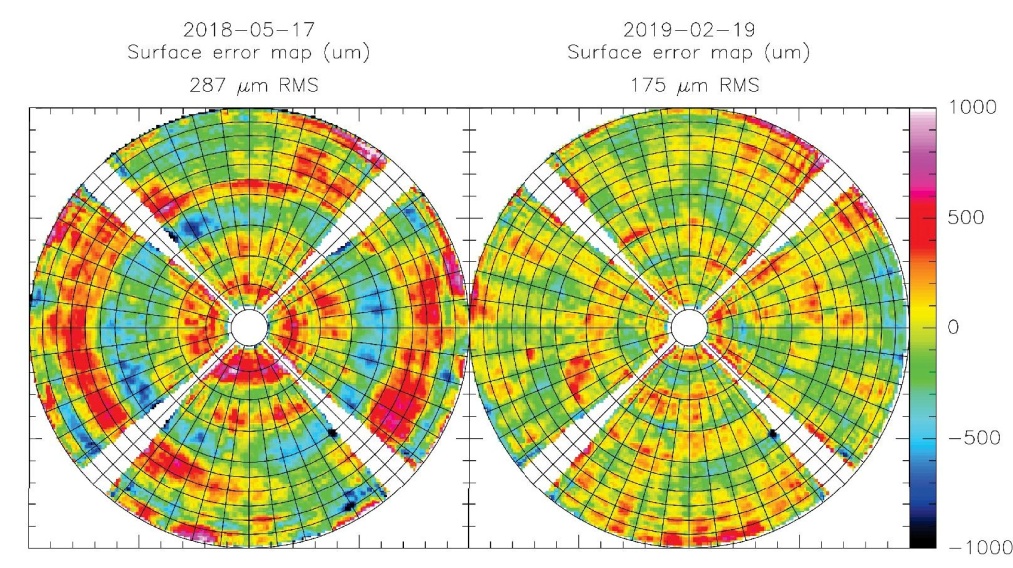}
  \centering
  \caption{The 40\,m surface accuracy before (left panel) and after (right panel) the panel adjustment performed during the summer of 2018. Large scale deformations were substracted and observations were performed at 43$^\circ$ elevation. The colour scale is given in microns.}
   \label{f.holography}
\end{figure}

The latest surface optimisation, performed via panel adjustment, was carried out in summer 2018. Figure\,\ref{f.holography} shows a comparison between the surface accuracy before, May 2018 (left panel) and after, February 2019 (right panel), both measured at 43$^\circ$ elevation. The surface accuracy (RMS) improved from 287\,$\mu$m to 175\,$\mu$m, once large scale deformations were substracted. Annual holographic measurements are foreseen to monitor changes in the main reflector. In addition we plan to install an array of temperature sensors at its back structure and at the tetrapod to study the effects of thermal gradients on the telescope's structure. 

\subsection{The receiver cabin}
\label{sect_cabin}

 The receivers, together with the optical elements described in the next sections, are installed in the receiver cabin, which is located in the turning head of the telescope. The cabin, a $\simeq$\,8.8\,m\,$\times$\,9.2\,m\,$\times$\,3.5\,m room, has a circular aperture 3.2\,m in diameter that communicates with the vertex of the main reflector and through which the radiation passes. The vertex is protected with 8 aluminium petals that are remotely operated and open outwards.
 In the aperture there is a 9\,mm thick membrane made of polyethylene foam  which isolates the cabin from the outside. The membrane forms an angle of 25 degrees with the axis of the paraboloid. To avoid water condensation on its inner surface two fans constantly inject air towards the membrane. The membrane has been characterised at the laboratory and the losses are 0.1\,dB across the whole W band and below 0.1\,dB for lower frequencies.
 
\begin{figure*}
  \includegraphics[width=14cm]{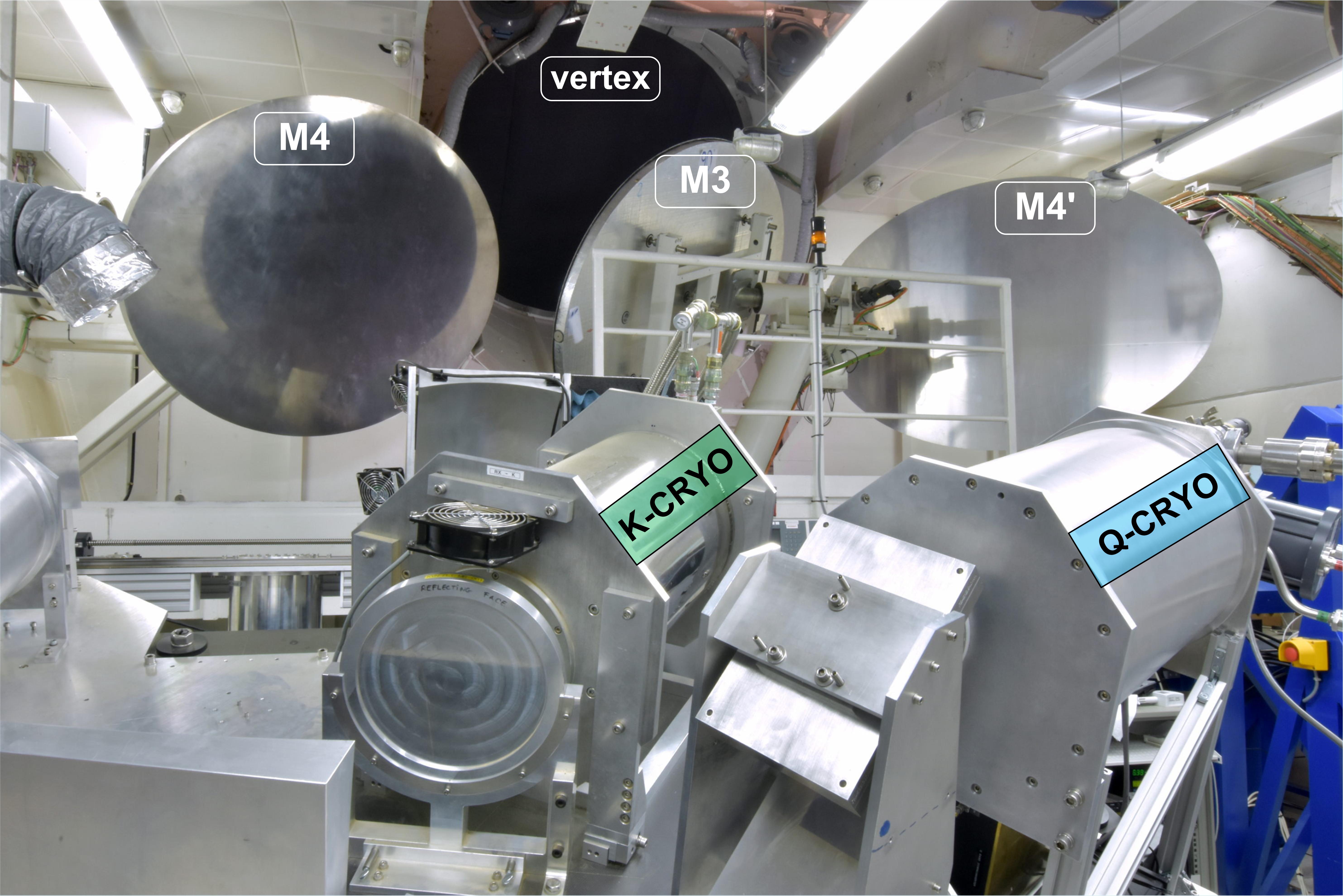}
  \centering
  \caption{Receivers cabin. View from the K, Q, W optical table towards the vertex (black circle on the back). From left to right: Nasmyth mirrors M4, M3 and M4'. M3 is pointing towards M4.}
   \label{f.rxcabin}
\end{figure*}

 The receiver cabin hosts a mirror (M3) that moves synchronously with the antenna in elevation and directs radiation to either one of the two M4 mirrors. This allows having two branches in the cabin to which different receivers are associated. The Nanocosmos receivers presented in this paper are all in branch M4 whereas the other branch is devoted to receivers for frequencies below 20 GHz. Figure\,\ref{f.rxcabin} shows a view of the membrane and the large Nasmyth mirrors M3, M4, and M4' from the K, Q, and W optical table (see Sect.\,\ref{sect_optics}).

 The receiver cabin is thermalized using two air conditioning systems. Currently the temperature is kept between 18$^{\circ}$C and 22$^{\circ}$C during most of the year except in summer, when the temperature may oscillate between 21$^{\circ}$C and 26$^{\circ}$C inside the cabin. The compressors of the cryostats are installed in the receiver cabin but in a separate room with forced ventilation outwards to avoid extra heating. A better control of the temperature of the cabin is under way and will probably be implemented in the next months.
 
\subsection{Optics}
\label{sect_optics}
The tertiary optics in the focal zone is a beam wave guide (BWG) system which drives and transforms the high directivity beams at the Cassegrain focus to smaller directivity beams at the cryostat side. The beam is redirected through several mirrors to the K, Q, and W band receivers.

\begin{figure}
  \includegraphics[width=9cm]{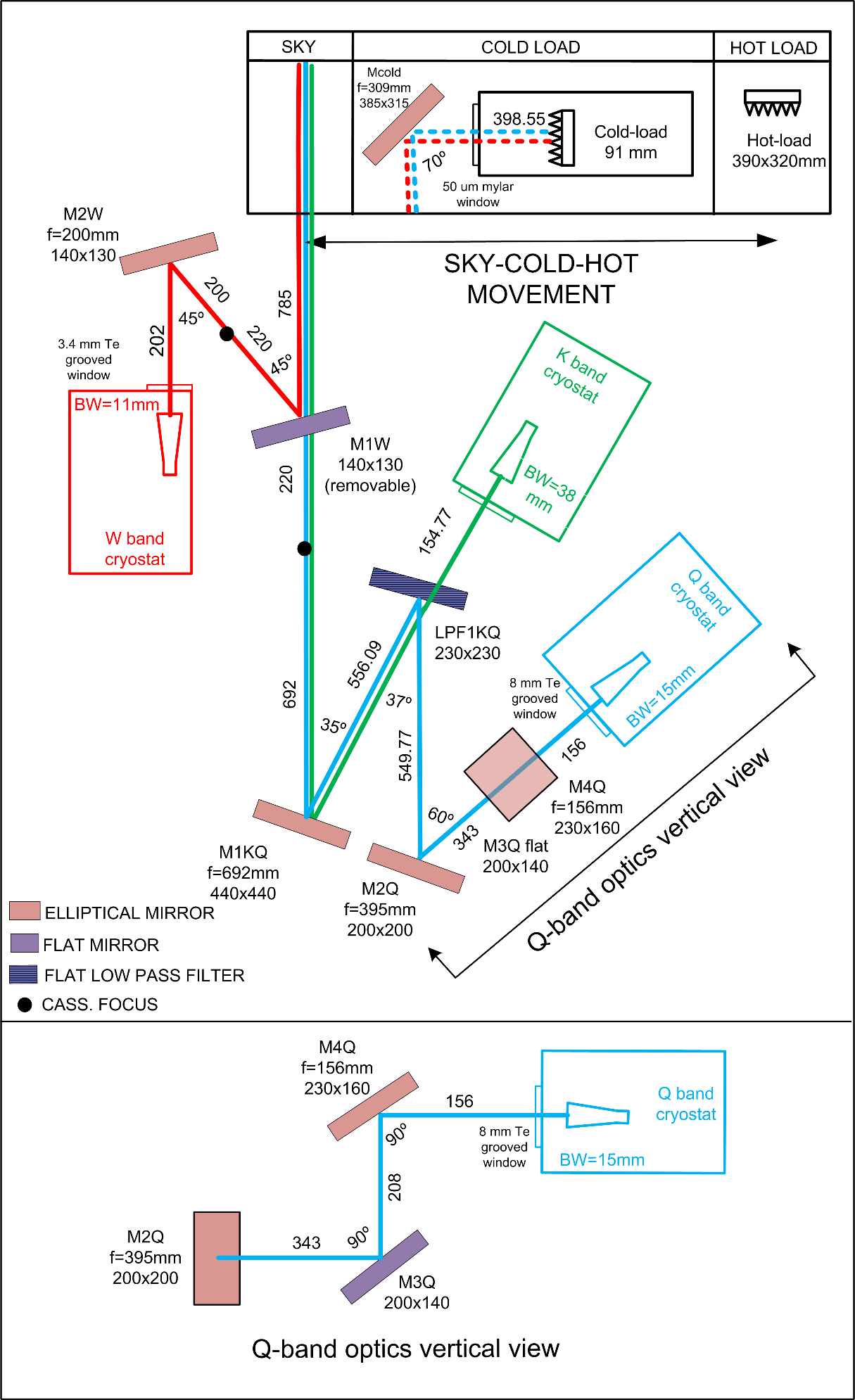}
  \centering
  \caption{Layout of the new optical system for the K, Q, and W band receivers at Yebes 40\,m radio telescope.}
   \label{f.opticsL}
\end{figure}

\begin{figure}
  \includegraphics[width=9cm]{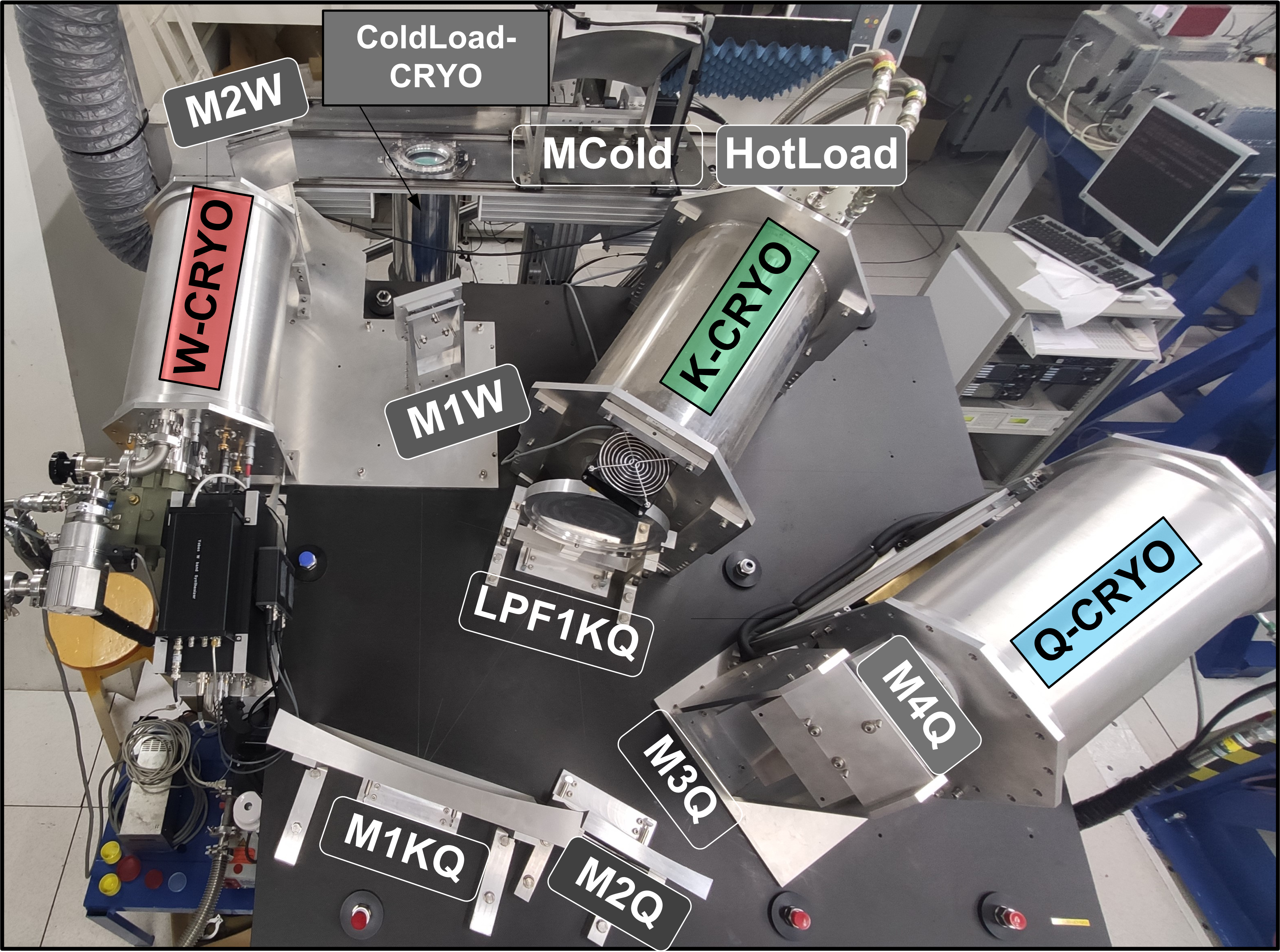}
  \centering
  \caption{Photograph of the optical table where the K, Q, and W band receivers are placed. The view from above allows to identify each optical element, the hot-cold load calibration system and the cryostats.}
   \label{f.opticsP}
\end{figure}

The feed system for each frequency is independently calculated using the procedure described in \citet{chu1983} and \citet{tercero2019}, where an image of the sub-reflector is created using a mirror. The image can be transformed using a cascade of mirrors that preserves the frequency independence of the image. 
Figure\,\ref{f.opticsL} shows the beam transformation, as it goes through the optical system,
for the K, Q, and W bands. The K and Q band optical paths share mirror M1KQ. This mirror creates a frequency independent image of the sub-reflector in the K band feed position with a beam radius ($wf$) of 38\,mm . Figure \ref{f.opticsP} shows a photograph of the optical table where the K, Q, and W receivers are placed. The mirrors, hot-cold load calibration system, and the cryostats can be easily identified.

To carry out Q band observations, filter LPF1KQ (in case of K and Q band simultaneous observations) or a total reflection mirror at the same position (for single Q band observations) is required (see Fig.\,\ref{f.opticsL} and \ref{f.opticsP}). For the Q band beam, the sub-reflector image stands at 154.77\,mm after reflecting on LPF1KQ or on the total reflection mirror. A Gaussian beam telescope (GBT), with M2Q and M4Q elliptical mirrors, transforms the subreflector image from LPF1KQ ($wf=38$\,mm at 154.77\,mm) to a $wf=15$\,mm subreflector image at 156\,mm from M4Q. An additional flat mirror (M3Q) is used to redirect the beam coming from the feed in vertical direction.

The W band optical circuit consists of a set of two mirrors. M1W flat mirror diverts the Cassegrain focus, shifting it 220\,mm away from M1W, in the direction of the M2W elliptical mirror. M2W has been designed to provide a focal distance which transforms the beam to $wf=11$\,mm. This distance is large enough to install the mechanical structure of the W receiver. M1W flat mirror will be replaced in the short term by a dichroic filter with W band being reflected and K and Q band being transmitted, allowing simultaneous observations at the three frequency bands.

We have used two approaches to design and analyse the complete optical system, and to estimate the theoretical illumination efficiency of the antenna: Quasioptical Gaussian beam fundamental mode (QO-GBFM; see e.g. \citealt{Goldsmith1998}) and physical optics\footnote{For this approximation we used General Reflector Antenna Software Package (GRASP), TICRA, Copenhagen, Denmark, \texttt{www.ticra.com}.} (PO). The PO method provides a more accurate estimation of the illumination efficiency based on numerical procedures, whereas QO-GBFM uses an analytical method that considers mirrors as thin lenses. Both methods provide very similar values for the illumination efficiency \citep{tercero2019}, as summarised in Table\,\ref{t.optmethods}.

\begin{table}
	\caption{Efficiency due to illumination (assuming no blockage and a primary reflector without surface errors) derived from QO-GBFM and PO methods at Q and W bands. The QO-GBFM method uses the theoretical designed feed patterns and an analytical method. The PO method uses the real feed patterns and a numerical procedure.}
\label{t.optmethods}
\centering
\begin{tabular}{c c c}
\hline\hline
Method & Q band (43 GHz) & W band (87 GHz) \\
\hline
QO-GBFM & 0.81 & 0.81 \\
PO      & 0.80 & 0.82 \\
\hline
\end{tabular}
\end{table}

The hot-cold load system is located at a distance of 1005\,mm from the Cassegrain focus. A servo-motor switches between three states: hot load, cold load, and sky. The hot load is a low reflectivity microwave absorber from ECCOSORB HPY series VHP-2, composed of pyramidal structures with a 40\,mm square base and tilted 35$^\circ$ with respect to the incident beam to minimise standing waves. The cold load is an array, 91\,mm in diameter, formed by small (4\,$\times$\,4\,$\times$\,20\,mm) pyramidal structures made of ECCOSORB MF117 inside a small cryostat and cooled down to 17\,K. 
The measured return loss of the structure is better than $-$22\,dB in the W band. The cryostat window uses a 50\,$\mu$m Mylar membrane that bulges with vacuum. In order to have a small cryostat and a reduced window, an optical system reduces the Cassegrain beam waist of the telescope from 41\,mm to 20\,mm at Q band while keeping this parameter similar in W band (from 18~mm to 20~mm).

\subsection{The receivers}

\subsubsection{Cryogenics}
\label{sect_cryostats}
\begin{figure}
  \includegraphics[width=8cm]{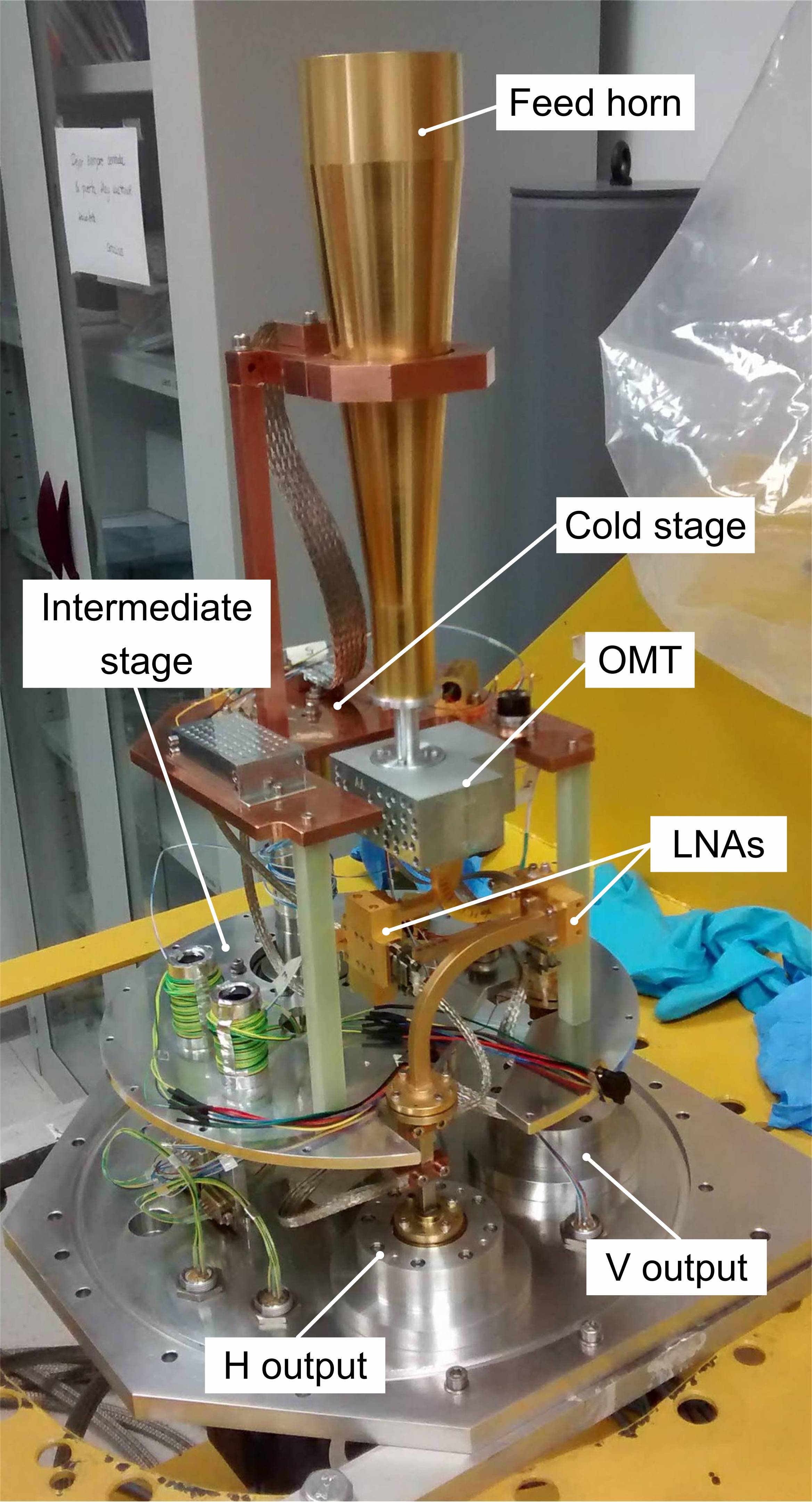}
  \centering
   \caption{Photograph of the Q band cryostat. The most important parts are tagged to facilitate their identification}
   \label{f.cryoQ}
\end{figure}

\begin{figure}
 \includegraphics[width=8cm]{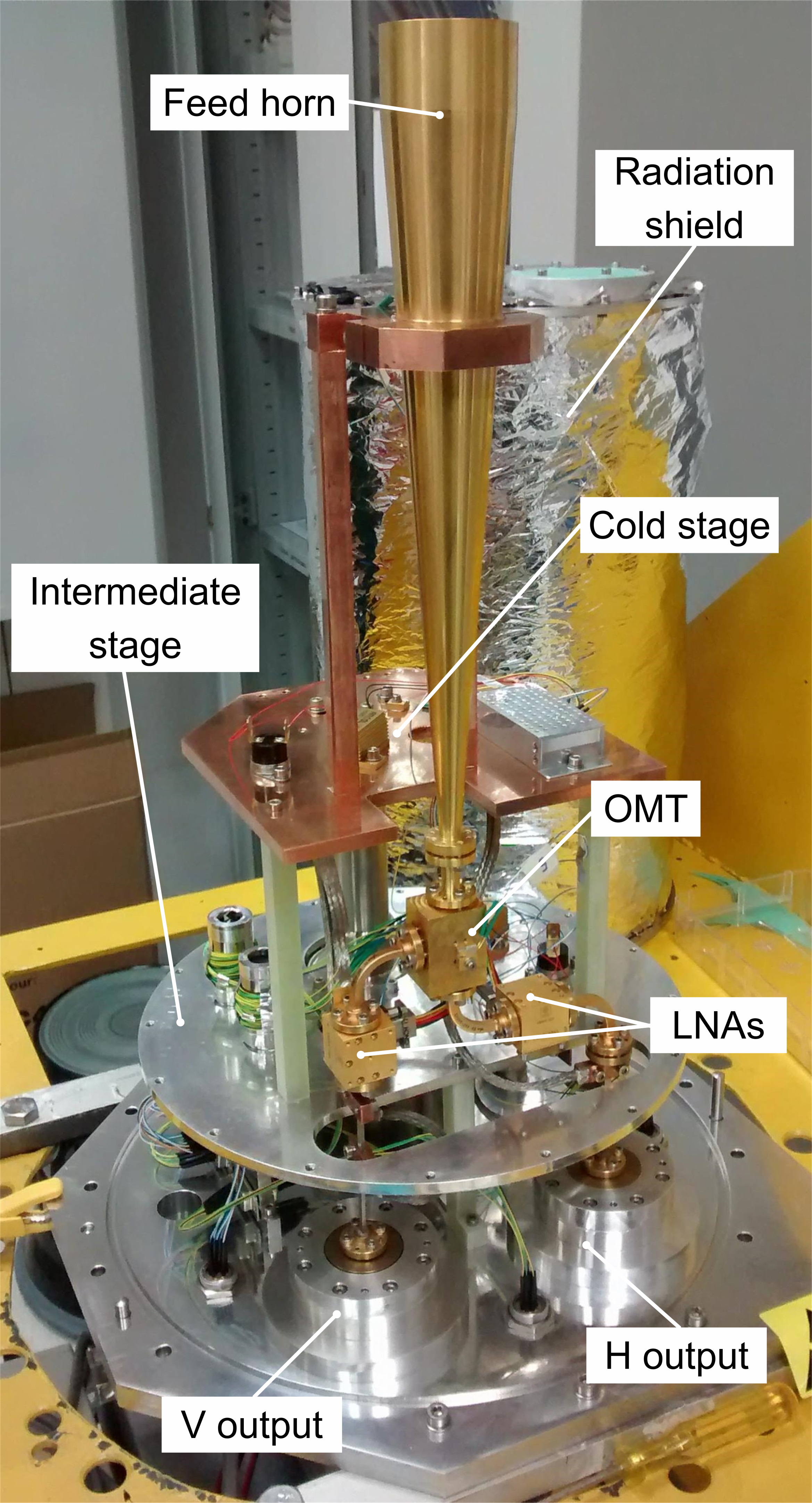}
  \centering
   \caption{Photograph of the W band cryostat. The most important parts are tagged to facilitate their identification}
   \label{f.cryoW}
\end{figure}

Both Q and W band cryostats use a two-stage GM cryo-cooling system, model 350 from CTI-Cryogenics. The dewar is a 1\,cm-thick hollow cylinder made of aluminium alloy. The external dimensions of the dewar are 514\,mm-length and 310\,mm-diameter and its internal volume is about 30\,L. The cylinder is closed with two covers. The front cover has a corrugated polytetrafluorethylene (PTFE) window through which the radio frequency (RF) radiation passes towards the feed horn, while keeping the vacuum inside. The back cover is machined to provide interfaces for the vacuum system, input/output electric connections, two rectangular waveguide outputs (one per polarisation), DC connections for housekeeping, and LNA bias. 
Figures\,\ref{f.cryoQ} and \ref{f.cryoW} show a photograph of the elements inside the cryostats for Q and W bands, respectively.

Inside the dewar there is a thinner cylinder attached to the intermediate stage of the cooling head. This cylinder is a radiation shield covered with multilayer isolation (MLI) sheets to reduce the thermal radiation that arrives at the inner cold stage. Between the outer corrugated PTFE window and the feed there is an intermediate window covered with an infrared filter made of 3 layers of extruded polystyrene foam (3\,mm-thick each).

The intermediate and cold stages of the cooling head are rigidly connected to an intermediate aluminium alloy plate and to a cold copper plate respectively. The measured physical temperatures at the intermediate and cold stages are $\simeq$\,70\,K and 15\,K, respectively. The LNAs are kept at a physical temperature of 15\,K using flexible copper thermal straps attached to the cold plate. There is also a rigid copper structure attached to the cold plate that supports the horn feed and keeps it at 15\,K. Thin layers of indium are used to improve the thermal conductivity at the critical joints between the cold head, the cold and intermediate plates, the thermal straps, and the LNAs.

\subsubsection{RF Chain}
\begin{figure}
  \includegraphics[width=8cm]{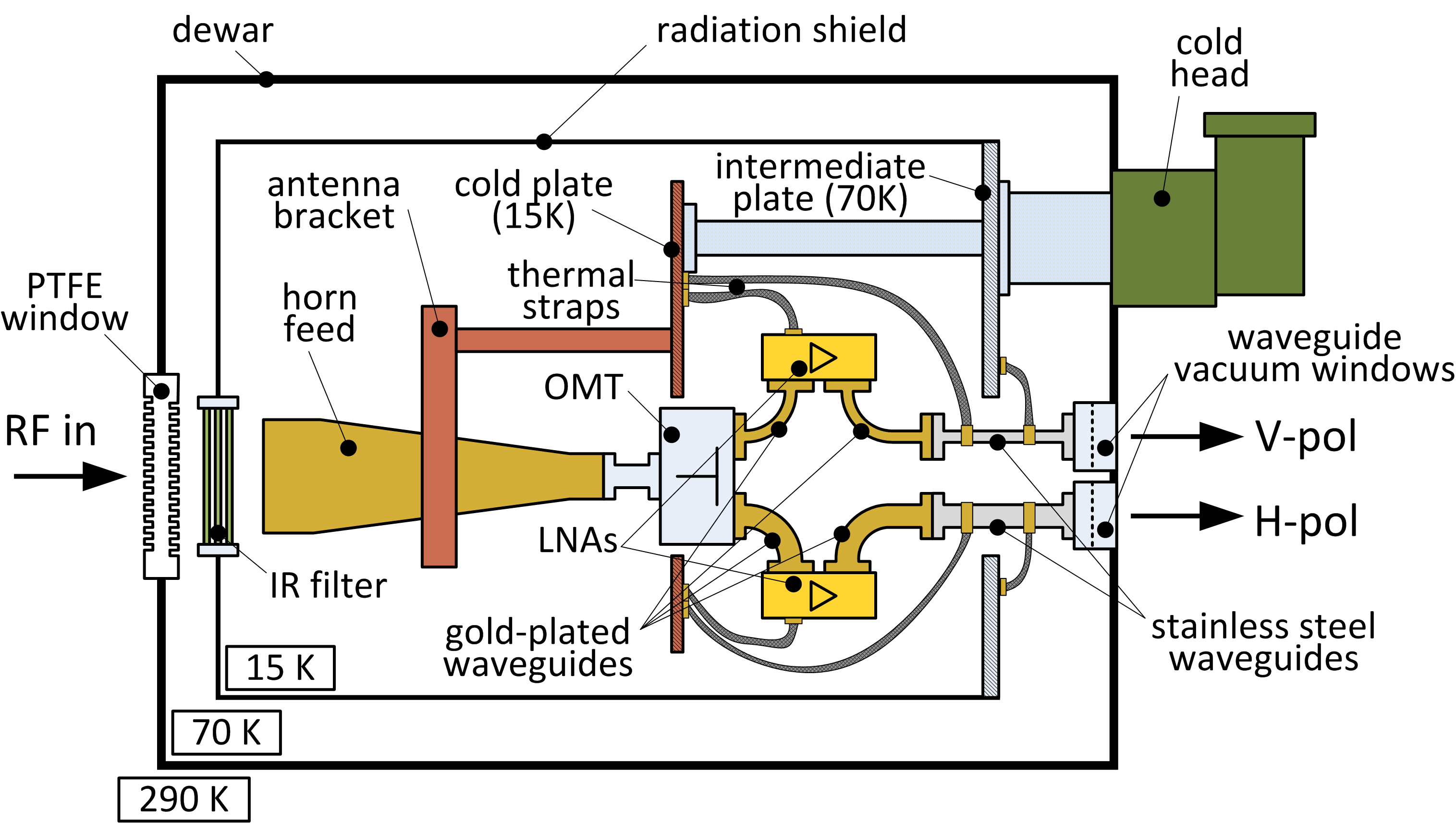}
  \centering
  \caption{Schematic diagram of the Q and W band cryogenic receivers showing the main elements inside and their connections to the two cold stages.}
   \label{f.insidecryo}
\end{figure}

The RF signal chain follows a similar layout in both bands: a corrugated PTFE vacuum window, a corrugated horn feed, a circular waveguide transition, an orthomode transducer (OMT), two LNAs (one per polarisation),
standard gold plated copper waveguides, a short section of thin wall stainless steel waveguide (thermal transition), and a waveguide vacuum window. The waveguide sizes are WR-22 and WR-10 for Q and W band, respectively. 
Figure \ref{f.insidecryo} shows a schematic diagram of these receivers.

The corrugated PTFE windows with rectangular profiled grooves providea theoretical insertion loss of 0.02\,dB in average and a return loss better than $-$20\,dB at Q and W bands. Such loss has been estimated from the permittivity 
and loss tangent measurements of the bulk material and the physical thickness of each window. The diameters of the windows are four times the beam waist to avoid truncation. Details of the design dimensions are shown in Table\,\ref{t.windowchars}.

\begin{table*}
\caption{Characteristics of the cryostat windows.}
\label{t.windowchars}
\centering
\begin{tabular}{cccccccc}
\hline\hline
Band & Diameter & Total thickness & Teeth depth & Teeth width & Groove cycle & Relative &  Loss \\
     &  (mm)    &   (mm)          &  (mm)       & (mm)        &   (mm)       & permittivity & tangent \\
\hline
Q & 88.5 & 8.0 & 1.5 & 1.5  & 3.0 & 2.05 & 5\,$\times$\,$10^{-4}$ \\
W & 72.1 & 1.9 & 0.7 & 0.75 & 1.4 & 2.05 & 5\,$\times$\,$10^{-4}$ \\
\hline
\end{tabular}
\end{table*}

The feeds are corrugated conical horn antennas designed to yield a frequency independent beam waist of 15~mm and 11~mm at Q and W bands, respectively. The feed profile follows a spline-type curve, which is optimised to minimise reflections and to achieve the required beam frequency stability. The final uncertainties for the beam radius parameters are below 1\,mm, the return loss is below $-$20\,dB, and the cross-polar peak is below $-$30\,dB in the whole bandwidth for both feeds \citep{Tercero2018,tercero2019}. The feeds were made from copper using the electroforming technique and the inner walls were gold plated. The transition between the horn antenna and the OMT required a smooth round-to-square custom transition made from gold-plated copper for W band, whereas an aluminium standard circular waveguide transition was used for Q band.

The Q band OMT is based on a turnstile waveguide design with a stepped cylindrical scattering element, in which the divided signals are combined in phase opposition via E-plane couplers \citep{Navarrini2006}. The component consists of four blocks, apart from the scattering element, made from aluminium alloy without any surface plating. The average insertion loss measured at ambient temperature is about 0.4\,dB, the reflection at every port is better than $-$18\,dB and the isolation is better than 25\,dB over the whole bandwidth, from 31.5\,GHz to 50\,GHz \citep{lopezruiz2017}. Although this OMT was not directly characterised at cryogenic temperatures, 
a reduction of the insertion loss is expected when cooled.

The W band OMT is based on a T-shaped waveguide in which the two orthogonal linear polarisations coming from a square input waveguide are split into two separated rectangular waveguides \citep{Dunning2009}. The OMT is not fully symmetric since one of the polarisations is frequency limited by the propagation of higher order modes. The final device, made from gold plated brass, is very compact and was manufactured with a simple split block technique. Its response is suitable for a moderate frequency bandwidth such as 1.25:1. The insertion loss measured at average ambient temperature is 0.6\,dB, the reflection at every port is better than $-$17\,dB, and the isolation is better than 35\,dB over the whole bandwidth, from 72\,GHz to 90.5\,GHz \citep{garciaperez2018}. The insertion loss was also measured at 20\,K yielding an improvement of $\simeq$ 0.3\,dB.

The OMT, LNAs, and the output waveguide vacuum windows are connected with commercial standard waveguides, made from gold-plated copper which minimises the electrical loss, except at the last part of the chain where stainless-steel straight waveguide sections about 60\,mm long are used. The waveguides are thermally connected to the 70\,K and 15\,K cryostat stages at both edges. This setup and the lower thermal conductivity of the steel reduces the heat conduction towards the LNAs although the lower electrical conductivity of the steel increases the electrical loss up to 0.4\,dB at Q band and 1.5\,dB at W band.

The output interfaces between the vacuum of the cryostat and the atmosphere (waveguide windows) consists of two short copper waveguide sections with a thin 10\,$\mu$m Mylar film in between. The loss introduced by the dielectric gap is lower than 0.1\,dB for both bands (Q and W) which is negligible compared with the losses of the previous stainless steel waveguide sections used for the thermal transition.

\subsubsection{Amplifiers}
\label{sect_amplifiers}
The cryogenic low noise Q band amplifiers used in the receiver front-end are based on a hybrid (discrete components) design developed at Yebes Observatory with four stages of $0.1\times50$\,$\mu$m InP high electron mobility transistor (HEMT) devices produced by Diramics\footnote{Diramics AG, Hinterbergstrasse 28, 6312 Steinhausen, Switzerland. http://diramics.com}. Each amplifier contains four transistor stages which allow individual biasing for optimizing the performance. The internal matching networks were produced in-house by laser milling on a 0.127\,mm thick DUROID$^{\rm TM}$ 6002 microwave substrate. The boxes were machined on 6082 aluminium alloy blocks and gold plated with soft gold. The modules were conceived to be configured either for 2.4\,mm coaxial input/output or for WR22 waveguides interface. The waveguide configuration was preferred for this application since it provides less losses and a direct interface to the other components of the receiver. These hybrid Q band amplifiers were selected for the receiver since they offer higher gain and lower noise than the previous microwave monolithic integrated circuit (MMIC) version used in the receiver described in \cite{Cernicharo2019a}. Additional details of the construction and measurements can be found in \cite{diez19a} and \cite{diez19b}.

\begin{figure}
  \includegraphics[width=8cm]{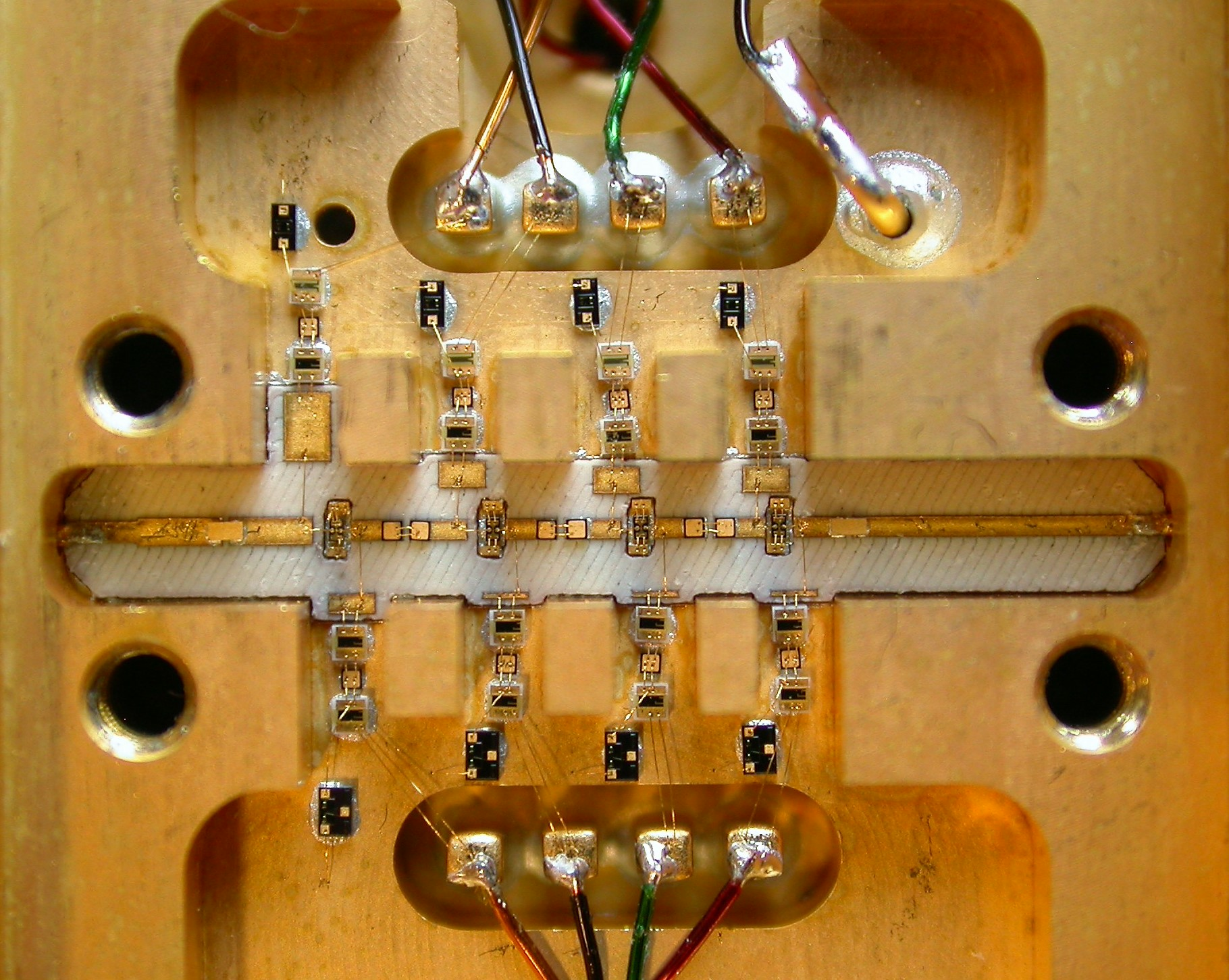}
  \centering
  \caption{Inner view of the Q band cryogenic amplifier module showing the four transistor stages, the microstrip matching networks, and the biasing components. Overall block dimensions are $43\times29.5\times10$\,mm including the waveguide transitions.}
   \label{f.qamplifier}
\end{figure}

\begin{figure}
  \includegraphics[width=8cm]{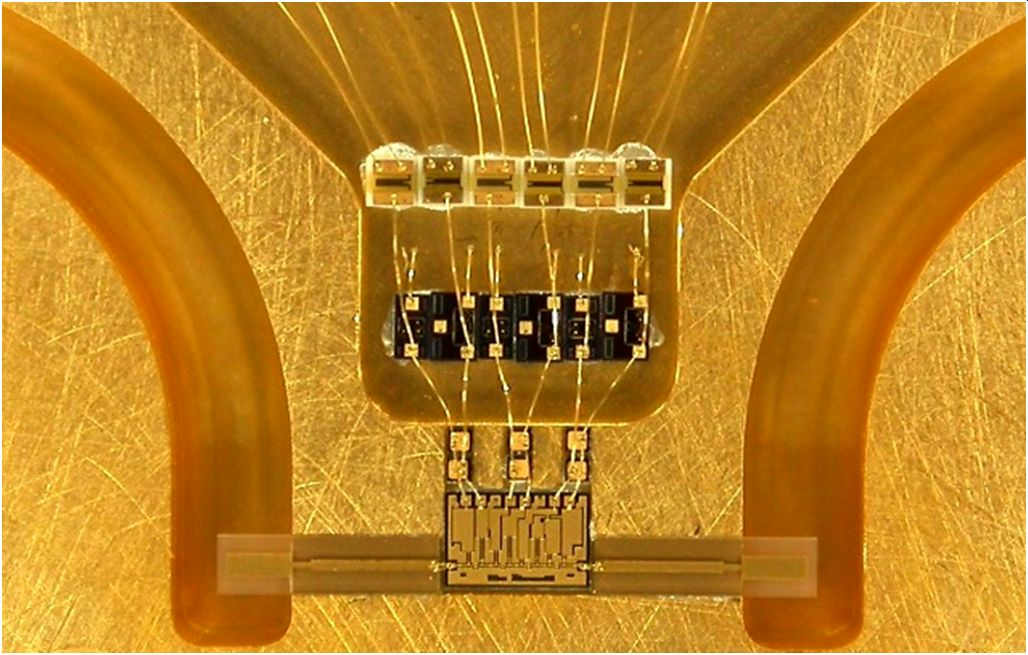}
  \centering
  \caption{Inner view of the W band cryogenic amplifier split block showing a detail of the MMIC, quartz microstrip to waveguide transitions, and DC bias protection components. MMIC size is $1.5\times1$\,mm. Overall block dimensions are $21.5\times21.5\times19.1$\,mm.}
   \label{f.wamplifier}
\end{figure}

The cryogenic low noise W band amplifiers are based on metamorphic GaAs millimetre-wave MMICs developed by cooperation agreements of Yebes Observatory with IAF\footnote {Fraunhofer Institut f\"ur Angewandte Festk\"orperphysik, Tullastrasse 72, 79108 Freiburg.} and UC\footnote{Universidad de Cantabria, Department of Communication Engineering, Avda. Los Castros, 39005 Santander.} and fabricated with the 50\,nm gate length IAF process \citep{leuther09}. They are identical to the ones used in the receiver previously reported in \cite{Cernicharo2019a}, although the band of interest is now limited to $72-90$\,GHz (the amplifiers were originally designed for the $72-116$\,GHz range). The performance of these units in the reduced band is reported here. Photographs of the amplifiers with their cover removed are shown in Figs.\,\ref{f.qamplifier} and \ref{f.wamplifier}.

\begin{figure}
  \includegraphics[width=9cm]{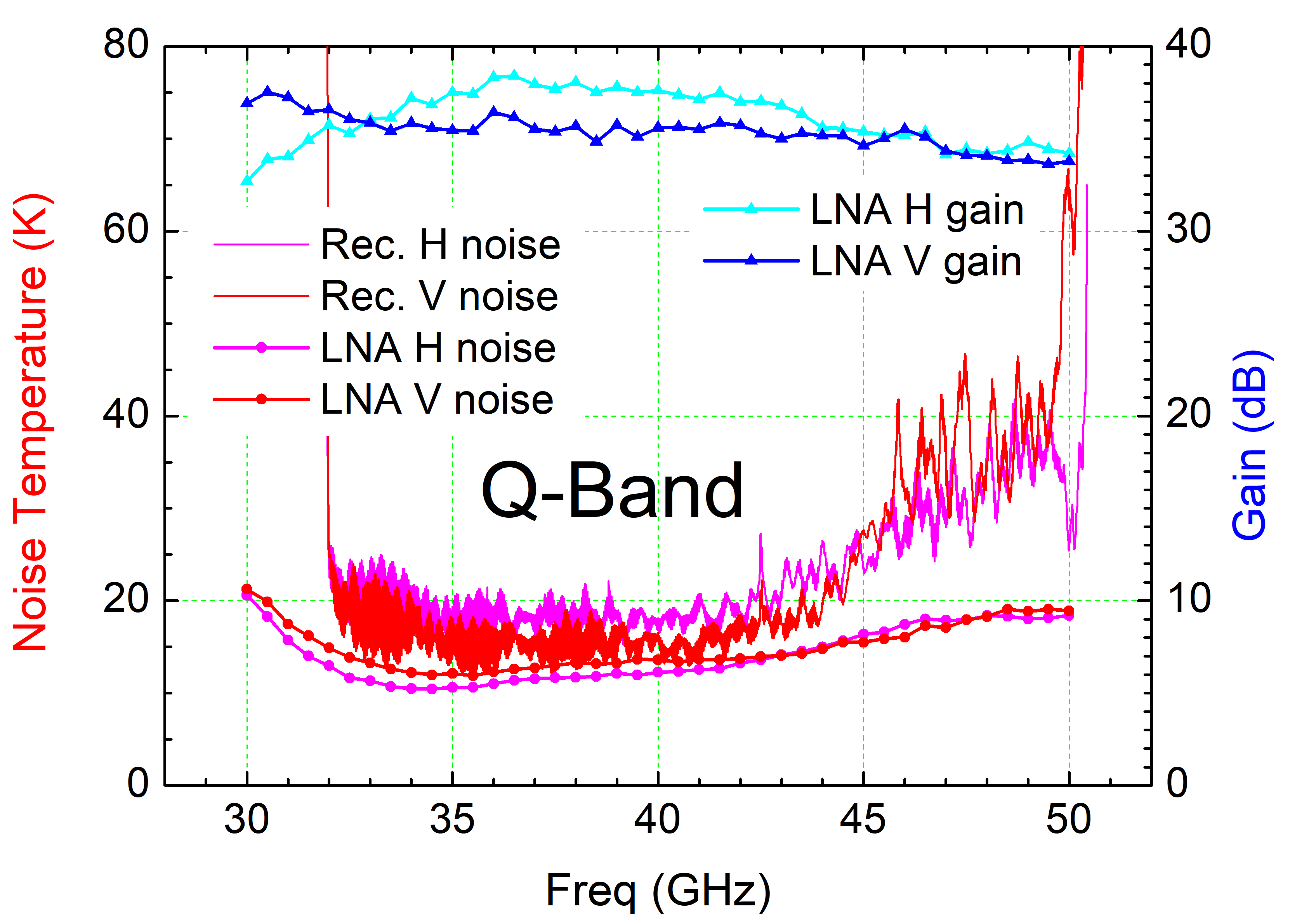}
  \centering
  \caption{Gain and noise temperature of the Q band cryogenic LNAs at 15\,K physical temperature (see Sect.\,\ref{sect_amplifiers}) compared with the total noise of the complete receiver (see Sect.\,\ref{sect_trec}).}
   \label{f.qgaintsys}
\end{figure}

\begin{figure}
  \includegraphics[width=9cm]{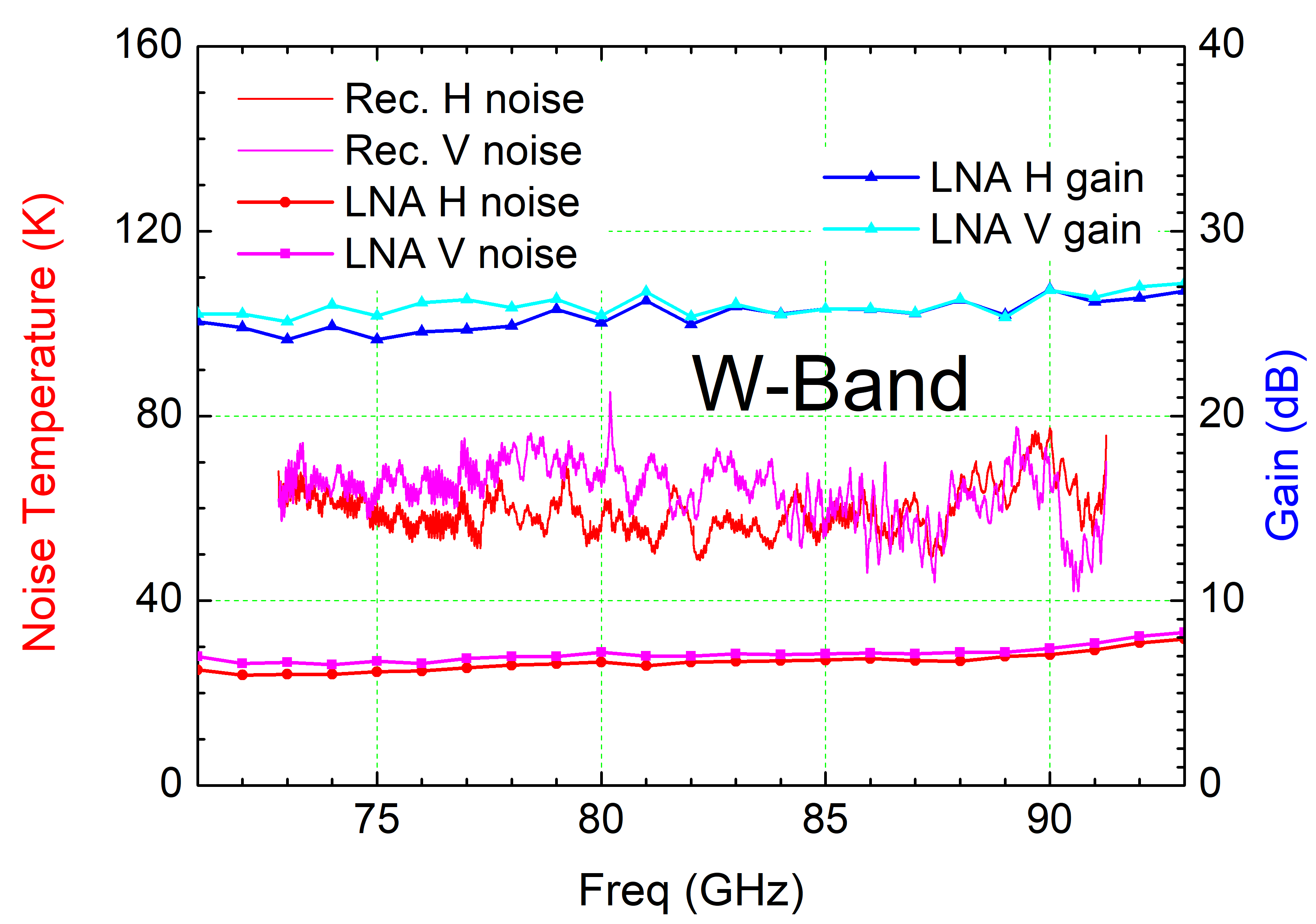}
  \centering
  \caption{Gain and noise temperature of the W band cryogenic LNAs at 15\,K physical temperature (see Sect.\,\ref{sect_amplifiers})  compared with the total noise of the complete receiver (see Sect.\,\ref{sect_trec}).}
   \label{f.wgaintsys}
\end{figure}

The bias point of each LNA was optimised for a good compromise between noise, reflection, gain, and flatness. Noise and gain were measured at 15 K physical temperature using a variable temperature cryogenic waveguide load \citep{malo16} connected to the input. The average noise temperatures obtained in the Q and W band at the waveguide input reference plane are roughly 14\,K and 29\,K, respectively. Noise and gain plots are shown in Figs.\,\ref{f.qgaintsys} and \ref{f.wgaintsys} and compared with the total receiver noise temperature measured with the receiver installed at the antenna (see Sect.\,\ref{sect_trec}). The main results of the measurements of the stand-alone amplifiers are summarised in Table\,\ref{t.lnaPerformance}. It is worth noting that the noise from the amplifiers can be considered to be state of the art (see Figs.\,\ref{f.qgaintsys} and \ref{f.wgaintsys}).

\begin{table}
	\caption{Measured performance of the cryogenic LNAs at 15\,K physical temperature. The values of noise temperature ($T_{\rm n}$) and gain ($G$) are the average value in the band and $P_{\rm diss}$ is the total heat power dissipation at the optimum bias value.}
\label{t.lnaPerformance}
\centering
\begin{tabular}{c c c c c}
\hline\hline
	& Band & $T_{\rm n}$ (av) & $G$ (av) & $P_{\rm diss}$ \\
        & (GHz)& (K)     & (dB)   & (mW) \\
\hline
YMWN 2001 & $72-91$ & 27.9 & 25.9 & 23.2 \\
YMWN 2004 & $72-91$ & 26.2 & 25.3 & 23.8 \\
YQN 1005  & $31-50$ & 13.9 & 36.3 & 16.0 \\
YQN 1007  & $31-50$ & 14.7 & 35.3 & 16.0 \\
\hline
\end{tabular}
\end{table}

\subsection{Frequency conversion}

\subsubsection{RF to IF downconverters}
\begin{figure}
  \includegraphics[width=9cm]{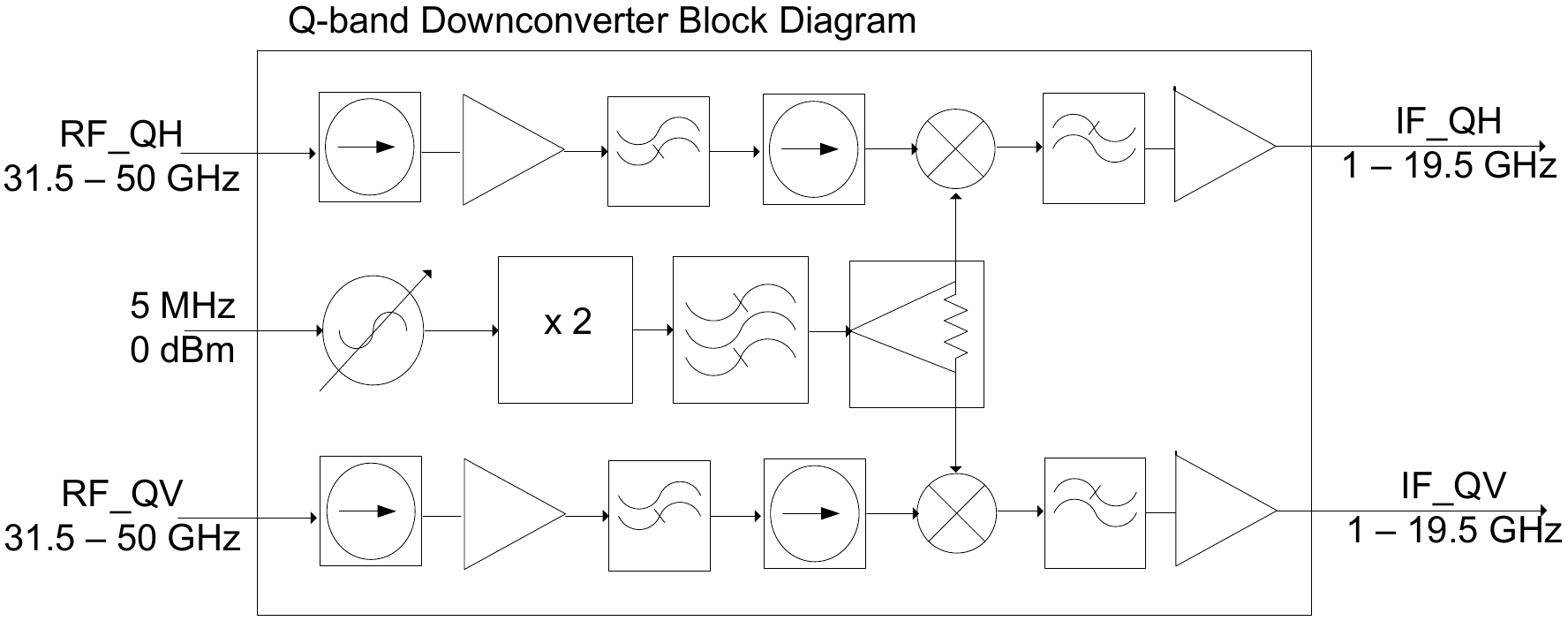}
  \includegraphics[width=9cm]{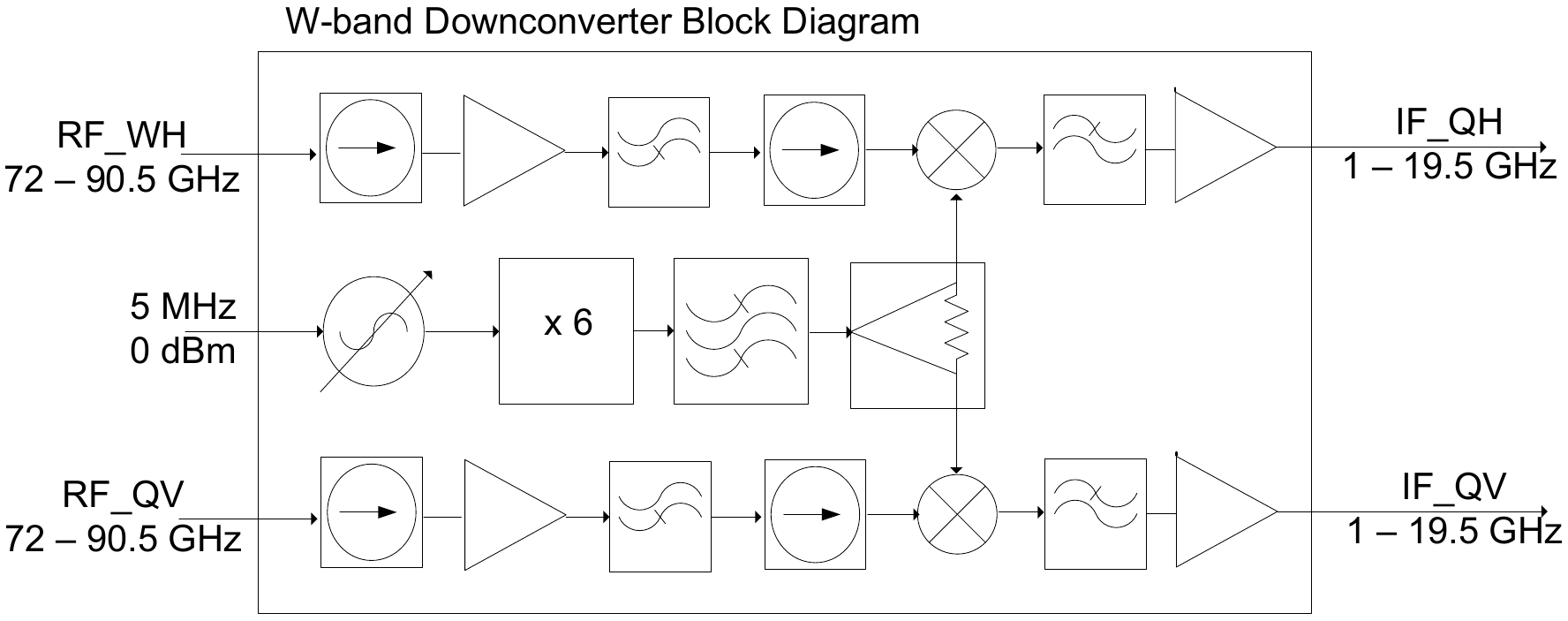} 
  \centering
  \caption{Schematic diagrams of the Q band (upper panel) and W band (lower panel) RF to IF downconverters.}
   \label{f.rf_if_int_qw}
\end{figure}

The Q and W receiver cryostats are connected via waveguides sections to two downconverters (see Fig.\,\ref{f.rf_if_int_qw} for a schematic diagram of both of them).

The Q band downconverter shifts both linear polarisation signals (horizontal, H, and vertical, V) in the range $31.5-50$\,GHz (RF) to an intermediate frequency (IF) of $1-19.5$\,GHz using a common computer-controlled synthesised local oscillator (LO) at 30.5\,GHz. Similarly, the W band downconverter transforms both polarisation signals in the range $72-90.5$\,GHz to an IF of $1-19.5$\,GHz using another common computer-controlled synthesised LO at 71\,GHz, obtained through a $\times6$ frequency multiplier. The frequency of both LOs is dynamically controlled to compensate the Doppler frequency shift due to the relative movement between the radio telescope and the observed source.

These downconverters provide an average gain of 25\,dB in Q band and 36\,dB in W band, with standalone noise temperatures in the ranges $1000-2500$\,K and $600-1200$\,K in Q and W band, respectively. The total contribution to the receiver noise of these 
units within the chain is lower than 1\,K. The gain stability of the Q band downconverter unit was independently tested in the laboratory with a vector network analyser using a continuous wave (CW) signal at the input (see Sect. \ref{sect_sngf}).

\subsubsection{IF signal transport and distribution}
\label{sect_IFtransport}
The four IF signals at the output of these downconverters are transported from the receivers room to the backends room at the 40\,m telescope using links of 80 m length. To avoid the large slope that would have been introduced by coaxial cables we used four single-mode RF-over-fibre optical links operating in the range $1-19.5$\,GHz.
The installation of the backends in the receiver cabin was discarded due to a) footprint requirements b) limited thermal stability of the receiver cabin, and c) to prevent that RF interferences generated by the backends 
deteriorated the observations of lower frequency receivers.

\begin{figure}
\includegraphics[width=7cm]{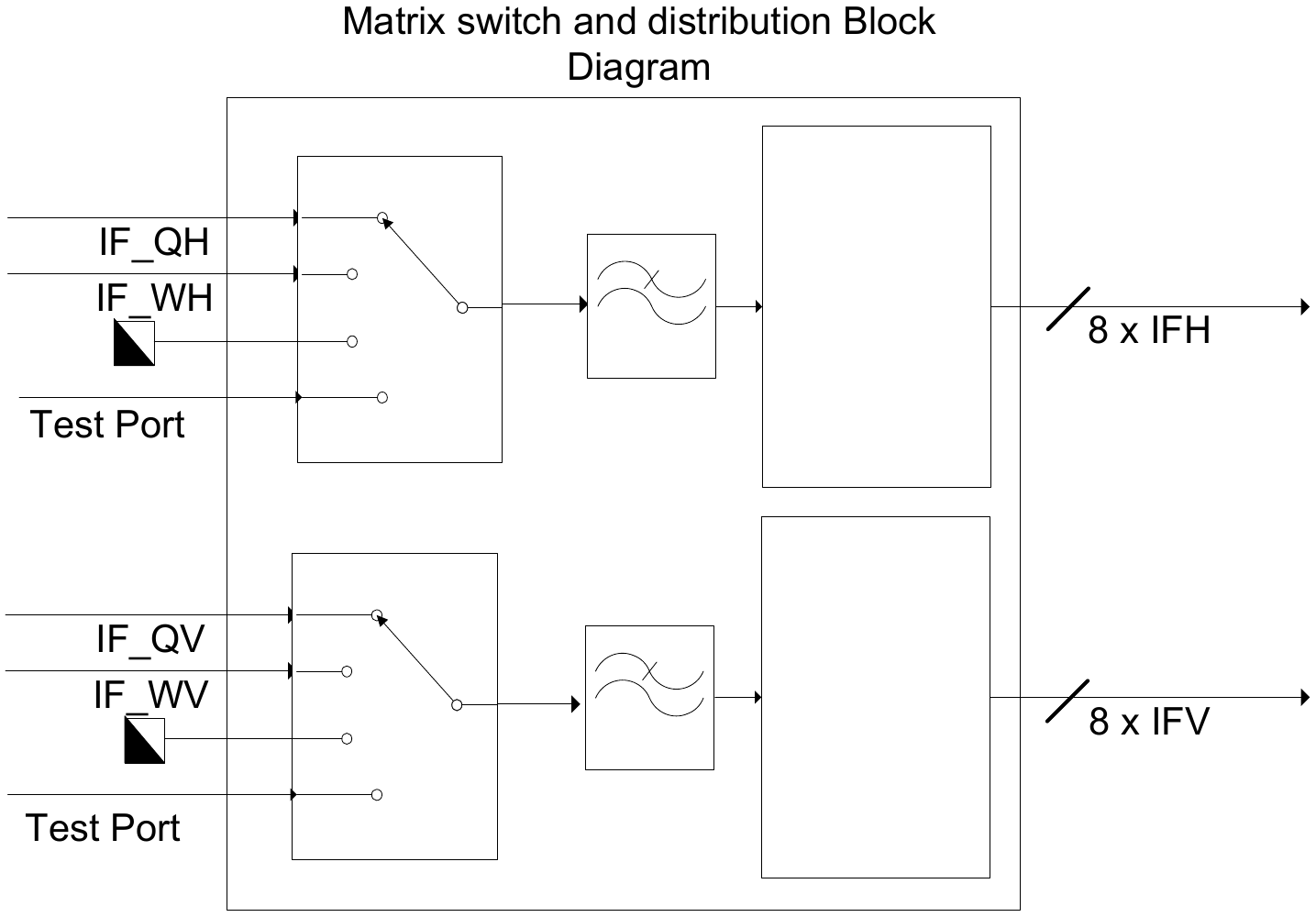}
 \centering
 \caption{Block diagram of the matrix that distributes the signals to the base band converters.}
 \label{matrix}
\end{figure}

The modules, from Microwave Photonics Inc., were modified at the factory to increase their maximum frequency from 18\,GHz to 19.5\,GHz.
The outputs of the optical links are sent to a matrix switch and a distribution unit (see Fig.\,\ref{matrix}).
This unit allows the selection of both polarisation signals in Q band or W band or one polarisation signal from each band. The selected IF signals are distributed to a bank of 8 IF to base band (BB) downconverters.

\subsubsection{IF to BB downconverters}
\begin{figure}
 \includegraphics[width=9cm]{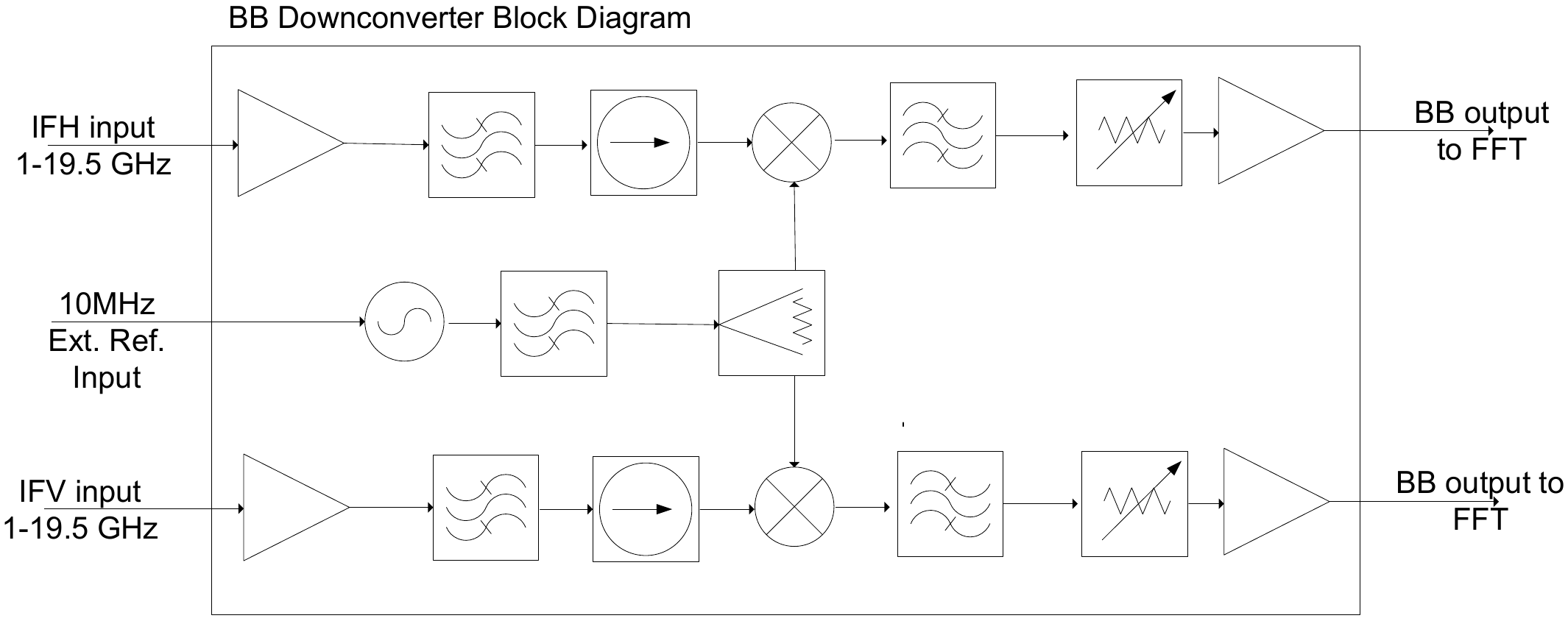}
 \centering
 \caption{Schematic diagram of the base band (BB) converters. Each BB uses a different LO frequency.}
 \label{f.rf_if_int}
\end{figure}

A bank of 8 IF to BB dual-channel downconverters splits the IF range ($1-19.5$\,GHz) into 8 sub-bands each delivering a DC\,$-2.5$\,GHz signal to feed the FFTSs with a suitably conditioned signal (see Fig.\,\ref{f.rf_if_int}). 
The gain of each BB downconverter is adjustable in 1\,dB steps from 20\,dB to 50\,dB to optimise the power level at the digitiser input. In addition, the measured minimum value of output power at 1\,dB compression point is +22\,dBm, when the maximum gain is set, and the average intermodulation point of 3$^{\rm rd}$ order (IP3) is +32\,dBm. The local oscillator of each BB downconverter is fixed in frequency and is common to both channels. The frequency of each LO follows the equation 
\mbox{$f_{\rm LO} ({\rm GHz}) = 2.3\times N - 1.4$}, where $N$ ranges from 2 to 8. For BB\#1, the LO frequency is 3.5\,GHz to avoid LO leakage signals in the final base band. As a result, consecutive sub-bands are overlapped by 100 MHz to ease the reconstruction of the full receiver band after a fast Fourier transform (FFT) of each sub-band.
Each BB downconverter and its local oscillator is integrated inside a 19”/2U rack. The monitor and control functions of each unit are performed by a micro-controller board, developed in-house, with local area network (LAN) connectivity. The BB outputs (8 sub-bands $\times$ 2 pols = 16 bands) are connected to FFTS boards (see next section). The input power into each FFTS board ranges, in the most extreme case, between $-$13 and $-$23\,dBm for a hot and cold load, respectively. The whole RF chain down to the base band converters keeps its linearity well within this working regime.

\subsubsection{Backends: the FFT spectrometers}

The spectrum of each sub-band is computed in an FFTS board from Radiometer Physics GmbH, which contains an analogue-to-digital converter (ADC) with 10 bit resolution and a field-programmable gate array (FPGA) for real time signal processing. Each FFTS board accepts $0.1-2.5$\,GHz input range and provides a spectrum of up to 65536 channels which corresponds to a channel spacing of 38\,kHz and an equivalent noise bandwidth (ENBW; \citealt{Klein12}) of 44\,kHz. The FFTSs are arranged in two crates of 8 boards. Each crate is integrated inside a 19”/3U rack. The cabling between the BB downconverters and the FFTS boards is optimised to reduce the length of the connecting cables and to allow sufficient air flow for cooling.

A 19”/42U cabinet houses the RF-over-fibre link receivers, the matrix switch and distribution unit, the 8 BB downconverters, and the two FFTS crates. All LOs and FFTS clocks are phase locked to a 10\,MHz signal derived from the Yebes Observatory hydrogen maser ensuring the accuracy of the frequency scale.

The backend room is thermalized to 21.5$^{\circ}$C with a maximum peak to peak oscillation of 0.1$^{\circ}$C to avoid gain variations from electronics.

\section{Performance of the receivers}
\label{sect_performance_receivers}
\subsection{Receiver noise temperature}
\label{sect_trec}

\begin{figure}
\includegraphics[width=8cm]{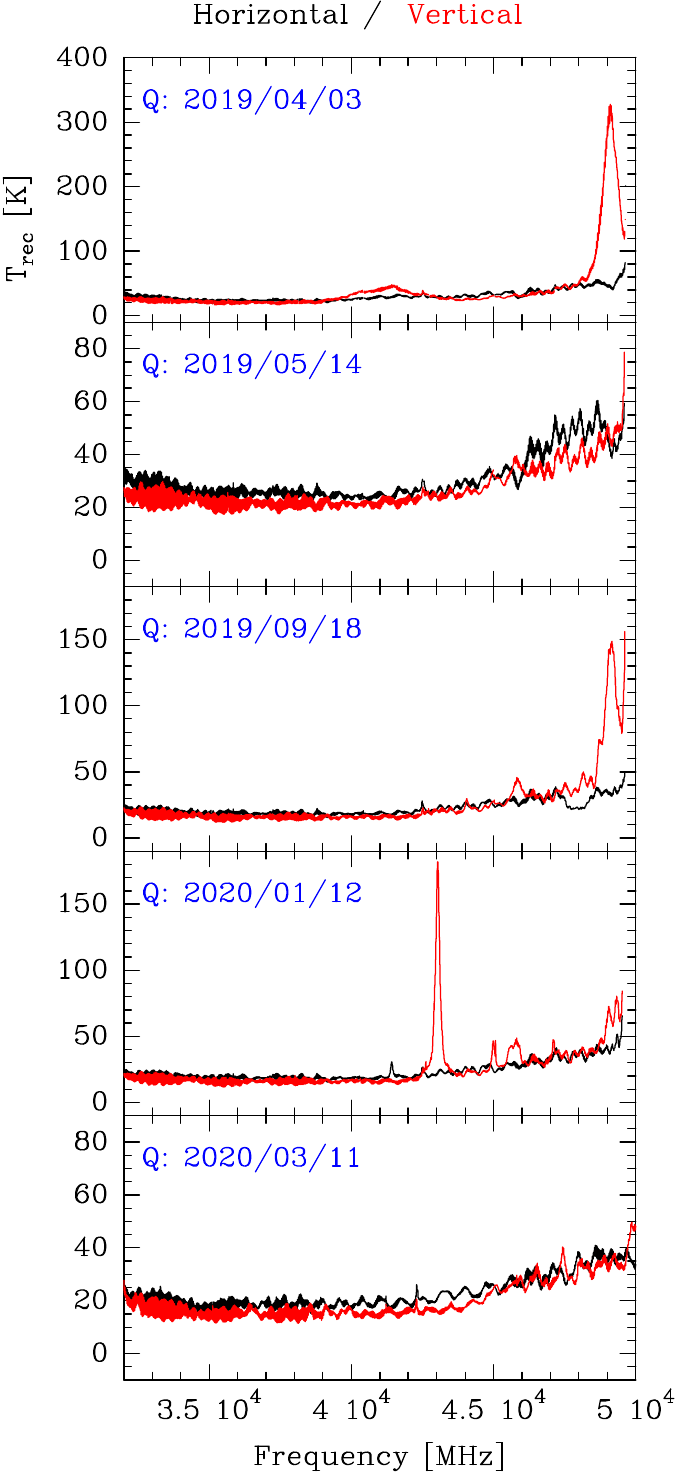}
\caption{Receiver temperature of the Q band receiver installed at the 40\,m telescope at different dates (see Sect.\,\ref{sect_trec}). These measurements alerted about the existence of problems in the system and helped to discover its origin.}
\label{f.trec_Q}
\end{figure}

\begin{figure}
\includegraphics[width=8cm]{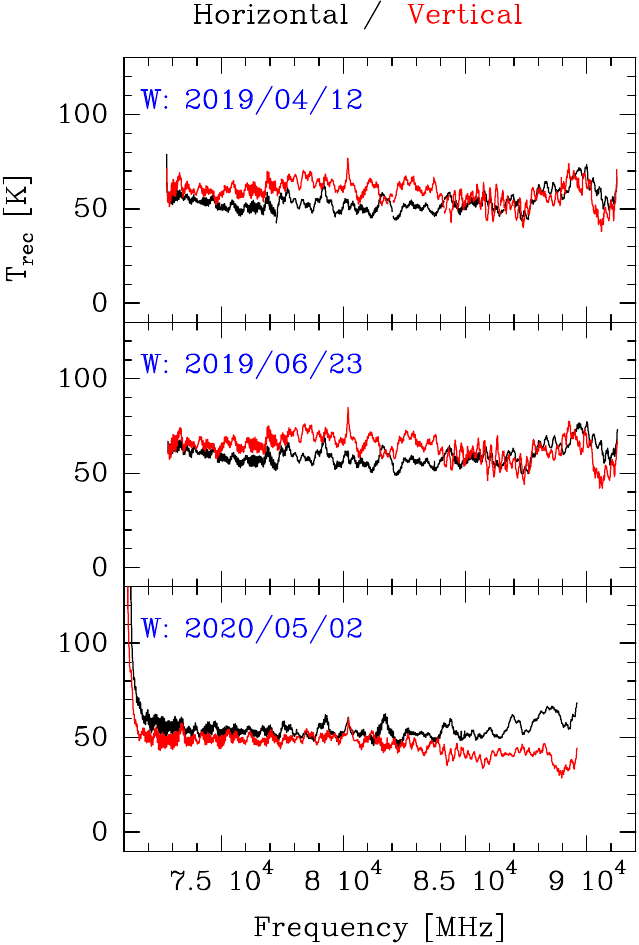}
\caption{Receiver temperature of the W band receiver installed at the 40\,m telescope at different dates (see Sect.\,\ref{sect_trec}). These measurements alerted about the existence of problems in the system and helped to discover its origin.}
\label{f.trec_W}
\end{figure}

The hot-cold load calibration system (see Sect.\,\ref{sect_optics}) allows continuous monitoring of the receiver noise temperature ($T_{\rm rec}$) at Q and W bands. The noise temperature obtained takes into account the contribution from all elements located between the loads and the final detectors (i.e. tertiary mirrors, OMTs, waveguides, amplifiers, downconversion elements, backends...) and therefore allows to check the performance of the complete system. $T_{\rm rec}$ can be calculated as,

\begin{equation}
T_{\rm rec} = \frac{T_{\rm hot} - Y T_{\rm cold}}{Y - 1}, Y = \frac{V_{\rm hot}}{V_{\rm cold}}.
\label{eq.Trec}
\end{equation}
where $V_{\rm hot}$ and $V_{\rm cold}$ are the backend counts towards the hot and cold load respectively, $T_{\rm hot}$ is assumed to be the ambient temperature at the receiver cabin during the calibration, and $T_{\rm cold}$ is the equivalent noise temperature of the cold load.

We calibrated $T_{\rm cold}$ using two absorbers at liquid nitrogen (LN2) and ambient temperatures. In average, we obtained an equivalent temperature of the cold load which ranges between $16.6-18.5$\,K (Q band) and $19.1-22.3$\,K (W band) while its physical temperature during these measurements was 17.8\,K,
according to the temperature sensor inside the cryostat.
Losses in the optics, windows, and in infrared filters should increase $T_{\rm cold}$ with respect to its physical temperature, but it seems that these losses are low. 
The temperature increments are in the range of $\Delta$$T$\,$\sim$\,$0.5-1$\,K and $\Delta$$T$\,$\sim$\,$1.5-3$\,K, for Q and W bands, respectively.
However, due to the uncertainties of the calibration method (primarily time variation of the assumed physical temperatures, ambient temperature, liquid nitrogen temperature and temperature of the LN2 absorbers in front of the receiver window) we estimate an uncertainty of $\pm$2\,K in the measurements. Therefore, no correction was applied to $T_{\rm cold}$ and the reading from the temperature sensor inside the cryostat is used to calibrate the astronomical data.  This may result in an overestimation of the measured receiver temperature of the same order of the estimated noise temperature excess of the cold load ($0.5-3$\,K).

Figures \ref{f.qgaintsys} and \ref{f.wgaintsys} present a typical measurement of the Q and W band receivers noise temperature obtained by the method described above compared with the noise temperature of the cryogenic LNAs (see Sect.\,\ref{sect_amplifiers}). As it can be seen, there is a significant contribution from other components in the chain, particularly prominent in the higher end of the Q band and through the entire W band. This contribution is larger than predicted by the models and its origin is not totally clear yet. 
We expect that this could be solved in future upgrades of the system.

The measurements of the receiver noise temperature show a combination of ripples with different periods which are originated by the reflections produced in some of the components (standing waves). As an example, a Fourier transform applied to the receiver noise measurement shown in Fig.\,\ref{f.qgaintsys} reveals peaks corresponding to reflections at approximately 0.20, 0.51, 3.9, 4.1, and 4.4\,m. The first two correspond to reflections within the components in the receiver cryostat while the last three are due to reflections between the cold load and the receiver. The longer distances correspond to the fine grain, noise-like structure clearly visible at frequencies below 40\,GHz in Fig.\,\ref{f.qgaintsys} 
and whose periodicity is too dense to be appreciated in that graph.

The systematic measurement of the noise temperature in the complete band allows us to find any problem which may appear in the system. For instance, Figs.\,\ref{f.trec_Q} and \ref{f.trec_W} show the receiver temperature at Q and W bands measured at different dates. Whereas we did not appreciate significant discrepancies between these measurements and the expected $T_{\rm rec}$ for the W band, we noticed unexpected values at some frequencies for the $T_{\rm rec}$ in the Q band, especially in the V polarisation. The high values of $T_{\rm rec}$ in the V polarisation around 41\,GHz and 49\,GHz in April 2019 were caused by a dichroic mirror located in front of the Q band receiver which allowed simultaneous K and Q band observations but
did not have the required performance for the new system since it was designed for a narrower band. We avoided this contribution by replacing the dichroic by a total reflection mirror to perform observations in Q band (see Sect. \ref{sect_optics}). From September 2019 to February 2020 we noticed some peaks in the $T_{\rm rec}$ at selected frequencies in the V polarisation. Moreover, the frequencies of the peaks, changed along this period. After some experiments which demonstrated that the peaks were indubitably originated within the cryostat, the Q band receiver was warmed up, opened, and revised in February 2020. No faulty component was found inside the cryostat. After the subsequent cool down, a significant reduction of the spurious signals was seen (see Fig.\,\ref{f.trec_Q}, March 2020). Although the cause was not clarified and it is still under investigation, 
one hypothesis is that the noise peak could be caused by a resonance in a component housing or in a waveguide junction.

\subsection{Stability}
\label{sect_stability}

The performance of a radio telescope is limited by the time stability in noise and gain of its receivers. The most common method of detecting weak signals is by long time integrations which require a stable receiver behaviour. The measurement of the stability addresses possible gain fluctuations of the receiver and mechanical coupling with elements in the receiver chain due to vibrations. In order to provide flat enough baselines for the astronomical data, we expect that
the gain fluctuation of the complete receiver chain has to be of the order of 10$^{-4}$ per second.  
To check if this requirement was fulfilled, we developed a software procedure which provided the averaged backend counts registered in each FFTS (2.5\,GHz of bandwidth) per 100 milliseconds.
Figures\,\ref{fig_estabilidad_Q} and \ref{fig_estabilidad_W} show the normalised backend counts of each FFTS section at Q and W bands observing towards the hot load for a total time of 40 minutes in January 2020. 
We measured a typical peak to peak value for the fast variation fluctuations of about 5$\times$10$^{-4}$ per second.
Slower drifts over $20-30$ minutes are due to temperature variations but they do not significantly affect astronomical data since these are usually collected using shorter integration times.

These measurements allowed us to significantly improve the stability of the Q band after the installation of the receivers in March 2019. For the Q band chain we had to address several operations in order to reach the current values. First, in June/July 2019, both commercial Quinstar\footnote{http://quinstar.com} pre-amplifiers (for H and V polarisations) installed in the first downconverted were replaced by in-house made amplifiers which significantly reduced the gain fluctuation. Besides, to reduce the frequency of the mechanical vibrations and increase the thermal inertia we increased the total mass of the downconverter.
Secondly, to increase the stability in the mechanical coupling between the IF and the cryostat, some sections of flexible waveguides were replaced by rigid in-house made waveguides. 

Currently, the element with the largest contribution to the gain fluctuations is the optical fibre (and/or its optical fibre links) which transfers the signal from the receiver cabin to the backend room 
(see Sect.\,\ref{sect_IFtransport}).
Moreover, the substantially noisier and unstable data found at the upper end of the Q and W bands (BB\#8, see Figs.\,\ref{fig_estabilidad_Q} and \ref{fig_estabilidad_W}) 
might be related to a large gain drop (20\,dB) at the IF frequencies between $17-19.5$\,GHz which cannot be completely compensated by levelling the IF power at the output of the BB converters.
Although, our goal is to reach a peak to peak value of less than 10$^{-4}$ per second, at this stage the system was found stable enough for astronomical observations (see Sect.\,\ref{sect_survey}). 

\subsection{The stability of the Q and W band receivers: a quantitative view}
\label{sect_sngf}

\begin{figure}
 \includegraphics[width=9cm]{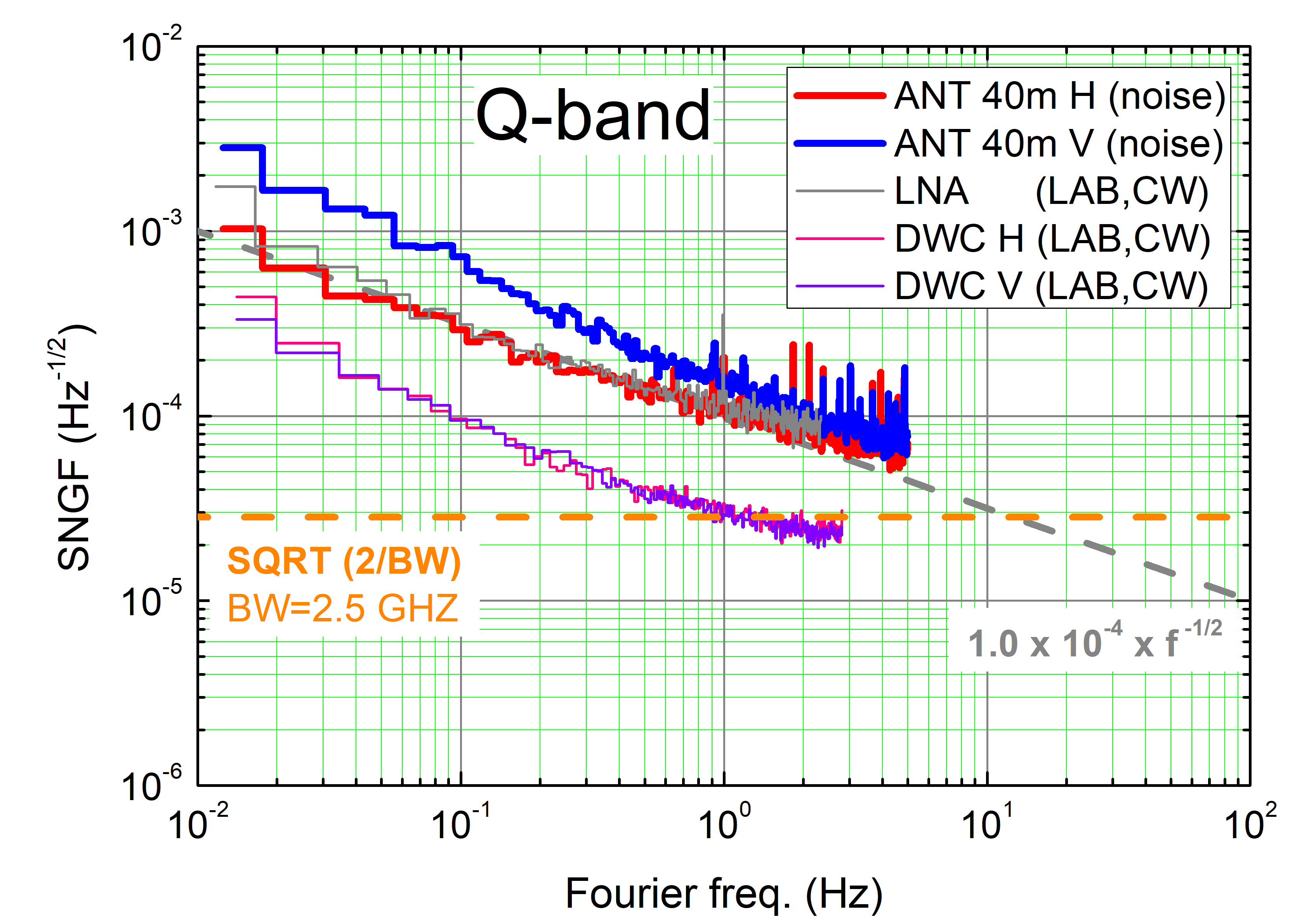}
 \centering
 \caption{Spectrum of normalised gain fluctuations(SNGF) of the complete Q band receiver (red and blue curves), one of the cryogenic LNAs (grey curve),
 and the first downconverters (DWC; pink, and violet curves).
 Dashed grey line corresponds to an ideal $\propto f^{-1/2}$ fluctuation. Dashed orange line shows the fluctuation level for ideal radiometer white noise in a 2.5\,GHz bandwidth. See Sect.\,\ref{sect_sngf} for a more detailed explanation.}
 \label{f.spectrumSNGF_Q}
\end{figure}

A powerful tool for the characterisation of the receiver stability is the spectrum of normalised gain fluctuation (SNGF; \citealt{gallego2006}). 
The advantage of SNGF over Allan variance is that it allows the efficient detection of periodic effects such as mechanical vibrations or thermal oscillations caused by the cryogenic cooler cycle that are otherwise not easily visualised. 
Figure\,\ref{f.spectrumSNGF_Q} shows an example of the results obtained for some components of the Q band receiver chain (LNA and downconverters) derived using the methods described in detail by \cite{gallego2006}. 
These results are compared with the total noise power fluctuation measured for both polarisations of the complete receiver installed at the antenna in a 2.5\,GHz bandwidth centred at $\sim40$\,GHz (BB\#4) with the input beam terminated in an ambient temperature microwave absorber. 

The cryogenic amplifier and the first downconverter were measured with a CW signal while the complete Q band receiver was characterised with random thermal noise. Cryogenic millimetre wave LNAs built with small gate area HEMTs provide state of the art low noise temperatures, but are known to be an important source of gain fluctuations (multiplicative noise; \citealt{weinreb2014}). The prototype cryogenic Q band amplifier presented in Fig.\,\ref{f.qamplifier} contributes with a spectrum which follows almost perfectly the $\sim f^{-1/2}$ law and with a value of approximately $1\times10^{-4} {\rm Hz}^{-1/2}$ at 1\,Hz. This can be considered a typical value for a well behaved cryogenic LNA in this band and ideally this should dominate the fluctuation of the complete receiver. Figure\,\ref{f.spectrumSNGF_Q} presents the fluctuations measured with a CW signal for the first downconverter (at ambient temperature), which are very low ($\simeq 3\times10^{-5} {\rm Hz}^{-1/2}$ at 1\,Hz). This low value was possible thanks to the selection of very low fluctuations and low gain travelling wave topology amplifiers for the first stage of the downconverter.

\begin{figure}
 \includegraphics[width=9cm]{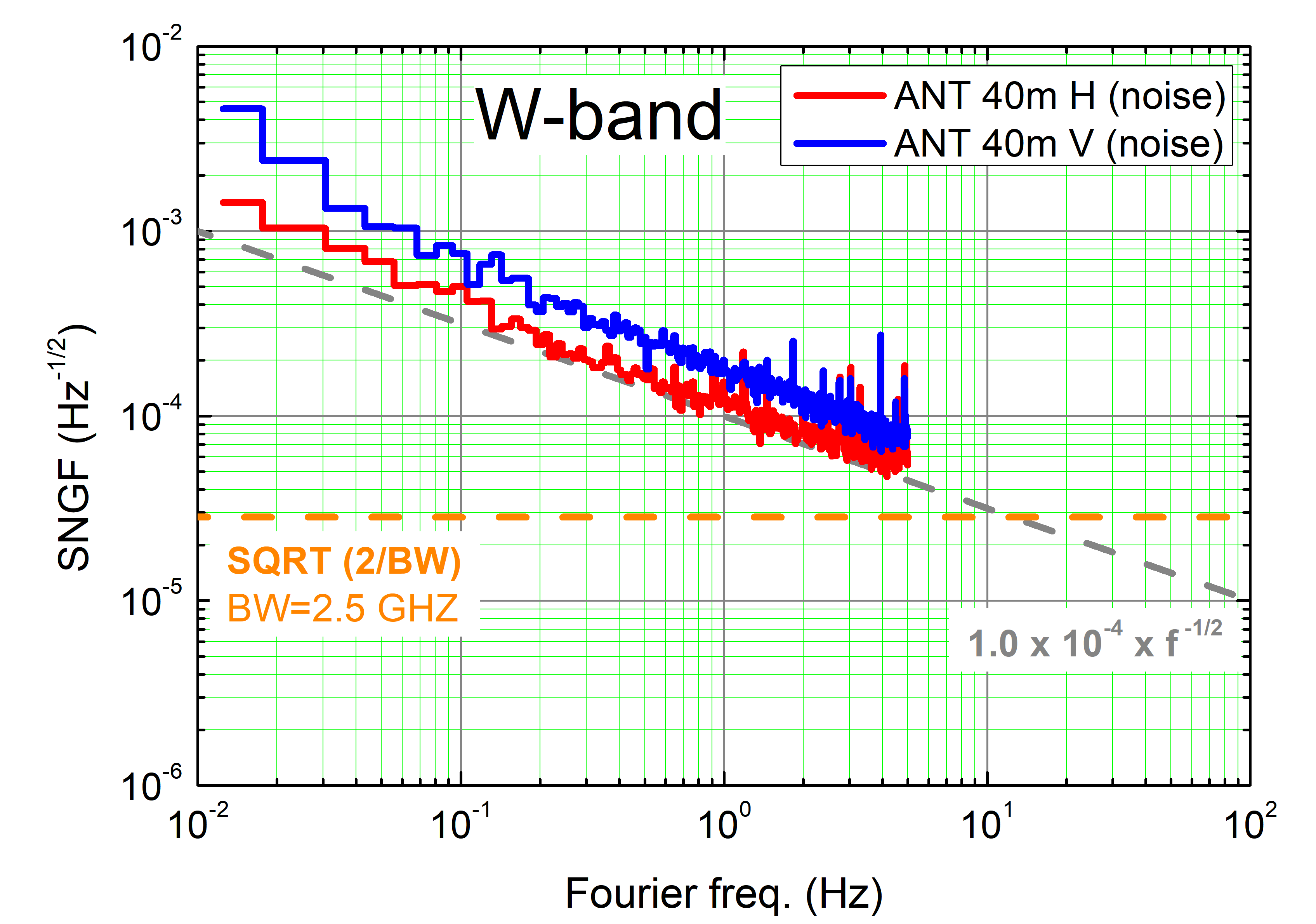}
 \centering
 \caption{Spectrum of normalised gain fluctuations (SNGF) of the complete W band receiver for both polarisations (red and blue curves). Dashed grey line corresponds to an ideal $\propto f^{-1/2}$ fluctuation. Dashed orange line shows the fluctuation level for ideal radiometer white noise in a 2.5\,GHz bandwidth. See Sect.\,\ref{sect_sngf} for a more detailed explanation.}
 \label{f.spectrumSNGF_W}
\end{figure}

The measured fluctuations of the H and V polarisations of the complete Q band receiver presented in 
Fig.\,\ref{f.spectrumSNGF_Q} show some interesting features. The H channel SNGF corresponds to what should be expected from the fluctuation of the cryogenic amplifier, but the V channel shows higher than expected values and a different Fourier frequency dependence.  
This misbehaviour of the V polarisation could be due to several causes, currently under investigation.  
Nevertheless, the optical fibre seems to be, for the time being, the main contributor (see Sect.\,\ref{sect_stability}).
Note that the SNGF of the H and V channels also show some discrete lines above 1 Hz which are believed to be induced by mechanical vibrations (motors, fans, etc.) in the receiver cabin. 
Although the cryostats are not provided with a mechanical decoupling between the cold head and the rest of the cryostat (Sect.\,\ref{sect_cryostats}), no prominent feature is found at 1\,Hz (cycle of the cryogenic cooler). 
The orange horizontal line in Fig.\,\ref{f.spectrumSNGF_Q} also presents the theoretical value of the white noise radiometer fluctuation for a 2.5\,GHz bandwidth.  The H and V channels measurement should converge asymptotically to that value at higher Fourier frequencies although the limited rate of the data acquisition system did not allow a better sampling of that region.

The fluctuations of the complete W band receiver with random thermal noise at the input are presented in Fig.\,\ref{f.spectrumSNGF_W}. In this case it was not possible to independently measure the components of the signal chain as in the Q band receiver, since the laboratory equipment available did not cover the W band frequency range. The results obtained are similar to the stability of the Q band receiver shown in Fig.\,\ref{f.spectrumSNGF_Q}. The same types of discrete lines above 1 Hz (mechanical vibrations) are also visible. As before, no effect of the 1\,Hz cryogenic cooler cycle is detected. As in the Q band receiver, the vertical polarisation shows worse gain stability.

\begin{figure}
 \includegraphics[width=9cm]{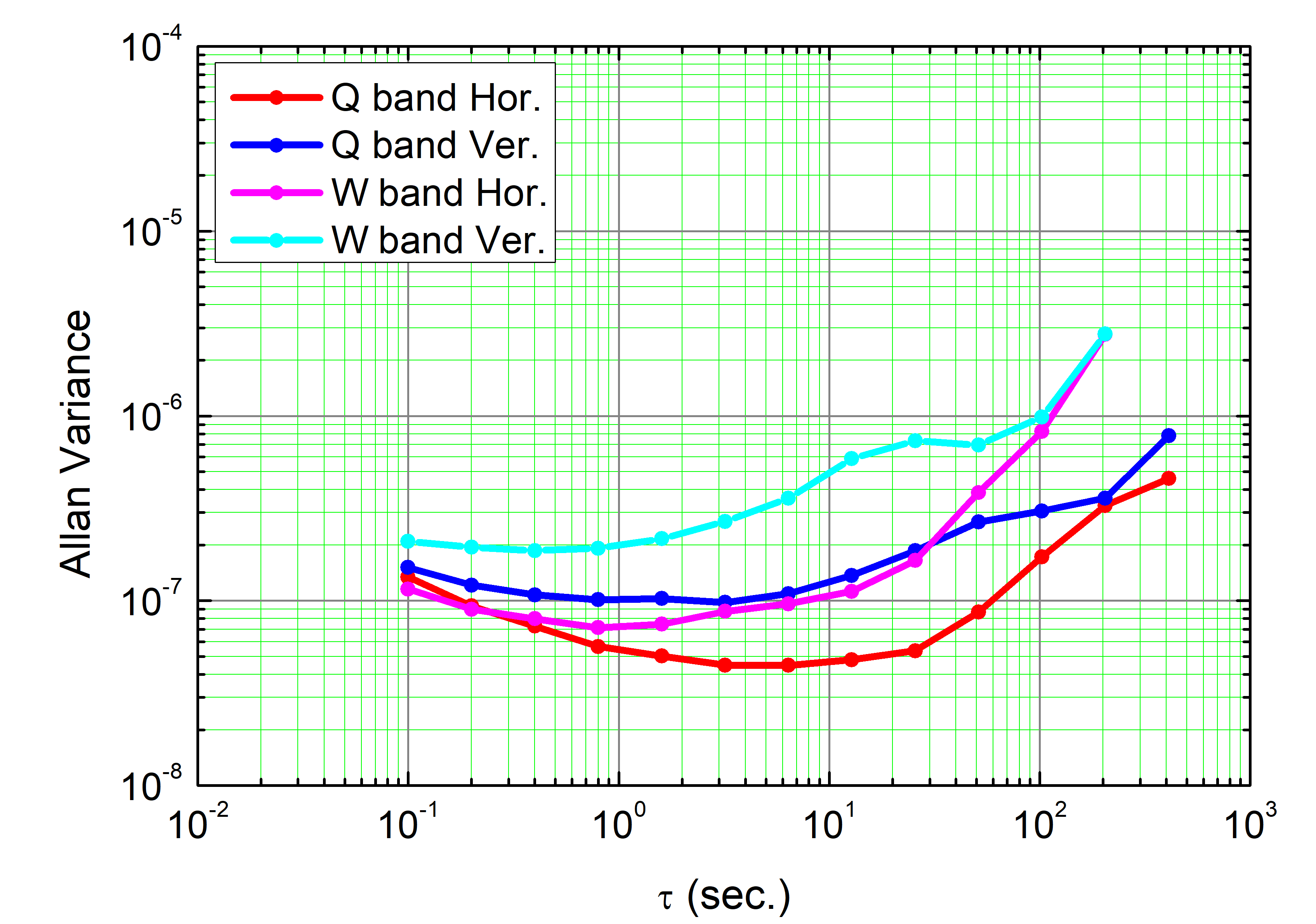}
 \centering
 \caption{Allan variance for both polarisations at Q and W band, obtained averaging all channels from BB\#4. The data were recorded every 100\,ms. See Sect.\,\ref{sect_sngf} for a more detailed explanation.}
 \label{f.avar}
\end{figure}

Figure\,\ref{f.avar} shows a measurement of the Allan variance at Q and W bands for both polarisations averaging all frequency channels from BB\#4. The data were recorded every 100\,ms. As already seen in Figs.\,\ref{f.spectrumSNGF_Q} and \ref{f.spectrumSNGF_W} the horizontal polarisation of both, Q and W band, show a much better behaviour than the vertical ones, although W band shows a poorer performance than Q band. The optimum integration times (2.5\,GHz bandwidth) for each case are: 3\,s (V, Q band), 4\,s (H, Q band), 0.5\,s (V, W band), and 0.8\,s (H, W band). Unfortunately the integration times for each band do not match, making the selection of a common basic integration time for the backends difficult. Given these results, integration times lower than 1 s should be used in case of having a chopper or a wobbler for continuum measurements, but neither of these elements are still present at the radio telescope. For practical reasons we have selected 1\,s as the common basic integration (dump) time from the FFTS, both for continuum drifts and for spectral observations (position switch and frequency switch, see Sect.\,\ref{sect_calibration}).

The results shown in Figs.\,\ref{f.spectrumSNGF_Q}, \ref{f.spectrumSNGF_W}, and \ref{f.avar} suggest that the gain stability of the Q band receiver's H-polarisation is good and behaves well, and that the other channels can be improved. In the future an attempt will be made to locate the components causing this degradation in order to solve this problem.

\section{Software control}
\label{sect_software}
The software control of the 40~m radio telescope is a distributed system which uses as basic infrastructure the ALMA common software (ACS), see \cite{devicente11} for further details. The ACS implements a component-container paradigm using 
 common object request broker architecture (CORBA) as the communication layer hiding its complexity to the software developer. Each hardware element, like for example the receiver, is implemented as a class which lives in the network (a component) and exposes its interface to the rest of the components. The components can run on different computers or embedded systems which are connected to a private local area network at 1\,Gbps.  Any component can call any other component to get its state or ask for an operation. Notification channels are also supported in ACS. This method allows components to continuously report the status of hardware, such as the position of the antenna, in an easy and powerful way. Most of the hardware devices use Ethernet interfaces and those which have serial or GPIB connections use Ethernet converters.

All the monitor and control software has been written in C++, Python, and Java and runs on Linux Debian computers. Time synchronisation in the whole network is achieved via the network transfer protocol (NTP), being the stratum one server a global positioning system (GPS) receiver at Yebes Observatory.

The interface with the scientist is a command line based on IPython, fully programmable, which allows complex operations such as repetitive macros, but keeping an easy usage. The interface allows to set the scan specifying the type of observation, source to be observed, integration time, observing frequency, and backend specification. Each scan can be composed of several phases fully programmable. Position and frequency switches are implemented. Drifts and on-the-fly (OTF) maps are also available both for spectral and continuum operations (see Sect.\,\ref{sect_calibration}). 

The status of the full radio telescope, antenna, frontend, backends, and auxiliary equipment is monitored continuously and displayed on an application developed in Java which is used by the telescope operators and observers. 

As explained in a previous section, the spectral backends at the 40\,m radio telescope are FFTSs. Given their large bandwidth these FFTSs are used for continuum and pseudocontinuum observations, mainly for pointing and focusing (see Sect.\,\ref{sect_calibration} for an explanation). The FFTSs, used as continuum detectors, offer the possibility to tag off groups of bad channels due to, for example radio frequency interferences (RFIs), thus providing a better signal to noise ratio (SNR).

Data from the backends are written in FITS files following MBFITS format (\citealt{muders2011}). The control system implements a pipeline that reads the FITS files and creates \texttt{CLASS} (software of the \texttt{GILDAS}
package\footnote{\texttt{http://www.iram.fr/IRAMFR/GILDAS/}}) compatible observations, both for continuum and spectral observations. However skydip and focus scans are written on ASCII files and analysed with graphical tools manually. 
The spectra are written in 2.5\,GHz observations with a common frequency value to allow \texttt{CLASS} stitching several spectra and generate a single 18\,GHz bandwidth spectrum with the correct Doppler corrections (the channel width changes with frequency, see \citealt{diazpulido14}).

\section{Calibration and observational procedures}
\label{sect_calibration}
Calibration at Q and W bands is carried out using two loads at different temperatures. A hot load at room temperature and a cold load cooled down to $\simeq$17\,K (see Sects.\,\ref{sect_optics} and \ref{sect_trec}),
were used to calibrate the data (see e.g. \citealt{Kramer1997} and \citealt{Jewell2002} for further details on the process).
For a perfectly linear detector, the output backend
counts of the hot and cold load signals are proportional to their physical temperatures which
are equivalent to their radiation temperatures in the Rayleigh-Jeans approximation. 
The system temperature ($T_{\rm sys}$), which gives the noise performance temperature of the receiver ($T_{\rm rec}$, see Sect.\,\ref{sect_trec}) plus the atmosphere temperature ($T_{\rm sky}$), can be derived from:

\begin{equation}
T_{\rm sys} = V_{\rm sky} \frac{T_{\rm hot} - T_{\rm cold}}{V_{\rm hot} - V_{\rm cold}}
\label{eq.Tsys1}
\end{equation}
where $V_{\rm hot}$, $V_{\rm cold}$, $T_{\rm hot}$, and $T_{\rm cold}$ are those defined in Sect.\,\ref{sect_trec}, and $V_{\rm sky}$ is the backend count towards the sky. During the commissioning of the receivers, we noted large ripples in the spectra of the cold load due to multiple reflections produced in some of the components (see Sect.\,\ref{sect_trec}). To avoid these ripples in the astronomical data, we averaged $T_{\rm sys}$ over the channels of each  FFTS section obtaining a single $T_{\rm sys}$ value per sub-band and polarisation. For a typical elevation of 45$^{\circ}$ and average weather conditions at Yebes Observatory (9.0\,mm of precipitable water vapour), the system temperature at the Q band ranges between 50\,K and 110\,K at 32.5\,GHz and 48.5\,GHz, respectively. For the W band the values range between 100\,K at 72\,GHz and 150\,K at 91\,GHz.

To correct the data from atmospheric attenuation, it is necessary to derive the atmospheric opacity of the site at the moment of the observation per sub band and polarisation during the calibration scans. By definition,

\begin{equation}
T_{\rm sys} = T_{\rm rec} + T_{\rm sky} = T_{\rm rec} + T_{\rm atm} \eta_{\rm F} (1-e^{-\tau A}) + (1 - \eta_{\rm F}) T_{\rm amb}
\label{eq.Tsys2}
\end{equation}
where $T_{\rm rec}$ is obtained from Eq.\,\ref{eq.Trec}
and $T_{\rm atm}$ is the effective temperature of the atmosphere at Yebes Observatory, estimated using the ATM code \citep{Cernicharo1985,Pardo2001} and local weather conditions, 
$\eta_{\rm F}$ is the forward efficiency (see Sect.\,\ref{sect_forward}), $\tau$ is the zenith opacity of the atmosphere, $A$ is the air mass, 
and $T_{\rm amb}$ is the ambient temperature.

Source antenna temperatures ($T_{\rm A}^*$) are the brightness temperatures of an equivalent source which fills the entire
forward beam pattern of the telescope. They are obtained using standard radioastronomical observational procedures which remove the instrumental and atmospheric responses, i.e. position switching and frequency switching methods (see e.g., \citealt{oneil02}, \citealt{Winkel2012}). 

When using the standard position switching method (ON-OFF), $T_{\rm A}$ (antenna temperature before corrections from atmospheric losses and forward efficiency) is given by,

\begin{equation}
T_{\rm A} = \frac{V_{\rm ON} - V_{\rm OFF}}{V_{\rm OFF}} T_{\rm sys}
\label{eq.ta1}
\end{equation}
assuming that \mbox{$T_{\rm A}=K$\,($V_{\rm ON}-V_{\rm OFF}$)}, $K$ = constant, and where
$V_{\rm ON}$ and $V_{\rm OFF}$ are the backend counts, for the same integration time, when pointing towards the source and towards a clean (free of emission) reference close to the source, respectively.

A typical position switching observation at the 40\,m RT is composed of one or two blocks of five ON-OFF scans, with 30\,seconds of integration time per phase (ON and OFF), followed by a calibration, usually done every 10 minutes, to update 
$T_{\rm sys}$. Nevertheless, the software allows flexible combinations of scans (ON-OFF and calibration) with different integration times to adapt the observation to the user requirements.

OTF mapping observations (see e.g. \citealt{Mangum2007}) are also available
at the 40\,m RT using the ON-OFF procedure described above. For instance, to map extended sources at 8\,mm (e.g. emission of the class I methanol maser at 36.9\,GHz
in giant molecular clouds) an optimised OTF setup consists of mapping a 400$''$\,$\times$\,400$''$ area with a sampling of 16$''$ (better than a Nyquist sample)
using one clean reference position per line and a raster scanning pattern. This setup yields $\simeq$0.5\,K RMS per spectral channel (38\,kHz) at each point of the mapped area and requires 
10\,minutes of telescope time. Calibrations between maps are possible when the observational goals require to map larger regions in the sky and/or a better SNR.

In frequency switching observations, the local oscillator swaps its frequency while the radio telescope points towards the source, by tens of MHz (usually $10-50$\,MHz) typically every two seconds. In this latter case, we derive
the antenna temperature scale according to

\begin{equation}
T_{\rm A} = (V_{\rm fre1} - V_{\rm fre2}) \frac{T_{\rm hot} - T_{\rm cold}}{V_{\rm hot} - V_{\rm cold}}
\label{eq.ta2}
\end{equation}
where $V_{\rm fre1}$ and $V_{\rm fre2}$ are the backend counts when pointing towards the source for each local oscillator frequency.
As the spectral line of interest is available in both frequency setups, the data are folded back together to achieve a higher SNR by changing the sign of one signal and shifting and averaging both signals \citep{Winkel2012}. Data acquired using frequency switching are affected by standing waves.
  Optimisation is often done by selecting a frequency throw that minimises
them. \citet{Winkel2012} indicate that, due to these fluctuations, the usual calibration approach
underestimates line intensities when their values are not negligible compared to the system temperature.
In order to minimise the negative effects of these potential fluctuations, they propose a least squares frequency switching procedure to
reconstruct the spectrum observed by this technique. At the Yebes 40m we have observed that the line intensity discrepancies between position switching and a classical frequency switching approach are always below 20\%. Since position switching provides much more accurate results, frequency switching is only recommended in special cases, where a clean reference position is too far away from the target source and where the spectrum shows sparse emission lines per band and narrow line widths.

Finally, to obtain $T_{\rm A}^*$ we correct for atmospheric attenuation and forward efficiency,

\begin{equation}
T_{\rm A}^* = T_{\rm A} \frac{e^{\tau A}}{\eta_{\rm F}},
\label{eq.ta*}
\end{equation}
and the main beam temperature ($T_{\rm MB}$), which is the brightness temperature of an equivalent source just filling the main beam, is derived by

\begin{equation}
T_{\rm MB} = T_{\rm A}^* \frac{\eta_{\rm F}}{\eta_{\rm MB}}
\label{eq.tmb}
\end{equation}
where $\eta_{\rm MB}$ is the main beam efficiency (see Sect.\,\ref{sect_mb}).

Continuum observations available at the 40~m RT are total power continuum drifts or pseudocontinuum drifts. In both cases the telescope moves continuously between two points on the sky during a certain time interval while integrating. The integration time is configurable as well as the total time for the drift. The integration time is usually chosen so that the movement of the antenna during that time is less than one beamwidth. Continuum scans are usually composed of two or four scans (back and forth) in azimuth and elevation although the direction of these drifts can also be set by the astronomer. Total power continuum drifts use a complete 2.5\,GHz FFTS backend in which all channels are summed. Simultaneous observations for the 2 polarizations $\times$ 8 sections (at different central frequencies along the Q or W bands) are allowed. Pseudocontinuum drifts also require a 2.5\,GHz FFTS backend and are used for focus or pointing calibration with sources with intense spectral emission like masers. The spectral channels in which no emission is detected are averaged and this number is substracted from the average of the channels where emission is detected. This latter type of observation, which allows a flexible choice of channels, is very convenient when used with masers, for example SiO $v=1$ $J=1-0$ and $J=2-1$, and provides very good baselines even with non-favorable weather conditions.

Continuum drifts are given in $T_{\rm sys}^*$ (see Eq.\,\ref{eq.tsys*}) units, that is, corrected for atmospheric attenuation and forward efficiency, while pseudocontinuum drifts are in arbitrary units and yield relative intensity values.

\begin{equation}
T_{\rm sys}^* = T_{\rm sys} \frac{e^{\tau A}}{\eta_{\rm F}},
\label{eq.tsys*}
\end{equation}

The data for spectral and continuum observations are generated in \texttt{CLASS} format (see Sect.\,\ref{sect_software}). The pipeline yields, for spectral observations, $T_{\rm A}^*$, as well as $V_{\rm sky}$, $V_{\rm hot}$, and $V_{\rm cold}$ during the calibration; all as a function of frequency.
The voltage values allow us to check the calibration procedure and to determine $T_{\rm rec}$ over the whole frequency range, i.e. channel by channel (see Sect.\,\ref{sect_trec}). The pipeline also generates continuum observations in $T_{\rm sys}^*$ as a function of position.
All calibration parameters are available at the \texttt{CLASS} header of each observation. 

\begin{table}[tbh]
\begin{center}
\caption{Yebes 40m Estimated efficiencies from different contributions at four observing frequencies.}
\label{t.efficiencies_mirrors}
\begin{tabular}{ccccc}
\hline \noalign {\smallskip}
Component & 32 GHz & 50 GHz & 72 GHz & 90 GHz \\
\hline \noalign {\smallskip}
$\eta_{\rm m1}$               & 0.95  & 0.87  &  0.75  & 0.65 \\
$\eta_{\rm m2}$               & 0.99  & 0.97  &  0.94  & 0.91  \\
$\eta_{\rm mn}$               & 1.00  & 0.97  & 0.97  &  0.95 \\
$\eta_{\rm ohm}$                & 1.00   & 1.00   &  1.00   & 1.00 \\
$\eta_{\rm block}$                & 0.92  & 0.92  &  0.92  & 0.92 \\
$\eta_{\rm mbr}$               & 0.98  & 0.98  &  0.97  & 0.97 \\
$\eta_{\rm i}$                  & 0.80  & 0.80  &  0.82  & 0.82 \\
\hline
Total $\eta_{\rm A}$$^{(a)}$ & 0.67  & 0.59  &  0.50  & 0.41 \\
\hline \noalign {\smallskip}
\end{tabular}
\end{center}
Notes. (a) Assuming $\eta_{\rm err}=1$.
\end{table}

\section{Telescope performance: efficiencies}
\label{sect_eficiencias}

\subsection{Aperture efficiency}
\label{sect_aperture_eff}

The aperture efficiency of a radio telescope ($\eta_\mathrm{A}$) is defined as the ratio between the effective antenna area and the geometrical area of the aperture.
It depends on the observing frequency and can be expressed as the product of several components:

\begin{eqnarray}
\eta_{\rm A} & = & \eta_{\rm m1} \eta_{\rm m2} \eta_{\rm mn} \eta_{\rm ohm} \eta_{\rm block} \eta_{\rm mbr} \eta_{\rm i} \eta_{\rm err} \\
\label{eq.eta1}
\end{eqnarray}
where $\eta_{\rm m1}$ and $\eta_{\rm m2}$ are the efficiency of the main reflector and of the subreflector, respectively, and $\eta_{\rm mn}$ includes the efficiency of all the other Nasmyth mirrors along the ray path. Additional errors which affect the total aperture efficiency are ohmic losses ($\eta_{\rm ohm}$), the efficiency of the membrane ($\eta_{\rm mbr}$) at the vertex (see Sect.\,\ref{sect_cabin}), the blockage by the tetrapod and subreflector ($\eta_{\rm block}$), and the illumination ($\eta_{\rm i}$) on the primary reflector (see Sect.\,\ref{sect_optics}). Optical effects such as defocusing and astigmatism are also important and usually depend on the elevation of the antenna and can be grouped under $\eta_{\rm err}$.

The efficiency from the mirrors can be estimated from Ruze's formula \citep{Ruze1966} assuming a given RMS error of their surface ($\sigma$).

\begin{eqnarray}
	\eta_{\rm mirror} = e^{-\left(\frac{4 \pi \sigma}{\lambda}\right)^2 }
\label{eq.ruze}
\end{eqnarray}
where $\lambda$ is the wavelength. Table\,\ref{t.efficiencies_mirrors} summarises the efficiency for the primary mirror (M1), subreflector (M2), the rest of Nasmyth mirrors, and the other components. 
We assumed that $\sigma$ is 175~$\mu$m for M1 (see Sect.\,\ref{sect_holography}), 50~$\mu$m for M2, 25~$\mu$m for M3 and M4, and 40~$\mu$m for the rest of Nasmyth mirrors.
The efficiency due to blockage is 0.92 at all frequencies \citep{devicente1998} following \cite{ruze1968} and assuming an illumination with a taper of $-$12\,dB at the edge of the dish. $\eta_{\rm mbr}$ values at different frequencies were measured using the same method as that described in \citet{Garcia2017} and 
$\eta_{\rm i}$ are those derived in Sect.\,\ref{sect_optics}. 
Table\,\ref{t.efficiencies_mirrors} shows the aperture efficiency derived assuming that $\eta_{\rm err}=1$.

In addition, the aperture efficiency can be measured from observations of a known flux source whose apparent size is relatively small compared to the telescope beam. In such a situation, according to \cite{baars07}:
\begin{equation} 
\eta_\mathrm{A} = \frac{2 k}{S A_\mathrm{g}} C_\mathrm{S} \eta_\mathrm{F} T_{\rm A}^*
\end{equation}
where $T_{\rm A}^*$ is the antenna temperature corrected for atmospheric opacity and forward efficiency (see Sect.\,\ref{sect_calibration}), $\eta_\mathrm{F}$ the forward efficiency (see Sect.\,\ref{sect_forward}), $k$ the Boltzmann constant, $S$ the flux density of the source, $A_\mathrm{g}$ the geometric aperture of the antenna, and $C_\mathrm{S}$ a correction factor arising from the deconvolution between the source and the telescope beam:

\begin{equation} 
C_\mathrm{S} = \frac{(\frac{x}{1.2})^2}{1-e^{-(\frac{x}{1.2})^2}}
\end{equation}
where, in turn, $x=\frac{\theta_\mathrm{S}}{\theta_\mathrm{b}}$ is the ratio of source to beam apparent size. This approximation is valid as long as the emission of the source can be reasonably approximated to a disk with homogeneous illumination. Such is the case of most planets and of Saturn in particular.

$\eta_\mathrm{A}$ was measured along the full bandwidth of the Q and W bands from observations towards Saturn on the 14th and 29th of August, 2019. The apparent size $\theta_\mathrm{S}$ of the planet and its flux $S$ at the central frequency of each FFTS section was obtained from \texttt{ASTRO} (part of \texttt{GILDAS} software) for the time and date of each individual observation. The resulting aperture efficiencies provided in Table\,\ref{t.etas} for each section correspond to the maximum value obtained across the observing session to reduce the effects of deformations in the main reflector due to thermal gradients. Therefore the efficiencies given in Table\,\ref{t.etas} are those measured at night in summer (23$^{\circ}$C average ambient temperature).

While Saturn allowed for a measurement of $\eta_\mathrm{A}$, its maximum elevation ($\sim$27$^\circ$ at the time of observations) prevented from measuring the aperture efficiency as a function of elevation. At Q band we observed 3C84, a stable (during the time of the observation) bright quasar with a size much smaller than the telescope beam, covering the whole range of elevations along its path on the sky. The relative values of the antenna gain were matched to the aperture efficiencies obtained towards Saturn at those elevations where both sources were observed.
Figure\,\ref{f.eficiencias} shows the resulting aperture efficiencies at Q band for different FFTS sections (i.e. different frequencies) as a function of elevation. A slight ($\sim$5\%) decrease in $\eta_\mathrm{A}$ with increasing elevation is apparent in every section of the Q band.

The deformation of the telescope structure due to temperature changes seems to have a large effect on 
$\eta_{\rm A}$ at W band. The estimated average efficiency, at summer night time ($\sim$23$^{\circ}$C ambient temperature), is 15\%. 
We plan to fulfill a a new campaign of measurements during winter night time to minimise the thermal effects and find out if 
$\eta_{\rm A}$ increases under better atmospheric conditions mainly at short wavelengths.

\begin{table*}[tbh]
\begin{center}
\caption[]{Half power beam width (HPBW) and efficiencies
of Yebes 40\,m radio telescope along the Q and W bands.}
\label{t.etas}
\begin{tabular}{cccccc}
\hline \noalign {\smallskip}
Frequency & Wavelength & HPBW$^{(a)}$ & $\eta_\mathrm{A}$ & $\eta_{\rm MB}$ & $\eta_{\rm F}$ \\
(GHz) & (mm) & ($''$) &  & & \\
\hline \noalign {\smallskip}
\hline \noalign {\smallskip}
\multicolumn{6}{c}{Q band}\\
\hline
32.4 & 9.2 & 54.4 & $0.50\pm0.05$ & $0.60\pm0.06$ & $0.92\pm0.03$ \\
34.6 & 8.6 & 50.9 & $0.48\pm0.05$ & $0.57\pm0.06$ & $0.92\pm0.03$ \\
36.9 & 8.1 & 47.8 & $0.46\pm0.05$ & $0.54\pm0.06$ & $0.92\pm0.03$ \\
39.2 & 7.6 & 45.0 & $0.44\pm0.04$ & $0.52\pm0.06$ & $0.92\pm0.03$ \\
41.5 & 7.2 & 42.5 & $0.42\pm0.05$ & $0.51\pm0.06$ & $0.93\pm0.03$ \\
43.8 & 6.8 & 40.2 & $0.40\pm0.05$ & $0.48\pm0.06$ & $0.93\pm0.03$ \\
46.1 & 6.5 & 38.2 & $0.39\pm0.06$ & $0.47\pm0.07$ & $0.92\pm0.03$ \\
48.4 & 6.2 & 36.4 & $0.36\pm0.07$ & $0.43\pm0.08$ & $0.91\pm0.03$ \\
\hline                       
\multicolumn{6}{c}{W band}\\
\hline
72.5 & 4.1 & 24.3 & $0.19\pm0.01$ & $0.23\pm0.02$ & $0.90\pm0.03$ \\
74.7 & 4.0 & 23.6 & $0.18\pm0.01$ & $0.22\pm0.02$ & $0.89\pm0.03$ \\
77.0 & 3.9 & 22.9 & $0.17\pm0.01$ & $0.21\pm0.02$ & $0.88\pm0.03$ \\
79.3 & 3.8 & 22.2 & $0.15\pm0.01$ & $0.18\pm0.02$ & $0.88\pm0.03$ \\
81.6 & 3.7 & 21.6 & $0.14\pm0.01$ & $0.17\pm0.02$ & $0.88\pm0.03$ \\
83.9 & 3.6 & 21.0 & $0.13\pm0.02$ & $0.16\pm0.03$ & $0.88\pm0.03$ \\
86.2 & 3.5 & 20.4 & $0.12\pm0.02$ & $0.14\pm0.03$ & $0.88\pm0.03$ \\
88.5 & 3.4 & 19.9 & $0.11\pm0.02$ & $0.13\pm0.03$ & $0.88\pm0.03$ \\
\hline \noalign {\smallskip}
\end{tabular}
\end{center}
Notes. (a) The HPBW is estimated as 1.14 $\lambda$/D according to the PO simulations of the illumination 
(Sect.\,\ref{sect_optics}.).
\end{table*}

There is a discrepancy between the measured aperture efficiency (Table \ref{t.etas}) and the estimated one (Table \ref{t.efficiencies_mirrors}). 
Using Eqs.\,\ref{eq.eta1} and \ref{eq.ruze} we can estimate the RMS of the primary reflector if $\eta_{\rm A}$ and the rest of the efficiencies are known. 
Assuming $\eta_{\rm err}=1$, the measured aperture efficiency is compatible with a $350-400$\,$\mu$m surface error for the primary reflector depending on the frequency.
It is worth noting that this value cannot be directly compared with that obtained from holography (175\,$\mu$m, see Sect.\,\ref{sect_holography}) 
since holography measurements substract the large scale deformations of the primary reflector surface whereas this estimated $\eta_{\rm A}$ includes such large scale deformations, like for example astigmatism, mirror misalignments, defocusing, gravitational deformation and encompasses them into a single larger RMS value. Besides, holography was performed in winter whereas these measurements were done in summer and the surface of the main reflector is not directly comparable.

Alternatively, if we assume that $\eta_{\rm m1}$, as derived from Table\,\ref{t.efficiencies_mirrors} (i.e. $\sigma=175$\,$\mu$m), and $\eta_{\rm A}$, as shown in Table\,\ref{t.etas}, are correct, 
then $\eta_{\rm err}$ should be the source of the decrease of the aperture efficiency. 
In that case, $\eta_{\rm err}$ should take a value between 0.7 and 0.6 along the Q band and between 0.4 and 0.2 for W band. As mentioned above, 
$\eta_{\rm err}$ accounts for effects such as defocusing or astigmatism which may contribute in a large percentage, especially at W band, to the decrease of the measured efficiency. 
However we believe that the discrepancy is due to a combination of those effects plus a larger surface error for the primary reflector than the estimated one using holography. 
This discrepancy is still under investigation and we plan to measure, among other tasks, thermal gradients at the structure as mentioned in Sect.\,\ref{sect_holography}.

\subsection{Main beam efficiency}
\label{sect_mb}

The main beam efficiency ($\eta_{\rm MB}$) is the fraction of power received which enters through the main beam. It is usually derived from continuum observations of planets whose angular diameter fills the main beam. This value is related to $\eta_{\rm A}$ via the antenna illumination. For a Gaussian beam \citep{Kramer1997}:

\begin{equation}
\eta_{\rm mb} = 2.092 \times 10^{-5} \eta_{\rm A} \theta_{\rm b}^2 \frac{D^2}{\lambda^2},
\label{eq_eta_mb}
\end{equation}
where $\theta_{\rm b}$ is the half power beam width (HPBW) [arcsec], D [m] is the diameter of the telescope, and $\lambda$ [mm] is the wavelength of observation. Table\,\ref{t.etas} lists $\eta_{\rm mb}$ values derived according to Eq.\,\ref{eq_eta_mb} along the Q and W bands.

\subsection{Forward efficiency}
\label{sect_forward}
The forward efficiency ($\eta_{\rm F}$) represents the coupling between the receiver and the sky, considered an extended source with a uniform brightness temperature. The usual procedure to estimate $\eta_{\rm F}$ are skydip observations. This technique measures the sky emissivity in $T_{\rm A}^*$ at different air masses ($A$). Observations at Q and W band were performed on May 29th 2019 and September 26th 2019, respectively, using all sections of the FFTS simultaneously. The total power measured in each FFTS was averaged and stored every 100 ms while the antenna moved from 6 to 85 degrees in elevation (9.5 and 1 airmasses, respectively).
These skydip drifts at Q and W bands were performed under good weather conditions, in early morning, towards a clear sky. 
We derived $\eta_{\rm F}$ for each band, polarisation, and section by fitting the data between 1.5 and 3.5 airmasses to Eq. \ref{eq.Tsys2}.
Results are summarised in Table\,\ref{t.etas}. $\eta_{\rm F}$ is close to 0.9 along the whole Q and W bands for both polarisations.

\subsection{Moon Coupling}

\label{sect_moon}

\begin{figure}
 \includegraphics[width=9cm]{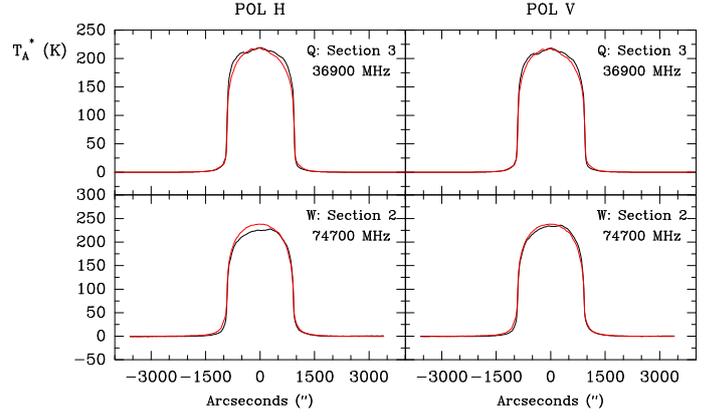}
 \centering
 \caption{Selected pointing drifts at 36900\,MHz and 74700\,MHz. Azimuth (elevation) drift is depicted in black (red).}
 \label{fig_luna}
\end{figure}

\begin{figure}
 \includegraphics[width=9cm]{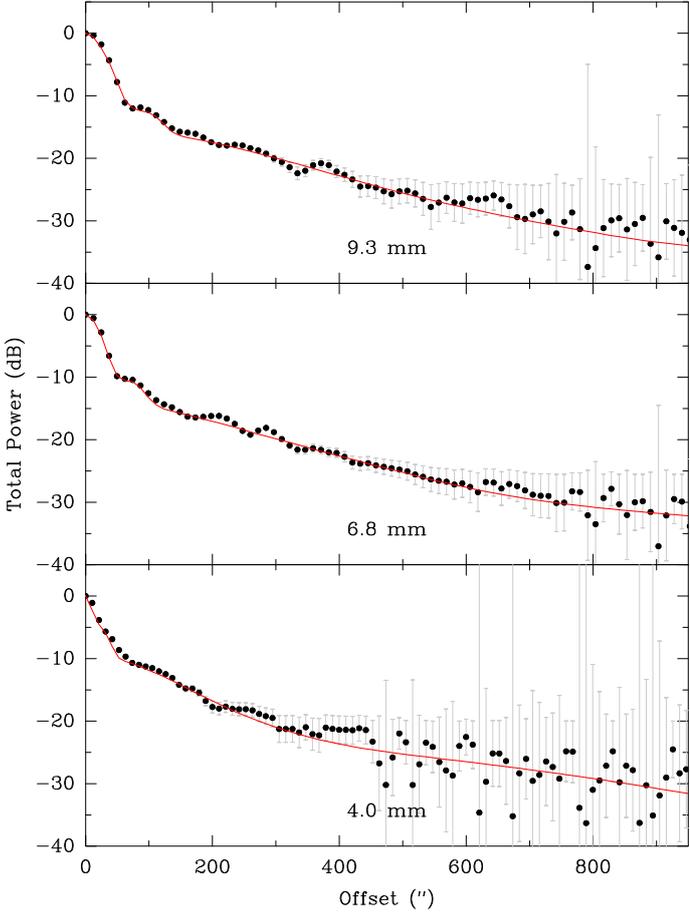}
 \centering
 \caption{Composite profiles observed around Full Moon (black points) together with the synthetic best-fit profiles (red curves). For a more detailed explanation see Sect.\,\ref{sect_moon}.}
 \label{fig_luna_beam}
\end{figure}

\begin{figure}
 \includegraphics[width=7cm]{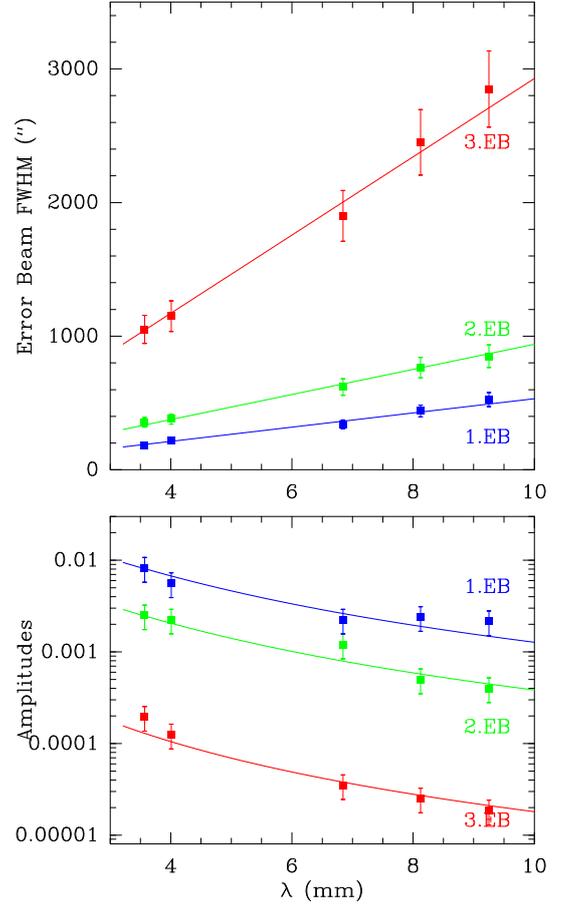}
 \centering
 \caption{Best-fit values (solid squares) of the widths (upper panel) and amplitudes (bottom panel) of the error beams (EB). Solid lines fit these values to Eq.\,\ref{eq_L} (upper panel) and Eq.\,\ref{eq_a} (bottom panel) in order to obtain the correlation length ($L$) and the RMS value of the surface errors ($\sigma$). For a more detailed explanation see Sect.\,\ref{sect_moon}.}
 \label{fig_luna_L_a}
\end{figure}

To evaluate the percentage of power lost through the error beam of the telescope, we have quantified the coupling between the Moon and its beam pattern ($\eta_{\rm Moon}$). Most of the astronomical sources are rather compact and their sizes in the sky are well below $\sim0.5^{\circ}$ (the apparent size of the Moon in the sky) so that $\eta_{\rm Moon}$ is a useful measurement for such cases. The Moon has been extensively used as calibrator and its brightness temperature at submillimetre, millimetre and centimetre wavelengths is well known (see \citealt{Pardo2005}, \citealt{Krotikov1987}).

To estimate $\eta_{\rm Moon}$ we performed pointing drifts at different frequencies towards the Moon, and close to full moon, on June 19th and October 17th, 2019 at W and Q bands, respectively. The spatial resolution of the scans was $\sim$\,10$''$. We used all FFTS sections, each of them operating as a continuum detector.

Since we expect $\eta_{\rm Moon}$ to be relatively constant along each observing band, we only analysed pointing drifts at 36900\,MHz ($\lambda=8.1$\,mm) and 74700\,MHz ($\lambda=4.0$\,mm), as shown in Fig.\,\ref{fig_luna}. These frequencies were chosen to fit those from \citet{Krotikov1987} and avoid interpolations. For the optical phase of the Moon at the moment of the observations, we derived a Moon brightness temperature of 293\,K and 311\,K for 8\,mm and 4\,mm, respectively. 
From our observations, we obtained $T_{\rm A}^*\sim220$\,K at 8\,mm which gives 
\mbox{$\eta_{\rm Moon}$\,(Q band)\,=\,$0.75\pm0.11$} and $T_{\rm A}^*\sim239$\,K at 4\,mm, yielding \mbox{$\eta_{\rm Moon}$\,(W band)\,=\,$0.77\pm0.12$}. 

The comparison between the measured antenna temperature, at full moon, and the estimated brightness temperature  allows us to estimate the fraction of power lost by the telescope over extended sources at different wavelengths and provides an effective ``forward'' efficiency. In any case we calibrate following the classical approach which uses the forward efficiency as determined from Sect.\,\ref{sect_forward}.

\begin{table*}[tbh]
\label{table_beam}
\begin{center}
\caption[]{Beam parameters of Yebes 40\,m radio telescope.}
\begin{tabular}{rllll}
\hline \noalign {\smallskip}
  & Main beam & 1st error beam & 2nd error beam & 3rd error beam \\ 
\hline \noalign {\smallskip}
\hline \noalign {\smallskip}
Origin & Diffraction & Large-scale & Panel frame & Panel\\
       & pattern & deformations & misalignments & deformations\\
Correlation length & & $L_{\rm 1}=4.1\pm0.2$\,m &  $L_{\rm 2}=2.3\pm0.1$\,m & $L_{\rm 3}=0.75\pm0.03$\,m\\
RMS-value & & $\sigma_{\rm 1}=253\pm20$\,$\mu$m & $\sigma_{\rm 2}=245\pm19$\,$\mu$m & $\sigma_{\rm 3}=164\pm12$\,$\mu$m \\
\hline \noalign {\smallskip}
\bf{9.2\,mm / 32.4\,GHz} & & & & \\
\cline{1-1} \noalign {\smallskip}
FWHM ($\theta_{\rm i}$) ($''$) & $56\pm5$ & $525\pm52$ & $850\pm85$ & $2850\pm285$ \\ 
Power amplitude ($a_{\rm i}$) & 1 & $0.0022\pm0.0006$ & $0.00040\pm0.00012$ & $(1.8\pm0.6)\times10^{-5}$ \\
Relative Power$^{(a)}$ ($P_{\rm i}$)  & 0.75 & 0.14 & 0.07 & 0.04 \\
\hline \noalign {\smallskip}
\bf{8.1\,mm / 36.9\,GHz} & & & & \\
\cline{1-1} \noalign {\smallskip}
FWHM ($\theta_{\rm i}$) ($''$)& $49\pm5$ & $440\pm44$ & $765\pm76$ & $2450\pm245$ \\ 
Power amplitude ($a_{\rm i}$) & 1 & $0.0024\pm0.0007$ & $0.0005\pm0.0002$ & $(2.5\pm0.8)\times10^{-5}$ \\
Relative Power$^{(a)}$ ($P_{\rm i}$)  & 0.73 & 0.14 & 0.09 & 0.04 \\
\hline \noalign {\smallskip}
\bf{6.8\,mm / 43.8\,GHz} & & & & \\
\cline{1-1} \noalign {\smallskip}
FWHM ($\theta_{\rm i}$) ($''$) & $42\pm4$ & $340\pm34$ & $620\pm62$ & $1900\pm190$ \\ 
Power amplitude ($a_{\rm i}$) & 1 & $0.0022\pm0.0007$ & $0.0012\pm0.0004$ & $(4\pm1)\times10^{-5}$ \\
Relative Power$^{(a)}$ ($P_{\rm i}$)  & 0.68 & 0.10 & 0.17 & 0.05 \\
\hline \noalign {\smallskip}
\bf{4.0\,mm / 74.7\,GHz} & & & & \\
\cline{1-1} \noalign {\smallskip}
FWHM ($\theta_{\rm i}$) ($''$) & $22\pm2$ & $220\pm22$ & $380\pm38$ & $1150\pm115$ \\ 
Power amplitude ($a_{\rm i}$) & 1 & $0.006\pm0.002$ & $0.0022\pm0.0007$ & $(1.2\pm0.4)\times10^{-4}$ \\
Relative Power$^{(a)}$ ($P_{\rm i}$)  & 0.38 & 0.21 & 0.26 & 0.13 \\
\hline \noalign {\smallskip}
\bf{3.6\,mm / 83.9\,GHz} & & & & \\
\cline{1-1} \noalign {\smallskip}
FWHM ($\theta_{\rm i}$) ($''$) & $18\pm2$ & $185\pm18$ & $355\pm36$ & $1050\pm105$ \\ 
Power amplitude ($a_{\rm i}$) & 1 & $0.008\pm0.002$ & $0.0025\pm0.0007$ & $(1.9\pm0.6)\times10^{-4}$ \\
Relative Power$^{(a)}$ ($P_{\rm i}$)  & 0.29 & 0.24 & 0.27 & 0.18 \\
\hline \noalign {\smallskip}
\end{tabular}
\end{center}
Notes. (a) The integrated relative power ($P_{\rm i}$) of each beam is estimated as 
$P_{\rm i}=a_{\rm i}\theta_{\rm i}^{2} / \sum_{\rm i=0,3} a_{\rm i}\theta_{\rm i}^{2}$, 
see Sect.\,6 of \citet{Greve1998}. 
\end{table*}

Total power scans towards the Moon at 9.2, 8.1, 6.8, 4.0, and 3.6\,mm were also used to derive the beam pattern of the 
Yebes 40\,m telescope down to approximately $-$30\,dB (0.1\%) and up to a width between $1000''-3000''$, depending on the wavelength (see \citealt{Greve1998} and \citealt{Kramer2013} for a detailed description of this method). Yebes 40\,m beam pattern can be described by the main beam and three error beams which are produced by large-scale deformations at the primary reflector such as astigmatism, panel frame misalignments, and panel deformations.

Full width at half maximum ($\theta$; FWHM) and amplitudes ($a$) of these error beams together with the correlation length of the deformations ($L$) and the RMS-value of the surface errors ($\sigma$) can be estimated following the method described in \citet{Greve1998}. First, the total power scans across the limb of the Moon are differentiated along the scan direction. Then, five functions (main beam, the first secondary lobe, and three error beams) with different widths and amplitudes are combined, convolved with the Moon shape and differentiated to obtain a composite synthetic profile which is compared with the differentiated Moon scans (see Sect.\,3.2 of \citealt{Greve1998}). Figure\,\ref{fig_luna_beam} shows the profiles obtained from azimuth total power scans together with composite synthetic profiles at three different wavelengths. The ``best-fit'' profiles were obtained by hand trying to minimise the RMS of the fit. See \citet{Pardo2017} for methods to measure error beams at shorter wavelengths.

Although the diffraction pattern of the antenna is properly described by [$J_{\rm 1}$($u$)/$u$]$^2$ with $J_{\rm 1}$ being the Bessel function of first kind and $u$ the spatial coordinate in the focal plane, we used independent Gaussian functions for the main beam and the first secondary lobe. On the other hand, no terms have been introduced in the composite synthetic beam accounting for panel buckling. We derived independent parameters of the 40\,m RT beam pattern for azimuth and elevation drifts. Nevertheless, the amplitude and $\theta$ results were similar (within uncertainties) in both directions.

Estimated beam parameters for the averaged results of azimuth and elevation scans are summarised in Table\,\ref{table_beam}. The percentage of power of the main beam varies between 75\% at 9.2\,mm and 29\% at 3.6\,mm. In our fits, the first secondary lobe contributes less than 1.5\% in power at all frequencies and we did not include it in Table\,\ref{table_beam}. We note that small errors of individual panels and panel frame misalignments, accounted in the third and second error beam respectively, are particularly important at shorter wavelengths. On the other hand, large scale deformations significantly contribute to power losses at all wavelengths. These large-scale deformations are mostly due to thermal gradients and should be addressed in the future by the thermalisation of the primary reflector. These deformations, in fact, can cause larger power losses during daytime. The panel frame misalignments are evaluated by holography techniques and are periodically corrected (see Sect.\,\ref{sect_holography}).

The correlation length of the deformations and the RMS of the surface errors associated with each error beam can be obtained according to
\begin{equation} 
\label{eq_L}
\theta_\mathrm{i} = \frac{0.53\lambda_\mathrm{i}}{(L_\mathrm{i}/2)}
\end{equation}
and
\begin{equation} 
\label{eq_a}
a_\mathrm{i} = (L_\mathrm{i}/D)^2 \frac{1-e^{\sigma_\phi^2}}{\eta_\mathrm{A}}
\end{equation}
where $\sigma_{\phi}=\eta_{\rm i} (4 \pi / \lambda) \sigma$ with $\eta_{\rm i}=0.8$ being the illumination efficiency (see Sect.\,\ref{sect_optics}) and $\eta_{\rm A}=0.52$ the aperture efficiency at long wavelengths (9.2\,mm, 
see Sect.\,\ref{sect_aperture_eff}). Figure\,\ref{fig_luna_L_a} shows the fits from which correlation lengths and RMS surface errors are obtained. The complete formalism can be found in Sect.\,2.3 of \citet{Greve1998}. We obtain deformation correlation lengths of 4.1, 2.3, and 0.75\,m for the first (large-scale deformations), second (panel frame misalignments), and third (panel deformations) error beam, respectively. These values are in agreement with the expected length of these inhomogeneities.

The associated RMS-values of the surface errors are 253\,$\mu$m (1st error beam), 
245\,$\mu$m (2nd error beam), and 164\,$\mu$m (3rd error beam) which gives a combined (root-sum-square, RSS) value of 390\,$\mu$m
which matches the estimated one in Sect.\,\ref{sect_aperture_eff}. As we only fit individual Gaussians for the main beam and the first secondary lobe, the individual estimation of the RMS for each error beam is affected by the contribution of the rest of secondary lobes of the diffraction pattern. This may result in overestimating the RMS surface errors by a difficult to estimate factor. Therefore it is not possible to make a direct comparison between the combined RSS value of $\sigma_{\rm 2}$ and $\sigma_{\rm 3}$ with the holography measurements which contain all the effects of blocking by the secondary, quadrupod and the full set of secondary beams from the main mirror and these optical elements.

\section{First astronomical observations} 
\label{sect_survey}

For the commissioning of the new Q band and W band Yebes 40\,m telescope receivers
we have undertaken very deep integrations of several evolved stars. In particular
we have chosen as first targets IRC+10216 (prototypical carbon-rich AGB star),
CRL 2688 (the Egg Nebula, a young protoplanetary nebula), and CRL 618
(a protoplanetary nebula). These well known C-rich stellar sources are among the richest in
molecular content, and strongest due to their distances. As
it is unfeasible to observe a single object evolving through
the post-AGB phase, a rough approach is to use the objects in this small
sample to represent the average characteristics in four stages
of this evolution. A fourth object, NGC 7027 (a young planetary nebula
object) was in the systematic study of \citet{Herpin2002} as
representative of the transition from carbon-rich AGB star to the Planetary
Nebula stage, but it has not been included in this series. 
We should keep in mind in this comparison
that IRC+10216 is $5-10$ times closer than the other three objects. Some results from 
these data have already been published such as the first detection of MgC$_3$N and MgH$_4$H \citep{Cernicharo2019b}
and vibrationally excited HC$_7$N and HC$_9$N \citep{Pardo2020}. 

Once the NANOCOSMOS Q band receiver was fully operational at the
40\,m telescope (May 2019) we started the very deep integrations that
resulted in the data presented in this paper. The deepest integration was
achieved on IRC+10216, and reached 128.5 hours of on source position switching
telescope time (see \citealt{Cernicharo2019b} and \citealt{Pardo2020} for more details).
For a typical system temperature of $\sim$60\,K at the centre
of the receiver bandpass ($\sim$41\,GHz), it translates into $0.85-0.90$\,mK RMS
at the ENBW of the backends used (44 kHz, FFTS).
As the range of frequencies simultaneously covered by this receiver is very large
and no tunings were necessary to move across the bandpass, the efficiency of the
integration has been very large. As a result, the sensitivity of our data is much
larger than the previous similar survey carried out with the Nobeyama 45\,m telescope
\citep{Kawaguchi1995}. The position switching integration times for CRL 2688
and CRL 618 were 98.75 and 45.63 hours, respectively. 

\begin{figure*}
\includegraphics[angle=0,width=\linewidth]{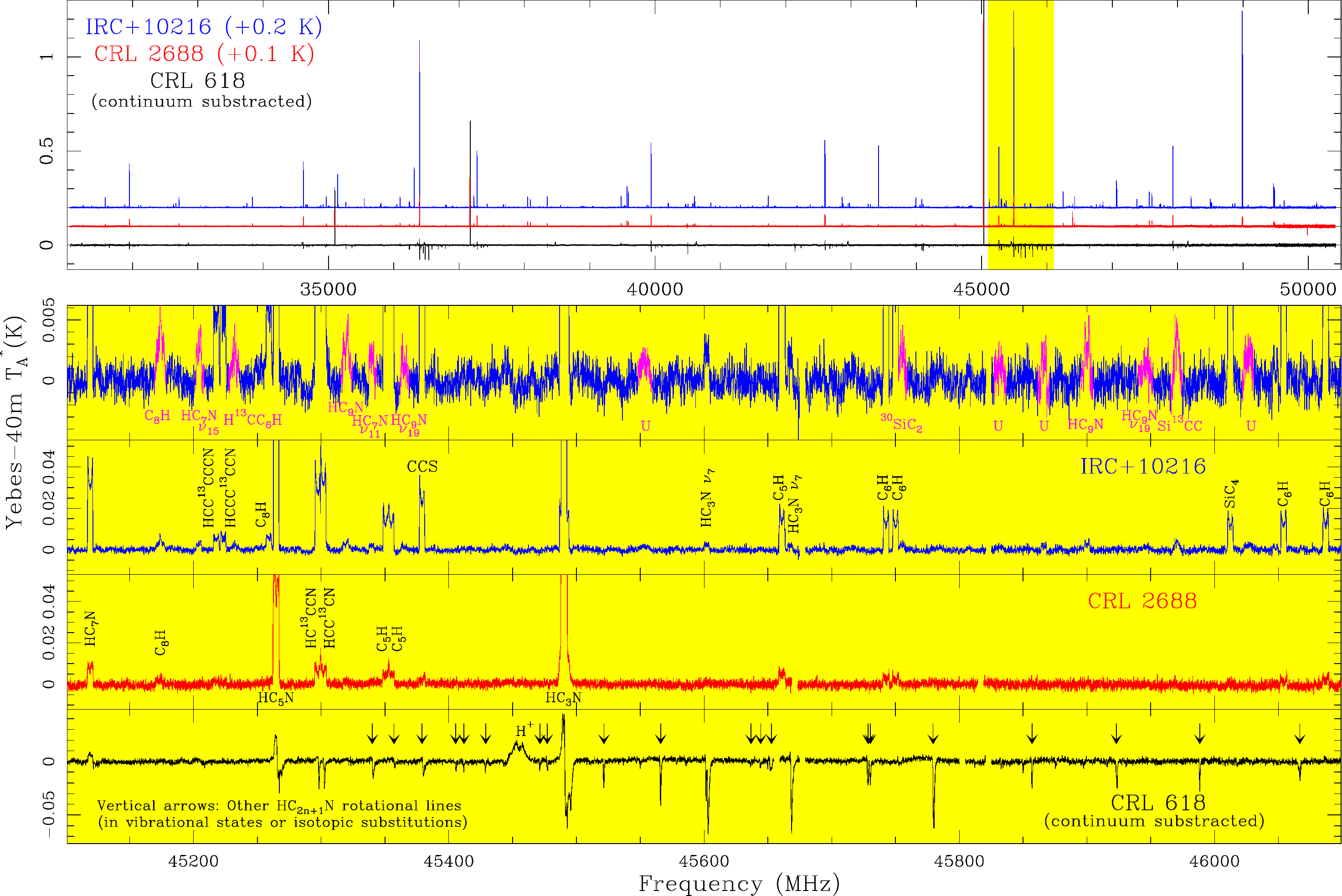}
\caption{First deep integrations on IRC+10216, CRL 2688 and, CRL 618 achieved with the
new Q band Nanocosmos receiver at Yebes 40\,m telescope.}
\label{fig-qband}
\end{figure*}

The data are presented in Fig.\,\ref{fig-qband} for IRC+10216, CRL 2688, and CRL 618.
The calibrated output spectra in \texttt{CLASS} format (see Sect.\,\ref{sect_software}) were reduced
  using \texttt{GILDAS}. Spikes were removed and the baseline was flattened in 0.5\,GHz sections using polynomial fits of order 3.    
We display the full frequency range covered by the Q band receiver, with identifications
of the strongest lines and a zoom to the $45.1-46.1$\,GHz range. The detailed spectra with
all line identifications and their discussion in the framework of the Nanocosmos project
and other line surveys will be published elsewhere.

\begin{figure*}
\includegraphics[angle=0,width=\linewidth]{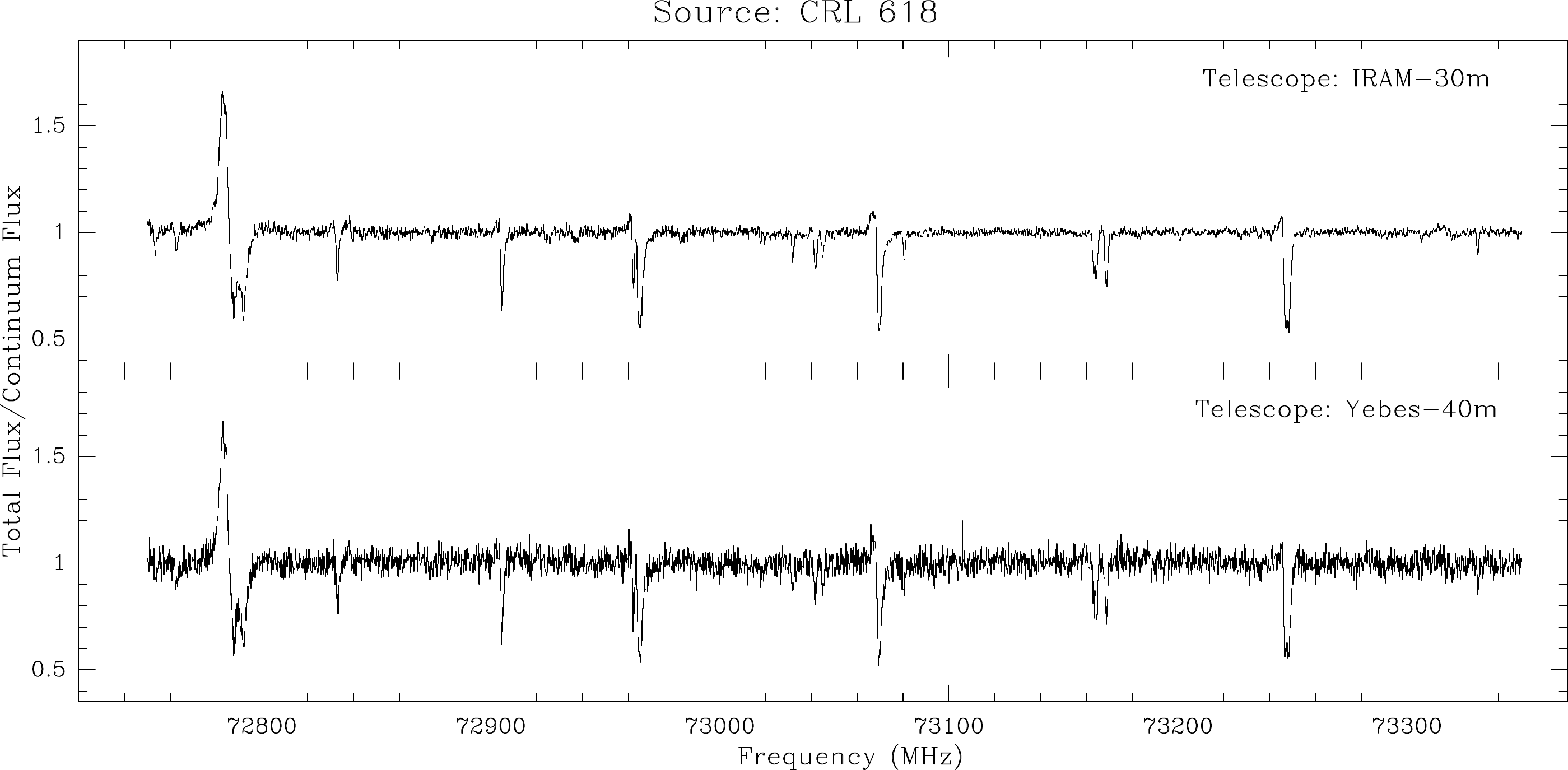}
\caption{Total flux/continuum flux spectra obtained the new W band Nanocosmos
receiver installed at Yebes 40\,m telescope and with the E090 receivers at
	the IRAM 30\,m telescope. Integration time at the 40\,m: 199 minutes (March 2019). Integration time at the 30\,m: 55 minutes (December 2019).}
\label{fig-wband}
\end{figure*}

A complementary commissioning task that we have carried out was to test the brand new
W band NANOCOSMOS receiver. For a non-thermally stabilised 40\,m antenna it is obvious
that the efficiency beyond 70\,GHz is expected to be small (see Sect.\,\,\ref{sect_eficiencias}).
Nevertheless, a large effort
has been done on upgrading the primary dish surface to get the best possible efficiency (see Sect.\,\ref{sect_holography}).
Taking into account the previous limitations we conducted a deep integration on 
CRL 618 (199 minutes integration time in March 2019 in both polarisations) because this object has a
continuum flux and the total flux divided by this continuum flux should be similar
for similar telescopes. In this case, the direct comparison can be done with the
IRAM 30\,m telescope. In parallel to our W band receiver commissioning work at Yebes,
we carried out IRAM 30\,m observations on the same source, CRL 618 (55 minutes integration time in December 2019 in both polarisations).
The continuum flux measured at Yebes is {$\sim$4} times weaker than that obtained in Granada at W band. However,
the total flux/continuum flux are very similar in both cases (see Fig.\,\ref{fig-wband}).
We can roughly say that we have a 8 times less efficient telescope at Yebes than in
Granada, and the efforts should focus on improving that efficiency as much as possible.

\section{Summary and conclusions}
In this paper we have presented the current capabilities of Yebes 40\,m telescope at Q and W bands.
The complete antenna-receiver system and the performance of the new NANOCOSMOS receivers and the telescope have been described.
Observations between $31.5-50$\,GHz and $72-90.5$\,GHz can be performed using an instantaneous band of 18\,GHz with an ENBW of 44\,kHz in both H and V linear polarisations. The receiver temperature at Q band is below 40\,K, being below 20\,K for the lower half of the band, and below 70\,K at W band. Both receivers provide the largest instantaneous bandwidths currently available at their bands and together with their receiver temperature demonstrate that the Q band receiver is state of the art while W band provides an excellent behaviour for very high sensitivity observations. The system stability has been improved to obtain good quality astronomical data (flat baselines) for long integration times allowing the detection of weak spectral lines.

A considerable effort has been undertaken to improve the efficiency at millimetre wavelengths.  
During the last two years the efficiencies were improved using holography techniques and checking the mirror aligments at the receiver cabin. We reached a surface accuracy of the primary reflector of 175\,$\mu$m after panel adjustment in 2018. The current values of $\eta_{\rm A}\simeq0.4$ and $\eta_{\rm A}\simeq0.2$ at 7\,mm and 3\,mm, respectively, although still low and requiring improvement, make the telescope competitive since it is equipped with sensitive and ultra-broad band receivers. 

The first astronomical observations using this facility have demonstrated the high capacity of the telescope. Examples that prove it are the recent observation of an extragalactic source at redshift 0.89 \citep{Tercero2020} and the detection of 
new molecular species in the interstellar medium using large integrations at 7\,mm in rich molecular sources \citep{Cernicharo2019b,Tercero2020,Pardo2020}.
The 7\,mm spectrum of IRC+10216 presented here has a sensitivity never seen before in this source at these wavelengths.
Moreover, the 40\,m data at 3\,mm in sources such as CRL618, are fully comparable to those of the IRAM 30\,m telescope although the 40\,m telescope is 8 times less efficient.
To summarise, thanks to the extremely large instantaneous bandwidth available, the high sensitivity and good gain stability of the new NANOCOSMOS receivers,
a 40\,m primary reflector with low surface errors, and the versatility of a Cassegrain antenna for 
flexible designs of the tertiary optics, Yebes 40\,m telescope demonstrates to be a new competitive facility to undertake the study of molecular line surveys minimising the required telescope time.

\begin{acknowledgements}
	We would like to thank the anonymous referee for a helpful report that led to substantial improvements in the paper.
	We also thank J. Pe\~nalver and C. Kramer for their prompt response to our questions regarding the analysis of the error beam using total power Full Moon scans. The research leading to these results has received funding from the European Research Council under the European Union’s Seventh Framework Programme (FP/2007-2013)/ERC-SyG-2013 Grant Agreement No. 610256 NANOCOSMOS. Part of this work was supported by \emph{Programa operativo pluriregional de Espa\~na, YDALGO}. The observations mentioned in this article have been done under commissioning time and within projects 19A010 and 20A017. The Yebes 40m radio telescope belongs to Instituto Geogr\'afico Nacional (IGN).
\end{acknowledgements}

\bibliographystyle{aa} 
\bibliography{references_nanocosmos} 

\begin{thebibliography}{48}
\expandafter\ifx\csname natexlab\endcsname\relax\def\natexlab#1{#1}\fi

\bibitem[{{Ag{\'u}ndez} {et~al.}(2015){Ag{\'u}ndez}, {Cernicharo}, {de
  Vicente}, {Marcelino}, {Roueff}, {Fuente}, {Gerin}, {Gu{\'e}lin}, {Albo},
  {Barcia}, {Barbas}, {Bola{\~n}o}, {Colomer}, {Diez}, {Gallego},
  {G{\'o}mez-Gonz{\'a}lez}, {L{\'o}pez-Fern{\'a}ndez},
  {L{\'o}pez-Fern{\'a}ndez}, {L{\'o}pez-P{\'e}rez}, {Malo}, {Serna}, \&
  {Tercero}}]{agundez15}
{Ag{\'u}ndez}, M., {Cernicharo}, J., {de Vicente}, P., {et~al.} 2015, \aap,
  579, L10

\bibitem[{{Baars}(2007)}]{baars07}
{Baars}, J. W.~M. 2007, {The Paraboloidal Reflector Antenna in Radio Astronomy
  and Communication}, Vol. 348

\bibitem[{{Cernicharo}(1985)}]{Cernicharo1985}
{Cernicharo}, J. 1985, Internal IRAM Report (Granada: IRAM)

\bibitem[{{Cernicharo} {et~al.}(2019{\natexlab{a}}){Cernicharo}, {Cabezas},
  {Pardo}, {Ag{\'u}ndez}, {Berm{\'u}dez}, {Velilla-Prieto}, {Tercero},
  {L{\'o}pez-P{\'e}rez}, {Gallego}, {Fonfr{\'\i}a}, {Quintana-Lacaci},
  {Gu{\'e}lin}, \& {Endo}}]{Cernicharo2019b}
{Cernicharo}, J., {Cabezas}, C., {Pardo}, J.~R., {et~al.} 2019{\natexlab{a}},
  \aap, 630, L2

\bibitem[{{Cernicharo} {et~al.}(2019{\natexlab{b}}){Cernicharo}, {Gallego},
  {L{\'o}pez-P{\'e}rez}, {Tercero}, {Tanarro}, {Beltr{\'a}n}, {de Vicente},
  {Lauwaet}, {Alem{\'a}n}, {Moreno}, {Herrero}, {Dom{\'e}nech}, {Ram{\'\i}rez},
  {Berm{\'u}dez}, {Pel{\'a}ez}, {Patino-Esteban}, {L{\'o}pez-Fern{\'a}ndez},
  {Garc{\'\i}a-{\'A}lvaro}, {Garc{\'\i}a-Carre{\~n}o}, {Cabezas}, {Malo},
  {Amils}, {Sobrado}, {Diez-Gonz{\'a}lez}, {Hernand{\'e}z}, {Tercero},
  {Santoro}, {Mart{\'\i}nez}, {Castellanos}, {Vaquero Jim{\'e}nez}, {Pardo},
  {Barbas}, {L{\'o}pez-Fern{\'a}ndez}, {Aja}, {Leuther}, \&
  {Mart{\'\i}n-Gago}}]{Cernicharo2019a}
{Cernicharo}, J., {Gallego}, J.~D., {L{\'o}pez-P{\'e}rez}, J.~A., {et~al.}
  2019{\natexlab{b}}, \aap, 626, A34

\bibitem[{{Chu}(1983)}]{chu1983}
{Chu}, T.~S. 1983, IEEE Transactions on Antennas and Propagation, 31, 614

\bibitem[{{de Vicente}(1998)}]{devicente1998}
{de Vicente}, P. 1998, OAN Technical Report 1998-10

\bibitem[{{de Vicente} {et~al.}(2011){de Vicente}, {Bola{\~n}o}, \&
  {Barbas}}]{devicente11}
{de Vicente}, P., {Bola{\~n}o}, R., \& {Barbas}, L. 2011, in Highlights of
  Spanish Astrophysics VI, ed. M.~R. {Zapatero Osorio}, J.~{Gorgas},
  J.~{Ma{\'\i}z Apell{\'a}niz}, J.~R. {Pardo}, \& A.~{Gil de Paz}, 697--702

\bibitem[{{de Vicente} {et~al.}(2016){de Vicente}, {Bujarrabal},
  {D{\'\i}az-Pulido}, {Albo}, {Alcolea}, {Barcia}, {Barbas}, {Bola{\~n}o},
  {Colomer}, {Diez}, {Gallego}, {G{\'o}mez-Gonz{\'a}lez},
  {L{\'o}pez-Fern{\'a}ndez}, {L{\'o}pez-Fern{\'a}ndez}, {L{\'o}pez-P{\'e}rez},
  {Malo}, {Moreno}, {Patino}, {Serna}, {Tercero}, \& {Vaquero}}]{devicente16}
{de Vicente}, P., {Bujarrabal}, V., {D{\'\i}az-Pulido}, A., {et~al.} 2016,
  \aap, 589, A74

\bibitem[{{D{\'\i}az-Pulido} \& {de Vicente}(2014)}]{diazpulido14}
{D{\'\i}az-Pulido}, A. \& {de Vicente}, P. 2014, CDT Technical Report 2014-10

\bibitem[{{Diez} \& {Gallego}(2019{\natexlab{a}})}]{diez19a}
{Diez}, C. \& {Gallego}, J.~D. 2019{\natexlab{a}}, CDT Technical Report 2019-2

\bibitem[{{Diez} \& {Gallego}(2019{\natexlab{b}})}]{diez19b}
{Diez}, C. \& {Gallego}, J.~D. 2019{\natexlab{b}}, CDT Technical Report 2019-3

\bibitem[{{Dunning} {et~al.}(2001){Dunning}, J., \& {Kerr}}]{Dunning2009}
{Dunning}, A., J., S.~S., \& {Kerr}, A. 2001, IEEE Transactions on Antennas and
  Propagation, 49, 1683

\bibitem[{{Fuente} {et~al.}(2019){Fuente}, {Navarro}, {Caselli}, {Gerin},
  {Kramer}, {Roueff}, {Alonso-Albi}, {Bachiller}, {Cazaux}, {Commercon},
  {Friesen}, {Garc{\'\i}a-Burillo}, {Giuliano}, {Goicoechea}, {Gratier},
  {Hacar}, {Jim{\'e}nez-Serra}, {Kirk}, {Lattanzi}, {Loison}, {Malinen},
  {Marcelino}, {Mart{\'\i}n-Dom{\'e}nech}, {Mu{\~n}oz-Caro}, {Pineda},
  {Tafalla}, {Tercero}, {Ward-Thompson}, {Trevi{\~n}o-Morales},
  {Rivi{\'e}re-Marichalar}, {Roncero}, {Vidal}, \& {Ballester}}]{fuente2019}
{Fuente}, A., {Navarro}, D.~G., {Caselli}, P., {et~al.} 2019, \aap, 624, A105

\bibitem[{{Gallego} {et~al.}(2006){Gallego}, {L{\'o}pez}, {Diez}, \&
  {Barcia}}]{gallego2006}
{Gallego}, J.~D., {L{\'o}pez}, I., {Diez}, C., \& {Barcia}, A. 2006, ALMA MEMO
  560

\bibitem[{{Garc{\'\i}a-P{\'e}rez} {et~al.}(2017){Garc{\'\i}a-P{\'e}rez},
  {Tercero}, \& {L{\'o}pez-Ruiz}}]{Garcia2017}
{Garc{\'\i}a-P{\'e}rez}, O., {Tercero}, F., \& {L{\'o}pez-Ruiz}, S. 2017, CDT
  Technical Report 2017-9

\bibitem[{{Garc{\'\i}a-P{\'e}rez} {et~al.}(2018){Garc{\'\i}a-P{\'e}rez},
  {Tercero}, \& {L{\'o}pez-Ruiz}}]{garciaperez2018}
{Garc{\'\i}a-P{\'e}rez}, O., {Tercero}, F., \& {L{\'o}pez-Ruiz}, S. 2018, CDT
  Technical Report 2018-9

\bibitem[{{Goldsmith}(1998)}]{Goldsmith1998}
{Goldsmith}, P.~F. 1998, {Quasioptical Systems}

\bibitem[{{Greve} {et~al.}(1998){Greve}, {Kramer}, \& {Wild}}]{Greve1998}
{Greve}, A., {Kramer}, C., \& {Wild}, W. 1998, \aaps, 133, 271

\bibitem[{{Herpin} {et~al.}(2002){Herpin}, {Goicoechea}, {Pardo}, \&
  {Cernicharo}}]{Herpin2002}
{Herpin}, F., {Goicoechea}, J.~R., {Pardo}, J.~R., \& {Cernicharo}, J. 2002,
  \apj, 577, 961

\bibitem[{{Issaoun} {et~al.}(2019){Issaoun}, {Johnson}, {Blackburn},
  {Brinkerink}, {Mo{\'s}cibrodzka}, {Chael}, {Goddi}, {Mart{\'\i}-Vidal},
  {Wagner}, {Doeleman}, {Falcke}, {Krichbaum}, {Akiyama}, {Bach}, {Bouman},
  {Bower}, {Broderick}, {Cho}, {Crew}, {Dexter}, {Fish}, {Gold}, {G{\'o}mez},
  {Hada}, {Hern{\'a}ndez-G{\'o}mez}, {Jan{\ss}en}, {Kino}, {Kramer}, {Loinard},
  {Lu}, {Markoff}, {Marrone}, {Matthews}, {Moran}, {M{\"u}ller}, {Roelofs},
  {Ros}, {Rottmann}, {Sanchez}, {Tilanus}, {de Vicente}, {Wielgus}, {Zensus},
  \& {Zhao}}]{issaoun19}
{Issaoun}, S., {Johnson}, M.~D., {Blackburn}, L., {et~al.} 2019, \apj, 871, 30

\bibitem[{{Jewell}(2002)}]{Jewell2002}
{Jewell}, P.~R. 2002, in Astronomical Society of the Pacific Conference Series,
  Vol. 278, Single-Dish Radio Astronomy: Techniques and Applications, ed.
  S.~{Stanimirovic}, D.~{Altschuler}, P.~{Goldsmith}, \& C.~{Salter}, 313--328

\bibitem[{{Kawaguchi} {et~al.}(1995){Kawaguchi}, {Kasai}, {Ishikawa}, \&
  {Kaifu}}]{Kawaguchi1995}
{Kawaguchi}, K., {Kasai}, Y., {Ishikawa}, S.-I., \& {Kaifu}, N. 1995, \pasj,
  47, 853

\bibitem[{{Klein} {et~al.}(2012){Klein}, {Hochg{\"u}rtel}, {Kr{\"a}mer},
  {Bell}, {Meyer}, \& {G{\"u}sten}}]{Klein12}
{Klein}, B., {Hochg{\"u}rtel}, S., {Kr{\"a}mer}, I., {et~al.} 2012, \aap, 542,
  L3

\bibitem[{{Kramer}(1997)}]{Kramer1997}
{Kramer}, C. 1997, Internal IRAM Report (Granada: IRAM)

\bibitem[{{Kramer} {et~al.}(2013){Kramer}, {Pe{\~n}alver}, \&
  {Greve}}]{Kramer2013}
{Kramer}, C., {Pe{\~n}alver}, J., \& {Greve}, A. 2013, Internal IRAM Report
  (Granada: IRAM)

\bibitem[{{Krotikov} \& {Pelyushenko}(1987)}]{Krotikov1987}
{Krotikov}, V.~D. \& {Pelyushenko}, S.~A. 1987, \sovast, 31, 216

\bibitem[{{Leuther} {et~al.}(2009){Leuther}, {Tessmann}, {Kalfass},
  {L{\"o}sch}, {Seelmann-Eggebert}, {Wadefalk}, {Sch{\"a}fer}, {Gallego Puyol},
  {Schlechtweg}, {Mikulla}, \& {Ambacher}}]{leuther09}
{Leuther}, A., {Tessmann}, A., {Kalfass}, I., {et~al.} 2009, in IEEE
  International Conference on Indium Phosphide and Related Materials, 188--191

\bibitem[{{L{\'o}pez Fern{\'a}ndez} {et~al.}(2006){L{\'o}pez Fern{\'a}ndez},
  {G{\'o}mez Gonz{\'a}lez}, \& {Barc{\'\i}a C{\'a}ncio}}]{lopezfernandez2006}
{L{\'o}pez Fern{\'a}ndez}, J.~A., {G{\'o}mez Gonz{\'a}lez}, J., \& {Barc{\'\i}a
  C{\'a}ncio}, A. 2006, {Radio Telescope Engineering: the 40m IGN Antenna}, ed.
  A.~{Ulla} \& M.~{Manteiga}, Vol.~2, 257--270

\bibitem[{{L{\'o}pez-P{\'e}rez}(2012)}]{lopezperez12}
{L{\'o}pez-P{\'e}rez}, J.~A. 2012, Thesis, Escuela Técnica Superior de
  Ingenieros de Telecomunicación (ETSIT). Universidad Politécnica de Madrid
  (UPM).

\bibitem[{L{\'o}pez-P{\'e}rez {et~al.}(2014)L{\'o}pez-P{\'e}rez, de~Vicente,
  L{\'o}pez-Fern{\'a}ndez, Tercero~Martinez, Barcia~Cancio, \&
  Galocha~Iraguen}]{lopezperez14}
L{\'o}pez-P{\'e}rez, J.~A., de~Vicente, P., L{\'o}pez-Fern{\'a}ndez, J.~A.,
  {et~al.} 2014, IEEE Transactions on Antennas and Propagation, 62, 2624

\bibitem[{{L{\'o}pez-Ruiz} {et~al.}(2017){L{\'o}pez-Ruiz}, {Tercero},
  {Nu{\~n}ez}, \& {L{\'o}pez-Fern{\'a}ndez}}]{lopezruiz2017}
{L{\'o}pez-Ruiz}, S., {Tercero}, F., {Nu{\~n}ez}, M.~G., \&
  {L{\'o}pez-Fern{\'a}ndez}, J.~A. 2017, CDT Technical Report 2017-1

\bibitem[{{Malo} {et~al.}(2016){Malo}, {Gallego}, {Amils}, {Garc{\'\i}a},
  {Diez}, {L{\'o}pez}, \& {Barcia}}]{malo16}
{Malo}, I., {Gallego}, J.~D., {Amils}, R., {et~al.} 2016, CDT Technical Report
  2016-6

\bibitem[{{Mangum} {et~al.}(2007){Mangum}, {Emerson}, \&
  {Greisen}}]{Mangum2007}
{Mangum}, J.~G., {Emerson}, D.~T., \& {Greisen}, E.~W. 2007, \aap, 474, 679

\bibitem[{{Muders} {et~al.}(2011){Muders}, {Polehapmton}, \&
  {Hatchell}}]{muders2011}
{Muders}, D., {Polehapmton}, E., \& {Hatchell}, J. 2011, APEX Internal report
  APEX-MPI-ICD-0002

\bibitem[{{Navarrini} \& {Plambeck}(2006)}]{Navarrini2006}
{Navarrini}, A. \& {Plambeck}, R.~L. 2006, IEEE Transactions on Microwave
  Theory Techniques, 54, 272

\bibitem[{{O'Neil}(2002)}]{oneil02}
{O'Neil}, K. 2002, in Astronomical Society of the Pacific Conference Series,
  Vol. 278, Single-Dish Radio Astronomy: Techniques and Applications, ed.
  S.~{Stanimirovic}, D.~{Altschuler}, P.~{Goldsmith}, \& C.~{Salter}, 293--311

\bibitem[{{Pardo} {et~al.}(2020){Pardo}, {Berm{\'u}dez}, {Cabezas},
  {Ag{\'u}ndez}, {Gallego}, {Fonfr{\'\i}a}, {Velilla-Prieto},
  {Quintana-Lacaci}, {Tercero}, {Gu{\'e}lin}, \& {Cernicharo}}]{Pardo2020}
{Pardo}, J.~R., {Berm{\'u}dez}, C., {Cabezas}, C., {et~al.} 2020, \aap, 640,
  L13

\bibitem[{{Pardo} {et~al.}(2001){Pardo}, {Cernicharo}, \&
  {Serabyn}}]{Pardo2001}
{Pardo}, J.~R., {Cernicharo}, J., \& {Serabyn}, E. 2001, IEEE Transactions on
  Antennas and Propagation, 49, 1683

\bibitem[{{Pardo} {et~al.}(2005){Pardo}, {Serabyn}, \& {Wiedner}}]{Pardo2005}
{Pardo}, J.~R., {Serabyn}, E., \& {Wiedner}, M. 2005, Icarus, 178, 19

\bibitem[{{Pardo} {et~al.}(2017){Pardo}, {Serabyn}, {Wiedner}, {Moreno}, \&
  {Orton}}]{Pardo2017}
{Pardo}, J.~R., {Serabyn}, E., {Wiedner}, M., {Moreno}, R., \& {Orton}, G.
  2017, Icarus, 290, 150

\bibitem[{{Ruze}(1966)}]{Ruze1966}
{Ruze}, J. 1966, Proceedings of the IEEE, 54, 633

\bibitem[{{Ruze}(1968)}]{ruze1968}
{Ruze}, J. 1968, Microwave Journal, 11, 76

\bibitem[{{Tercero} {et~al.}(2020){Tercero}, {Cernicharo}, {Cuadrado}, {de
  Vicente}, \& {Gu{\'e}lin}}]{Tercero2020}
{Tercero}, B., {Cernicharo}, J., {Cuadrado}, S., {de Vicente}, P., \&
  {Gu{\'e}lin}, M. 2020, \aap, 636, L7

\bibitem[{{Tercero} \& {Garc{\'\i}a-P{\'e}rez}(2018)}]{Tercero2018}
{Tercero}, F. \& {Garc{\'\i}a-P{\'e}rez}, O. 2018, CDT Technical Report 2018-08

\bibitem[{{Tercero} \& {Garc{\'\i}a-P{\'e}rez}(2019)}]{tercero2019}
{Tercero}, F. \& {Garc{\'\i}a-P{\'e}rez}, O. 2019, in International Conference
  on Electromagnetics in Advanced Applications (ICEAA), 0719--0724

\bibitem[{{Weinreb} \& {Schleeh}(2014)}]{weinreb2014}
{Weinreb}, S. \& {Schleeh}, J. 2014, IEEE Transactions on Microwave Theory
  Techniques, 62, 83

\bibitem[{{Winkel} {et~al.}(2012){Winkel}, {Kraus}, \& {Bach}}]{Winkel2012}
{Winkel}, B., {Kraus}, A., \& {Bach}, U. 2012, \aap, 540, A140

\end{thebibliography}

\begin{appendix}

\section{Complementary Figures}

\begin{figure*}
\includegraphics[angle=0,width=16cm]{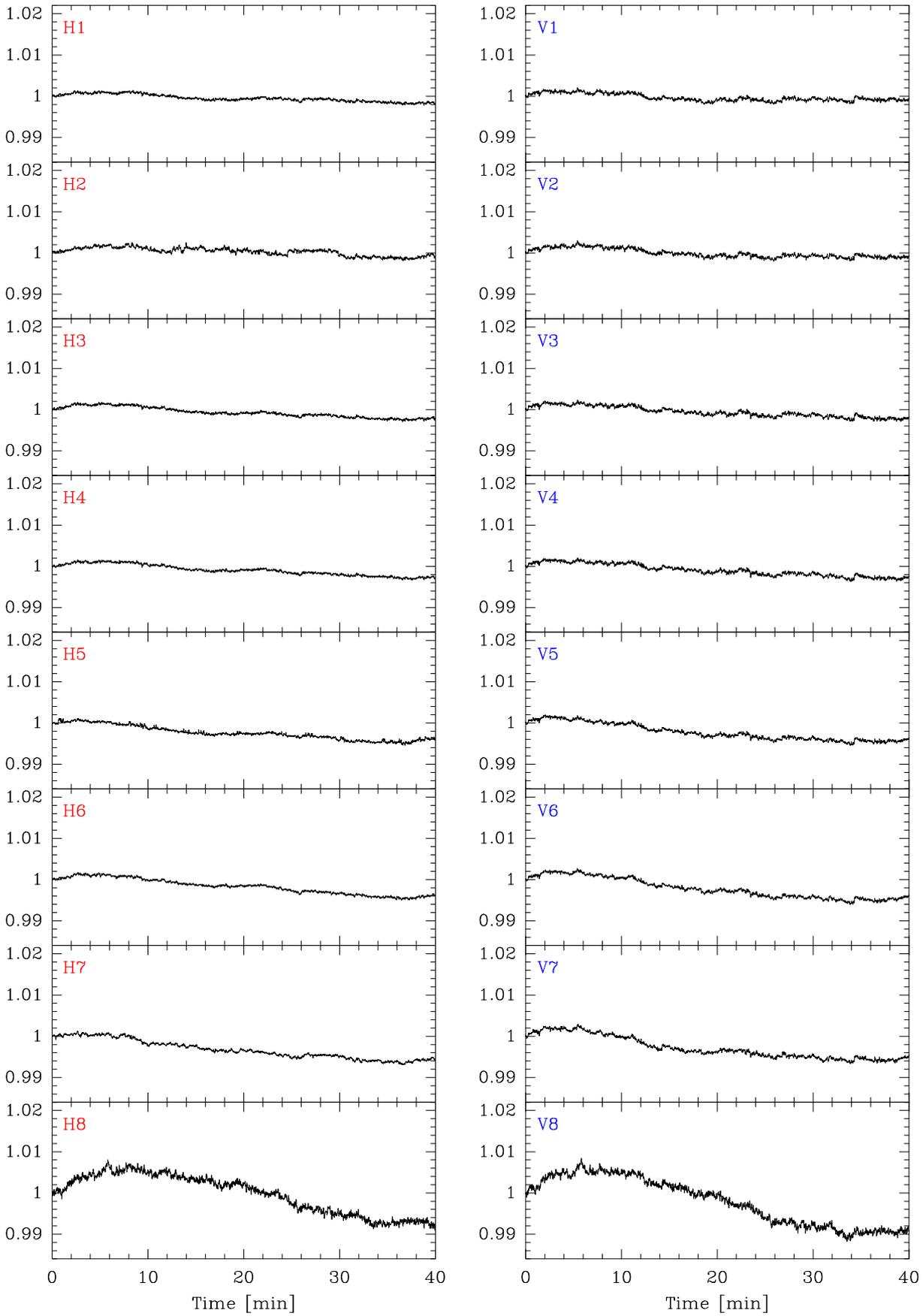}
\caption{Normalised averaged FFTS backend counts (2.5\,GHz) per time registered in observations
of the hot load at Q band.
The first column shows the values of the H polarisation whereas the second
depicts the V polarisation. Central frequencies for the different FFTS
sections are: 32.5, 34.7, 37.0, 39.3, 41.6, 43.9, 46.2, and 48.5\,GHz for
sections 1, 2, 3, 4, 5, 6, 7, and 8, respectively.}
\label{fig_estabilidad_Q}
\end{figure*}

\begin{figure*}
\includegraphics[angle=0,width=16cm]{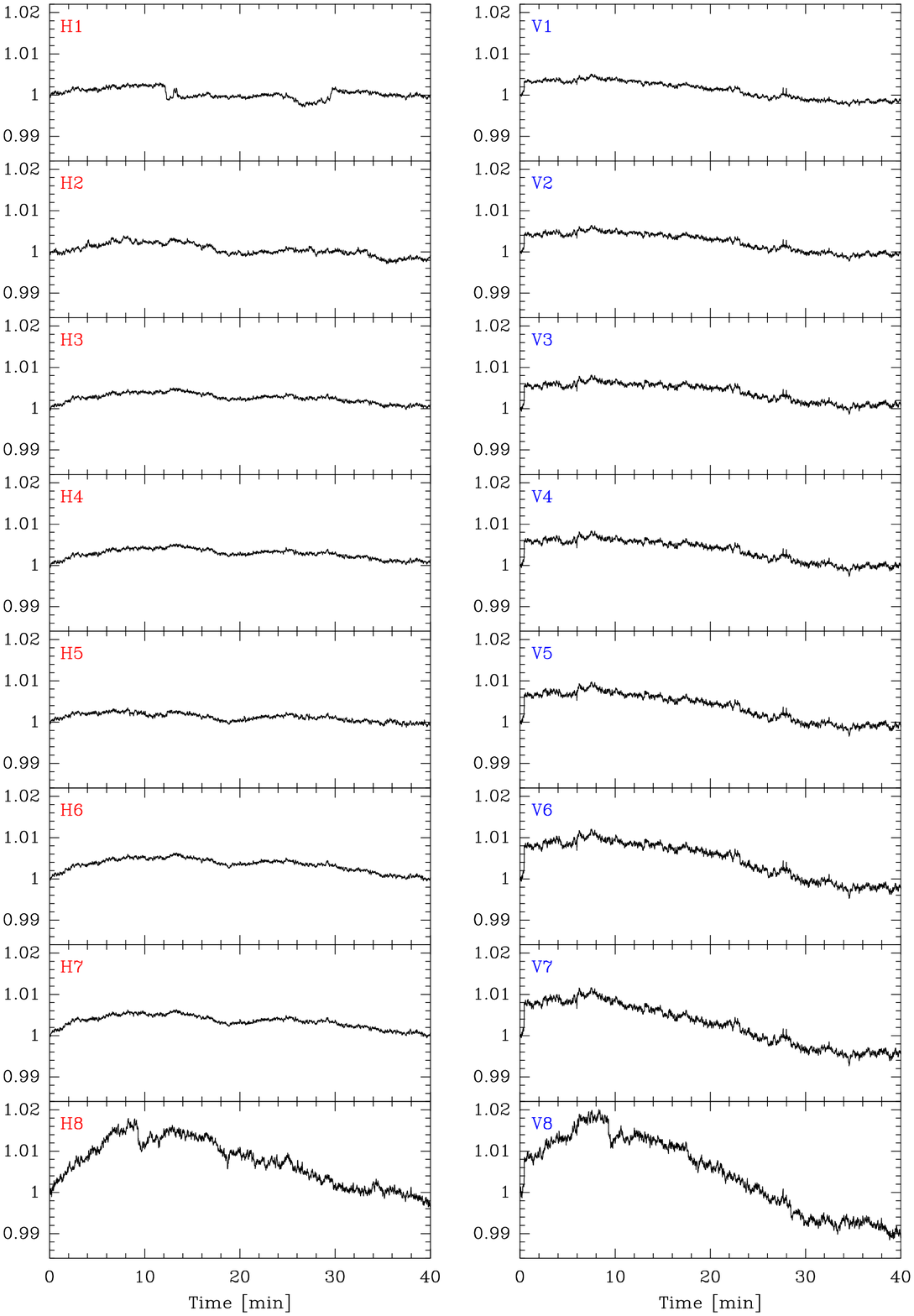}
\caption{Normalised averaged FFTS backend counts (2.5\,GHz) per time registered in observations
of the hot load at W band.
The first column shows the values of the H polarisation whereas the second
depicts the V polarisation. Central frequencies for the different FFTS
sections are: 72.5, 74.7, 77.0, 79.3, 81.6, 83.9, 86.2, and 88.5\,GHz for
sections 1, 2, 3, 4, 5, 6, 7, and 8, respectively.}
\label{fig_estabilidad_W}
\end{figure*}

\begin{figure*}[tbh]
\centerline{\includegraphics[scale=0.3]{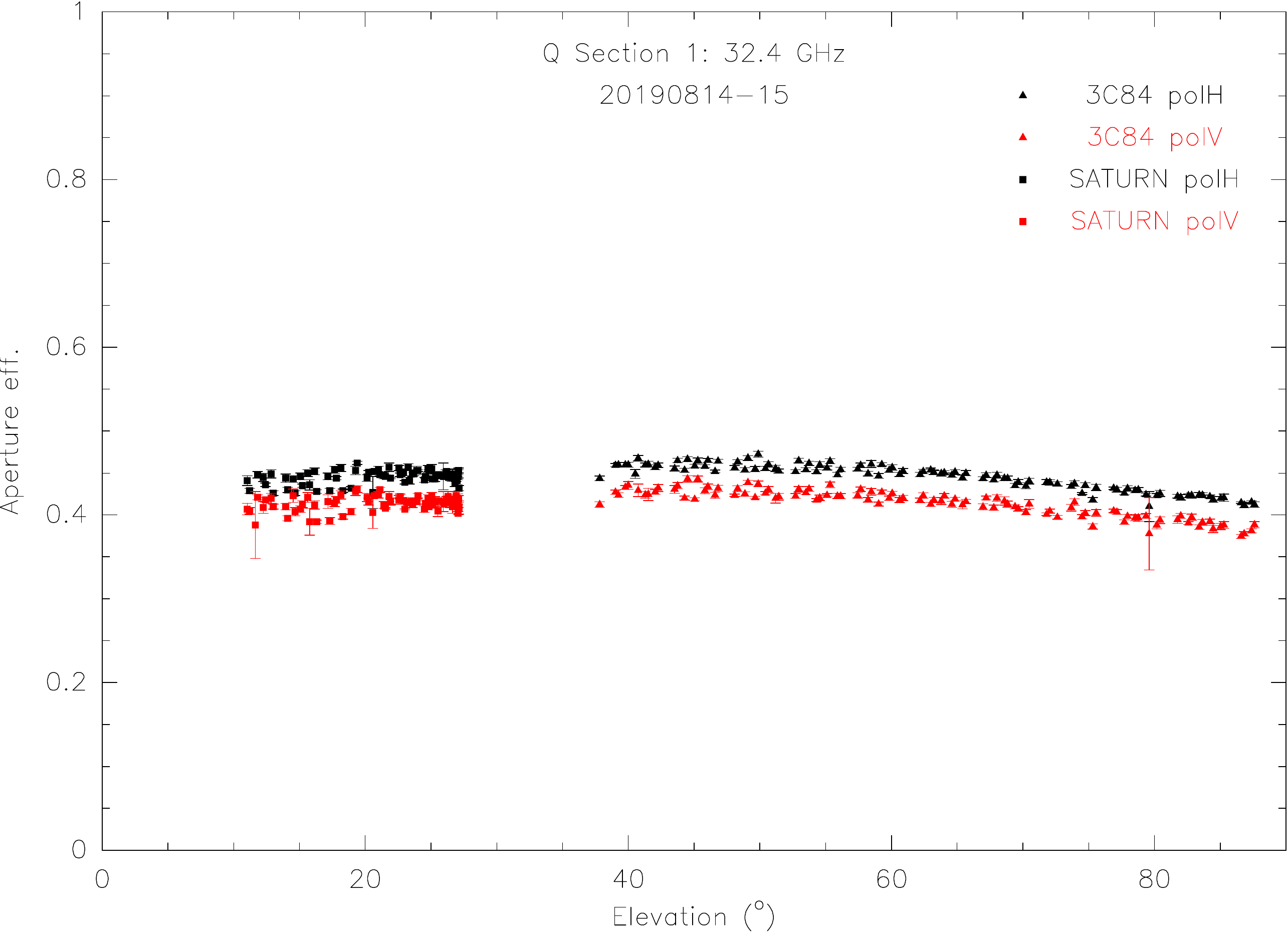}\includegraphics[scale=0.3]{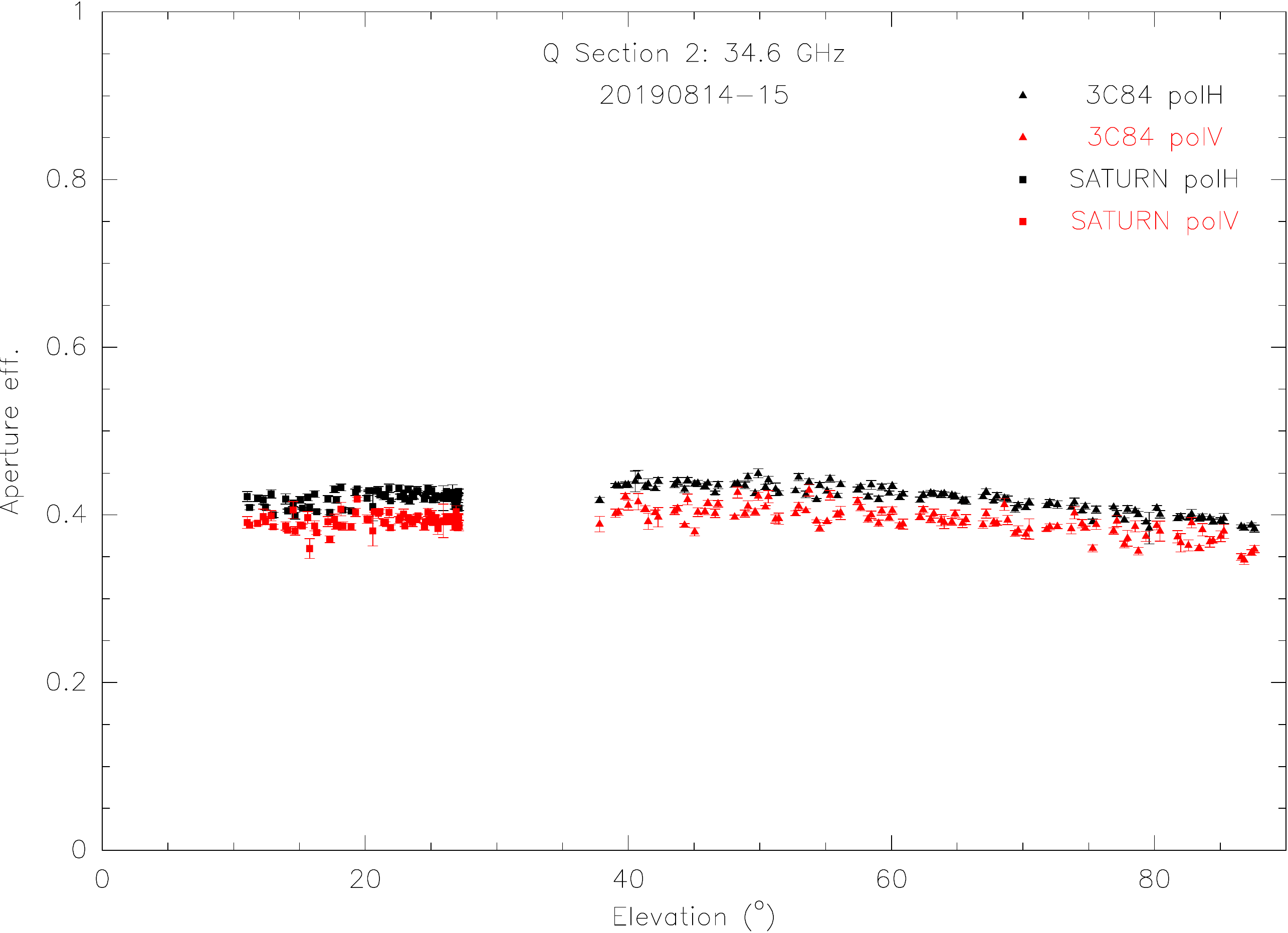}}
\centerline{\includegraphics[scale=0.3]{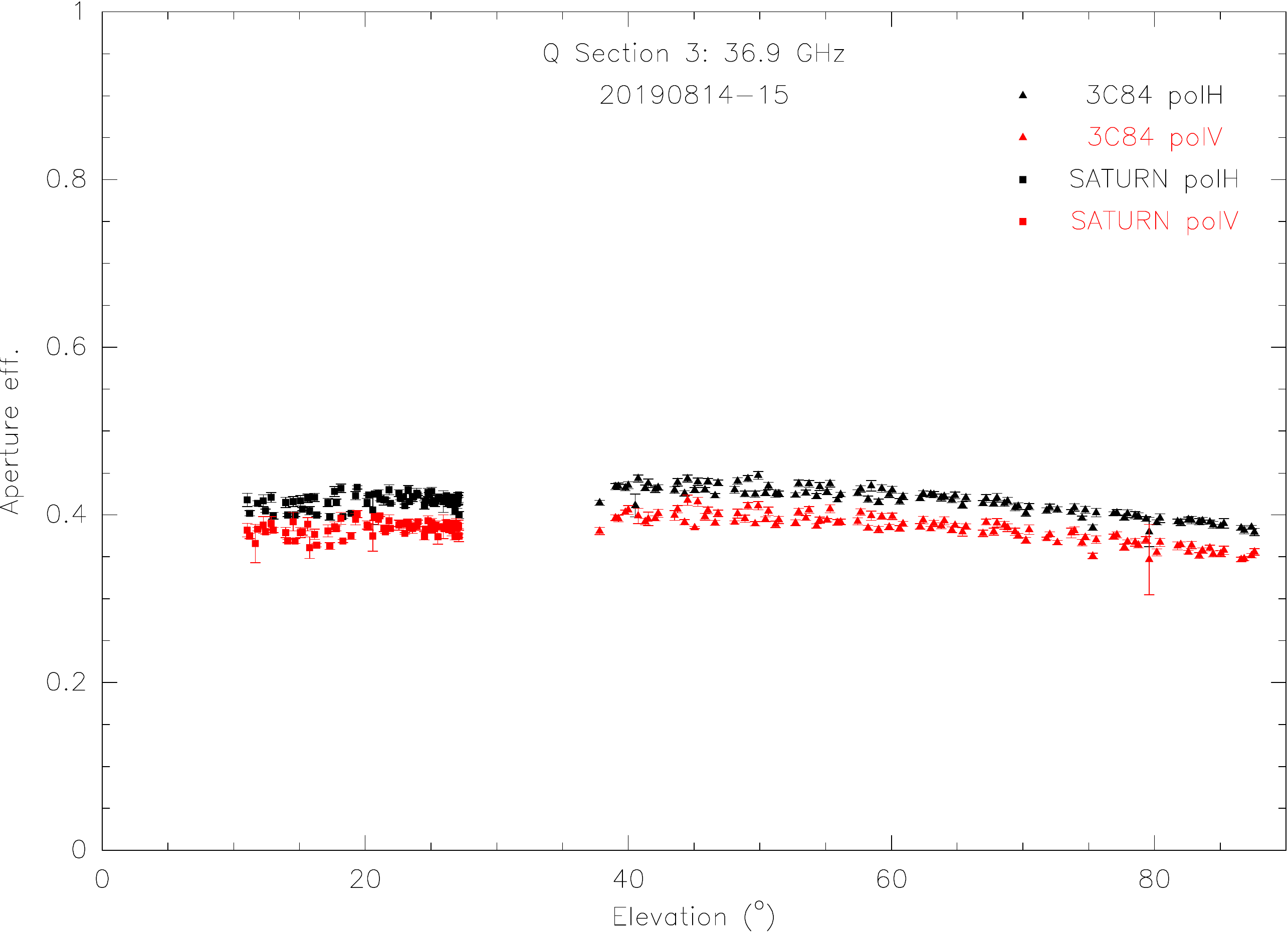}\includegraphics[scale=0.3]{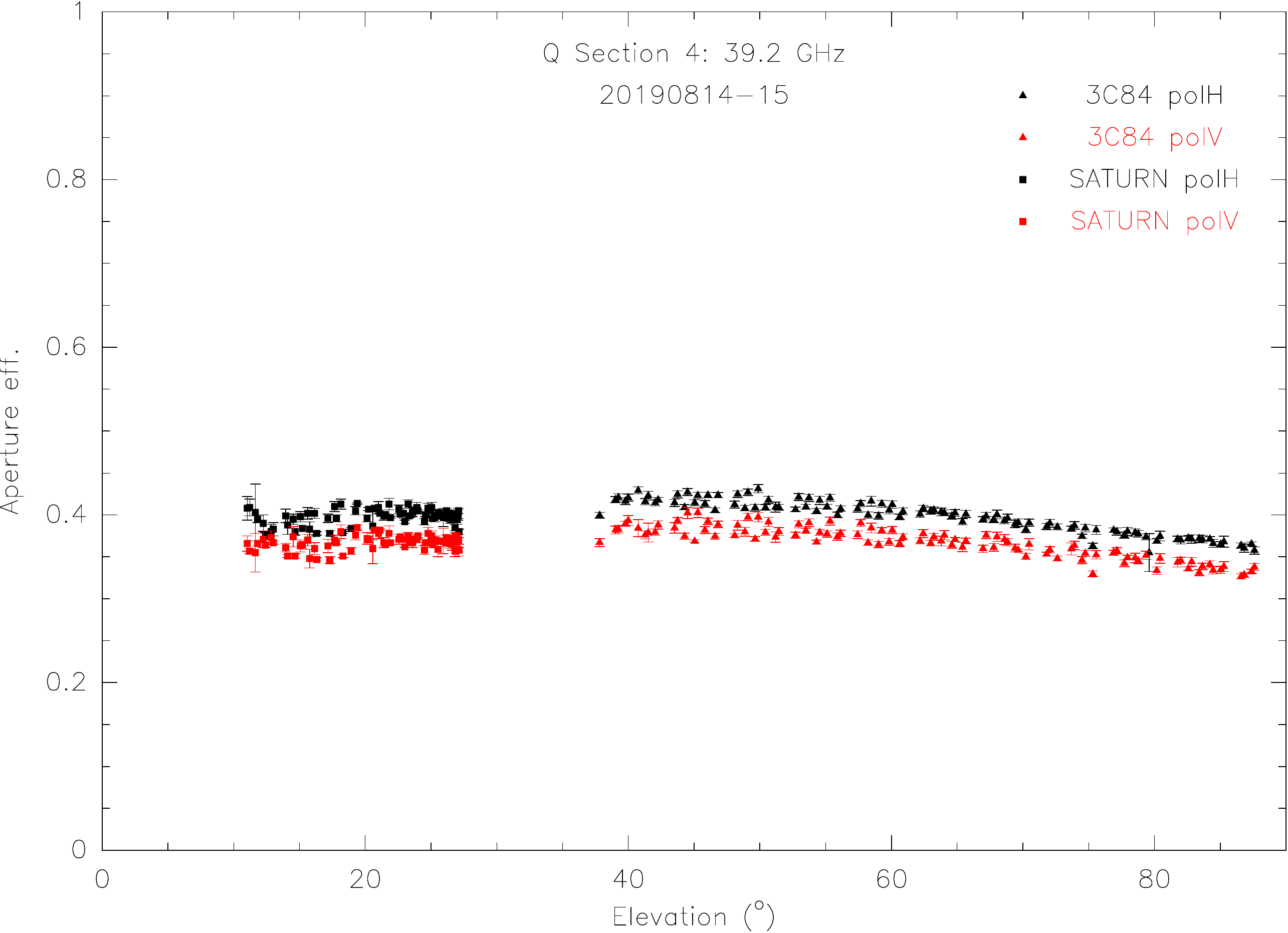}}
\centerline{\includegraphics[scale=0.3]{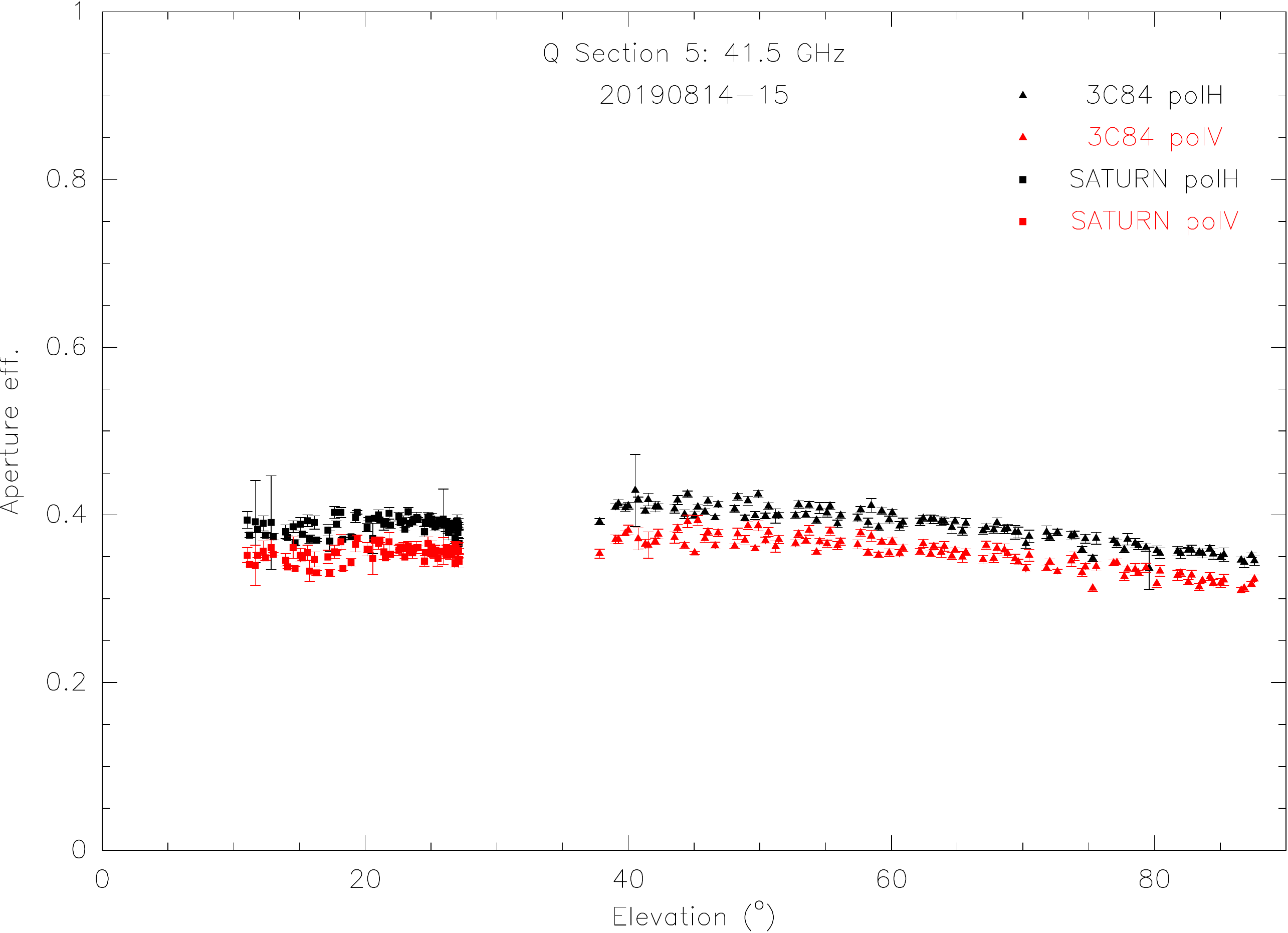}\includegraphics[scale=0.3]{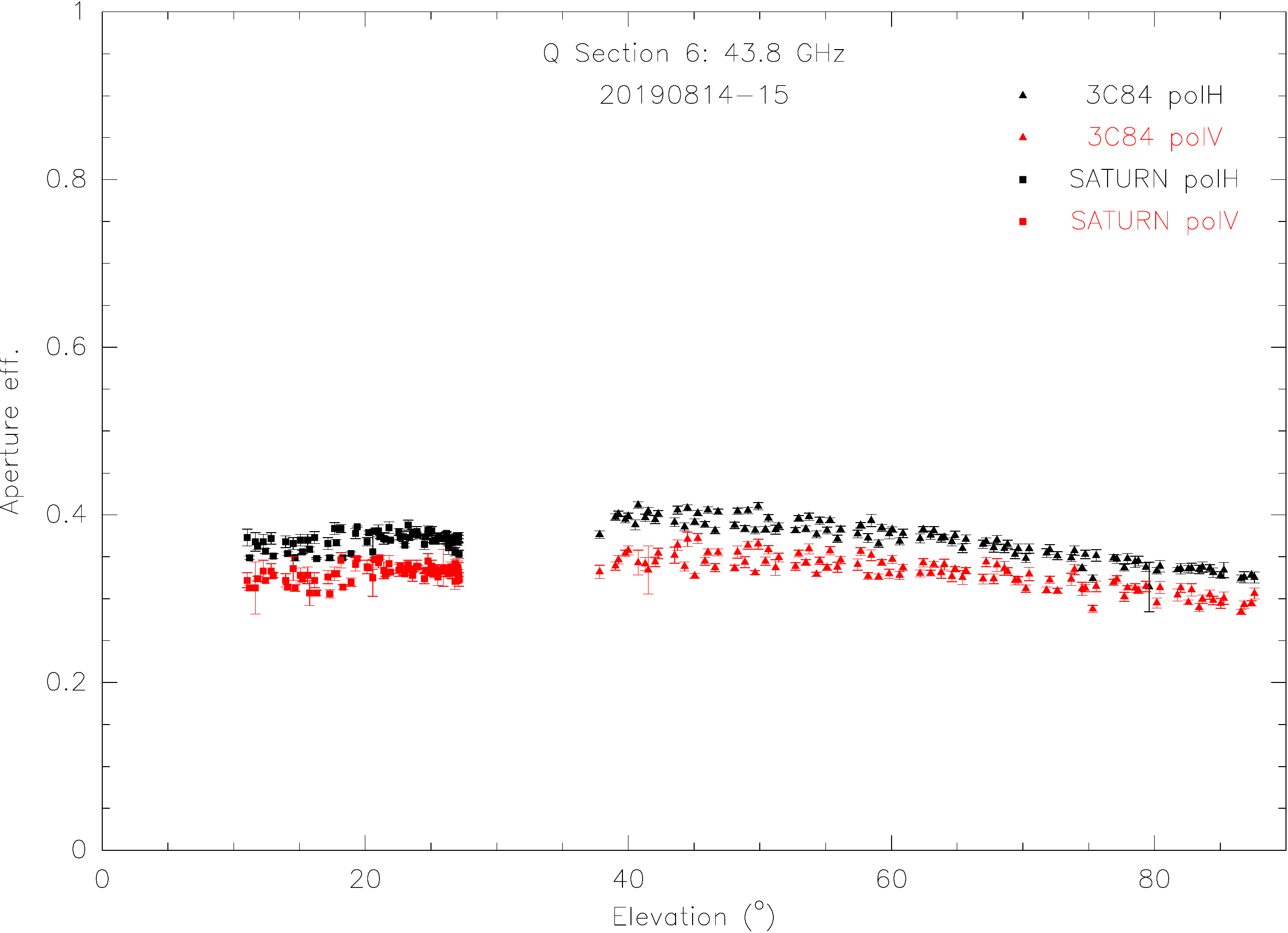}}
\centerline{\includegraphics[scale=0.3]{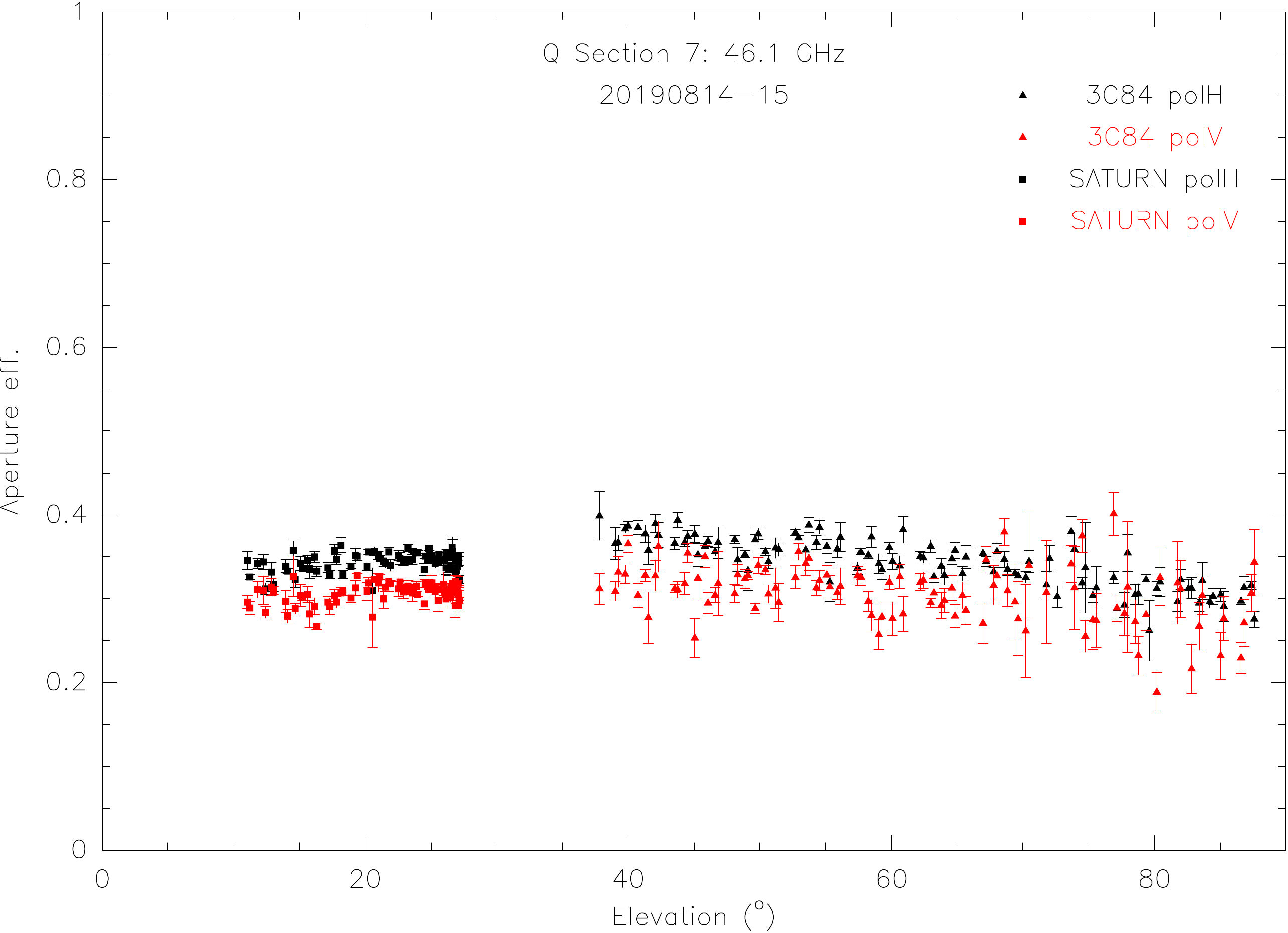}\includegraphics[scale=0.3]{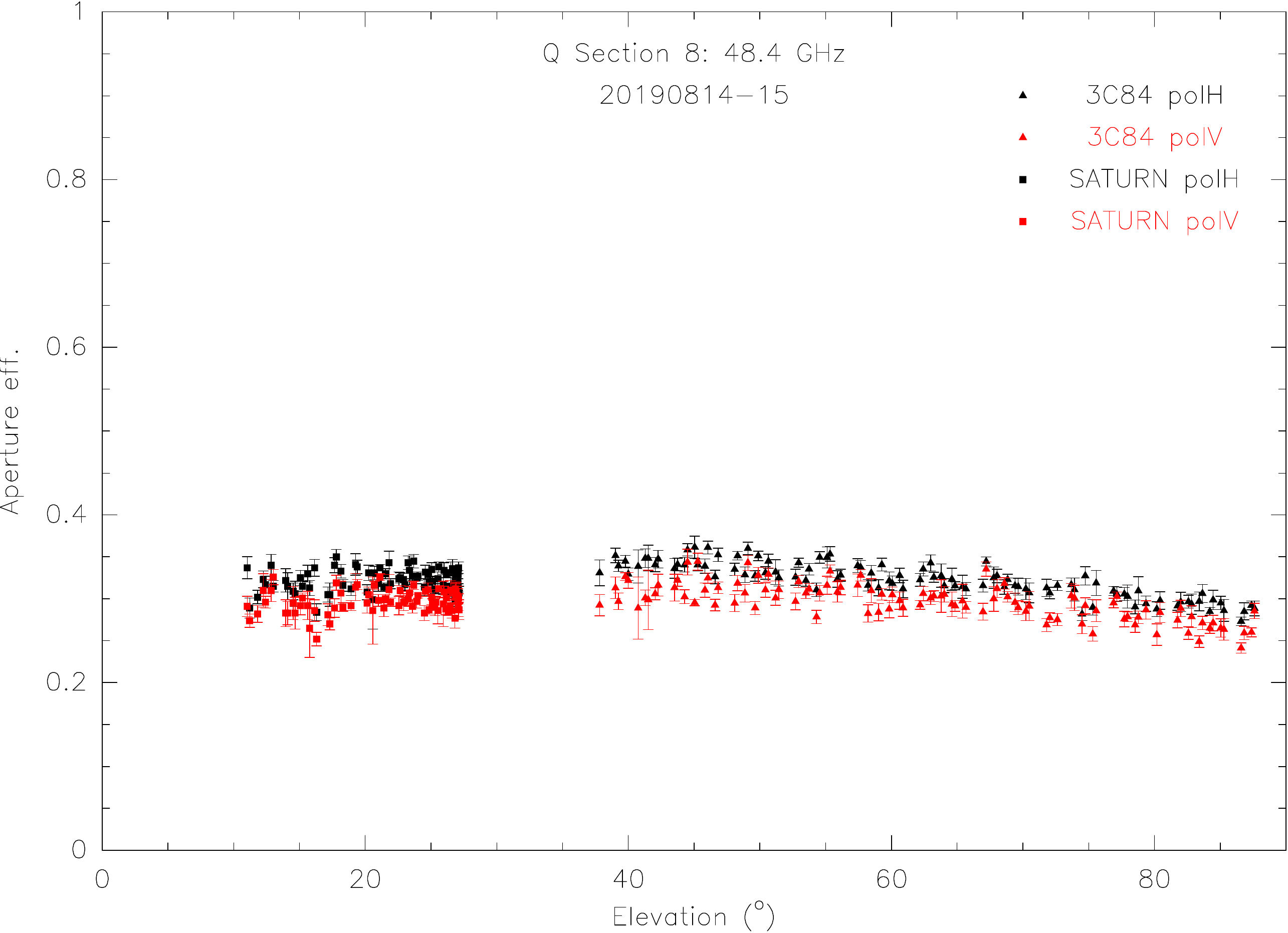}}
\caption{Aperture efficiency versus elevation in each of the FFTS sections along the Q band, using Saturn and 3C84 as calibrators. Measurements in the horizontal (vertical) polarisation are shown in black (red).
The central frequency of each section is displayed at the top of each graph.}
\label{f.eficiencias}
\end{figure*}

\end{appendix}

\end{document}